\documentclass[twocolumn,twocolappendix]{aastex631}

\usepackage{booktabs}
\usepackage{multirow}
\usepackage{graphicx}
\usepackage{amsmath}
\usepackage{hyperref}

\usepackage{xspace}
\usepackage{multirow}
\usepackage{array}
\usepackage{booktabs} 
\usepackage{CJK}
\hypersetup{colorlinks, linkcolor={blue}, citecolor={blue}, urlcolor={blue}}

\newcommand\nustar{\textit{NuSTAR}\xspace}

\newcommand\swift{\textit{Swift}\xspace}
\newcommand\swiftall{\textit{Neil Gehrels Swift Observatory}\xspace}
\newcommand\xmm{\textit{XMM-Newton}\xspace}

\newcommand\neowise{\textit{NEOWISE}\xspace}
\newcommand\srg{\textit{SRG}\xspace}
\newcommand\srge{\textit{SRG}/eROSITA\xspace}
\newcommand\rosat{\textit{ROSAT}\xspace}

\newcommand{\kms}{km s$^{-1}$\xspace}
\newcommand{\msun}{$M_{\odot}$\xspace}
\newcommand{\mstar}{$M_{*}$\xspace}
\newcommand{\ergcms}{erg cm$^{-2}$ s$^{-1}$\xspace}
\newcommand{\ergs}{erg s$^{-1}$\xspace}
\newcommand{\mbh}{$M_{\rm BH}$\xspace}
\newcommand{\lx}{$L_{\rm X}$\xspace}
\newcommand{\lbb}{$L_{\rm BB}$\xspace}

\newcommand{\hr}{HR\xspace}

\newcommand{\lbblx}{$L_{\rm BB}/L_{\rm X}$\xspace}

\shorttitle{X-ray emission in optical TDEs}
\shortauthors{Guolo et al.}
\graphicspath{{./}{figures/}}

\begin{document}


\title{A systematic analysis of the X-ray emission in optically selected tidal disruption events:\\ observational evidence for the unification of the optically and X-ray selected populations}

\author[0000-0002-5063-0751]{Muryel Guolo}
\affiliation{Bloomberg Center for Physics and Astronomy, Johns Hopkins University, 3400 N. Charles St., Baltimore, MD 21218, USA}

\author[0000-0003-3703-5154]{Suvi Gezari}
\affiliation{Space Telescope Science Institute, 3700 San Martin Drive, Baltimore, MD 21218, USA}
\affiliation{Bloomberg Center for Physics and Astronomy, Johns Hopkins University, 3400 N. Charles St., Baltimore, MD 21218, USA}

\author[0000-0001-6747-8509]{Yuhan Yao}
\affiliation{Cahill Center for Astrophysics, California Institute of Technology, MC 249-17, 1200 E California Boulevard, Pasadena, CA 91125, USA}
\affiliation{Miller Institute for Basic Research in Science, 468 Donner Lab, Berkeley, CA 94720, USA}
\affiliation{Department of Astronomy, University of California, Berkeley, CA 94720, USA}

\author[0000-0002-3859-8074]{Sjoert van Velzen}
\affiliation{Leiden Observatory, Leiden University, PO Box 9513, 2300 RA Leiden, The Netherlands}

\author[0000-0002-5698-8703]{Erica Hammerstein}
\affil{Department of Astronomy, University of Maryland, College Park, MD 20742, USA}
\affiliation{Astrophysics Science Division, NASA Goddard Space Flight Center, Mail Code 661, Greenbelt, MD 20771, USA}
\affil{Center for Research and Exploration in Space Science and Technology, NASA/GSFC, Greenbelt, MD 20771, USA}

\author[0000-0003-1673-970X]{S. Bradley Cenko}
\affiliation{Astrophysics Science Division, NASA Goddard Space Flight Center, Mail Code 661, Greenbelt, MD 20771, USA}
\affiliation{Joint Space-Science Institute, University of Maryland, College Park, MD 20742, USA}

\author[0000-0002-0430-5798]{Yarone M. Tokayer}
\affiliation{Department of Physics, Yale University, P.O. Box 208120, New Haven, CT 06520, USA}



\begin{abstract}

We present a systematic analysis of the X-ray emission of a sample of 17 optically selected, X-ray-detected tidal disruption events (TDEs) discovered between 2014 and 2021.
The X-ray light curves show a diverse range of temporal behaviors, with most sources not following the expected power-law decline.
The X-ray spectra are mostly extremely soft and consistent with thermal emission from the innermost region of an accretion disk, which cools as the accretion rate decreases. Three sources show formation of a hard X-ray corona, at late-times. 
The spectral energy distribution shape, probed by the ratio ($L_{\rm\,BB}/L_{\rm\,X}$) between the UV/optical and X-ray, shows a wide range $L_{\rm BB}/L_{\rm X}\,\in\,(0.5,\,3000)$ at early-times, and converges to disk-like values $L_{\rm\,BB}/L_{\rm\,X}\,\in\,(0.5,\,10)$ at late-times. 
We estimate the fraction of optically discovered TDEs with $L_{\rm\,X}\,\geq 10^{42}~\rm{erg}~\rm{s}^{-1} $ to be at least $40\%$, and show that X-ray loudness
is independent of black hole mass. 
We argue that distinct disk formation time scales are unlikely to fully explain the diverse range of X-ray evolutions.
We combine our sample with X-ray discovered ones
to construct an X-ray luminosity function, best fitted by a broken power-law, with a break at \lx $ \approx 10^{44}$ \ergs.
We show that there is no dichotomy between optically and X-ray selected TDEs, instead there is a continuum of early time $L_{\rm\,BB}/L_{\rm\,X}$, at least as wide as $L_{\rm\,BB}/L_{\rm\,X}\,\in\,(0.1,\,3000)$, with optical/X-ray surveys selecting preferentially, but not exclusively,
from the higher/lower end of the distribution. Our findings are consistent with unification models for the overall TDE population.

\end{abstract}



\section{Introduction} \label{sec:intro}

The occasional tidal disruption of a star that approaches close enough to a massive black hole (MBH) was predicted by theorists as a signpost for MBHs lurking in the center of galaxies \citep[][]{Hills1975,Rees1988,Ulmer1999}. These luminous events, called tidal disruption events (TDE), are  observed throughout the entire electromagnetic spectrum, and are now a well-established
class of transients \citep[see recent review by][]{Gezari2021}.
TDEs are a unique opportunity to probe the existence of quiescent black holes in the low mass end of the MBH's mass function ($< 10^{8}$ \msun).  At higher black hole masses, a TDE is not observable; stars are swallowed whole since the tidal disruption radius lies inside the event horizon \citep[][]{Hills1975,vanVelzen2018,Yao2023}. 

The first observational evidence for TDEs came from the detection of X-ray flares from the centers of quiescent galaxies during the \rosat all-sky survey (RASS) in 1990–1991 \citep[e.g.,][]{Bade1996,Grupe1999,Komossa1999,greiner2000}. The flares exhibited soft spectra with temperatures  $T \sim 10^6$ K \citep[for a review on X-ray selected TDEs, see][]{Saxton2020}. Since 2020, the Spektrum-Roentgen-Gamma (\srg) mission \citep{Sunyaev2021}, with its sensitive eROSITA telescope \citep{Predehl2021}, and six-month cadenced all-sky surveys, has become the most prolific discoverer of TDEs in X-rays, presenting thirteen new sources discovered during the first two All-sky scans \citep{Sazonov2021}.

The discovery of TDEs has increased dramatically in the last few years due to the operation of wide-field optical surveys, such as iPTF
\citep{Blagorodnova2017,Blagorodnova2019,Hung2017}, Pan-STARRS \citep{Gezari2012,Chornock2014,holoien2019}, ASASSN
\citep{Holoien2014,Holoien2016_14li,Holoien2016_15oi,Wevers2019_18fyk}, and ZTF \citep{vanVelzen2019,vanVelzen2021,Hammerstein2022,Yao2023};
with ZTF now dominating the number of discoveries with a rate of $\sim$ 10 per year \citep{Hammerstein2022,Yao2023}. Although, the
number of optically selected TDEs dominates over the ones discovered by means of high-energy observations, the nature of what is powering
their bright optical flares is uncertain. Unlike the soft X-ray component detected in some optically selected TDEs -- which is similar to the X-ray selected TDEs and is consistent with thermal emission from the inner radii of an accretion disk -- the UV/optical component seems, in most cases, not consistent with the direct emission from the Rayleigh-Jeans tail of the expected disk to form from the circularization of the stellar debris streams around a $10^{5}-10^{8}$ \msun black hole.
 This implies the existence of an unknown, larger emitting structure that competing interpretations invoke to be either produced as a result of reprocessing of the extreme UV and X-ray emission \citep{Miller2015,Dai2018,parkinson22_disk_wind,Thomsen2022}, or from shocks between intersecting debris streams themselves \citep{Shiokawa2015,Piran2015,Jiang2016_self_crossing_shock,Bonnerot2017}.

Besides the origin of optical emission, a second important aspect of optically-selected TDEs is that most of them are X-ray faint. In the unifying reprocessing scenario, the distinct classes of TDEs are given by the viewing angle with respect to the accretion disk and its associated reprocessing layer; the X-ray bright, `optically faint' TDEs are the ones observed more face-on, the optically bright, X-ray faint are edge-on, while the ones that show both emission components are seen at intermediate angles \citep{Guillochon2014,Dai2018,parkinson22_disk_wind}. In this scenario, X-rays from sources at intermediate angles can only break out after the reprocessing gas has expanded enough to become transparent to X-rays \citep{Metzger2016,Lu18,Thomsen2022}. However, in the stream-stream collision scenario, the intrinsically X-ray faint TDEs have been proposed to be a result of delayed accretion, due to the timescale required for the circularization of the debris into an appreciable accretion disk \citep{Gezari2017,Liu2021}. 

In addition to the study of nascent accretion disks, TDEs can be used to study the formation and evolution of other physical structures related to MBH accretion. Relativistic \citep[e.g.,][]{Cenko2012,Pasham2015,Zauderer2011,Brown2017,Pasham2023,Yao2023_22cmc} and non-relativistic \citep[e.g.,][]{Alexander2016,Bright2018,Stein2021} jets, as well as outflows with velocities varying from 200-600 \kms \citep{Miller2015,Krolik2016,Cenko2016,Blagorodnova2018,Kosec2023} up to $0.2 c$ \citep[e.g.,][]{Lin2015,Kara2018} have being detected in several sources. 

This paper presents an analysis of the X-ray light curves, X-ray spectral evolution, and broadband UV/optical/X-ray spectral energy distribution (SED) evolution of 16 X-ray detected, optically discovered TDEs between 2014 and December 2021, and one more simultaneous discovered by \srge (AT2020ksf) but with extensive UV/optical follow-up. We present new \xmm data for half of our sources and systematically reanalyze all publicly available ZTF, \swiftall and \xmm data presented in previous studies on individual sources. In \S\ref{sec:sample} we present our sample and its selection criteria.  In \S\ref{sec:data} we present the data and describe its basic analyses. In \S\ref{sec:X-ray_fit} we present a detailed discussion on the X-ray spectral fitting in TDEs and its general properties. In \S\ref{sec:results} we show our results and their interpretations, which are discussed in terms of the current literature in \S\ref{sec:discussion}; our conclusion are presented in \S\ref{sec:conclusions}.

We adopt a standard $\Lambda$CDM cosmology with matter density $\Omega_{\rm M} = 0.3$, dark energy density $\Omega_{\Lambda}=0.7$, and the Hubble constant $H_0=70\,{\rm km\,s^{-1}\,Mpc^{-1}}$. Optical and UV magnitudes are reported in the AB system. Uncertainties of X-ray model parameters are reported at the 68\% confidence level, and upper limits are reported at $3\sigma$.

\subsection{Sample Selection}\label{sec:sample}

\begin{figure*}

\centering
\includegraphics[width=0.95\textwidth]{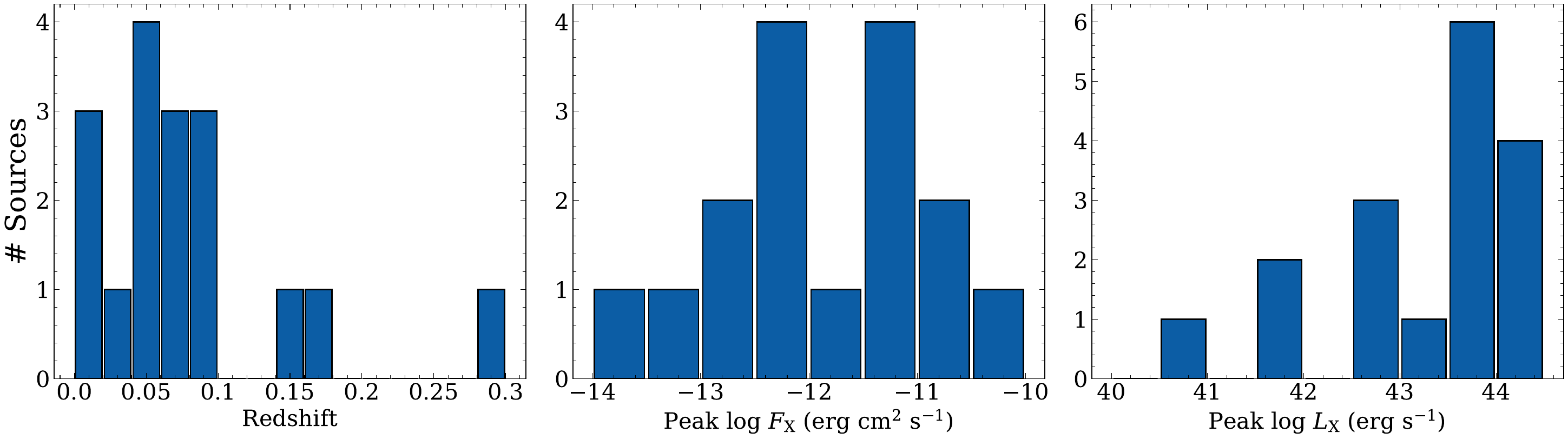}
\caption{ Histogram of the distributions of the properties of our main sample. Redshift ($z$, left), neutral absorption corrected peak 0.3-10 keV X-ray flux ($F_X$, middle), neutral absorption corrected peak 0.3-10 keV X-ray luminosity ($L_X$, right).}
\label{fig:hist_prop}
\end{figure*}

Aiming to explore the diversity of X-ray evolution in optically selected TDEs, we draw our sample from
sources discovered by optical time-domain surveys. We compile sources from ZTF sample papers \citep{vanVelzen2021,Hammerstein2023,Yao2023} as well as studies on individuals sources from other surveys \citep{Wyrzykowski2017, Holoien2016_14li,Holoien2016_15oi,Wevers2019_18fyk}.
We did not consider nuclear transients with pre-existing active galactic nuclei (AGN) from TDE candidates, these excludes those with host galaxies with AGN-like broad emission lines or
\neowise \citep{Mainzer2014} Mid Infrared (MIR) variability before the transient, as well
as $W1 - W2$ MIR color exceeds AGN selection criteria \citep[e.g.,][]{Stern2012}. We limited the sample
based on the epoch of discovery; given that the first optically discovered TDE to be systematically followed-up by X-ray telescopes
was ASASSN-14li, we delimited the sample with sources discovered after 2014. Our discovery epoch
criteria also exclude those sources discovered after December 2021, allowing us to have more than one year
of data for the entire sample. Finally we require every source to have at least one 3$\sigma$
\swift/XRT detection\footnote{For the standard 2~ks observations this translate to a 0.3-10 keV flux
greater than \textit{few}$\times 10^{-13}$ \ergcms},  yielding our final sample of 17 optically selected X-ray detected
TDEs. The complete sample and the basic information on the sources are shown upper portion of Table \ref{tab:sources}, in Fig.~\ref{fig:hist_prop}
we show some the distribution of some basic properties of our sample. 
Besides the observations on the 17 sources that make our main sample, we also present deep upper-limits based on \xmm observations of another 9 optically-selected TDEs that  never show detectable X-ray emission; these sources are shown the bottom portion of Table \ref{tab:sources}. 

\section{Observations, Data Reduction and Analyses}\label{sec:data}

\begin{figure*}
\centering
\includegraphics[width=0.9\textwidth]{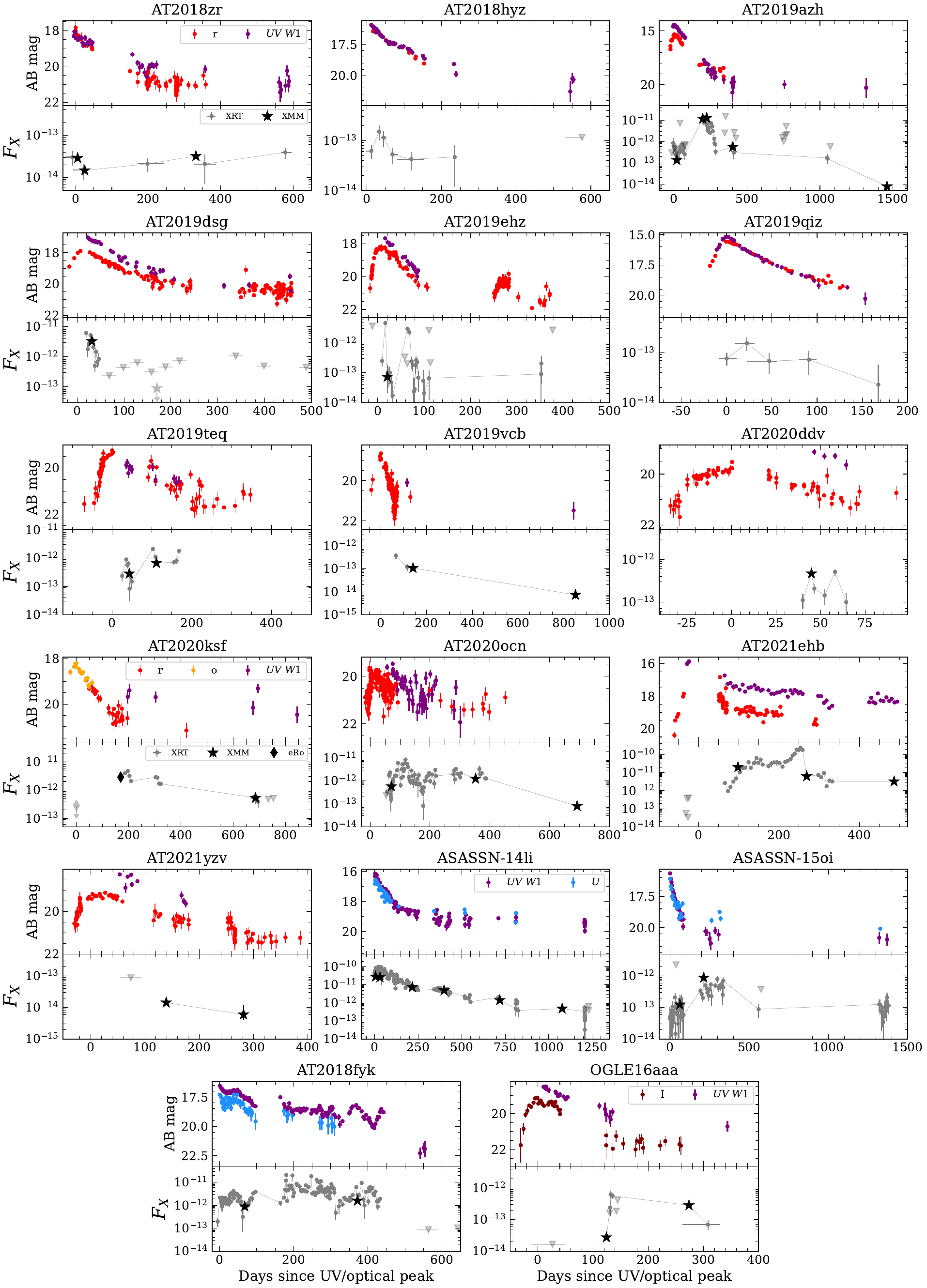}
\caption{UV/Optical (top panels) and X-ray (bottom panels) light curves from our sample of 17 optically selected X-ray detected TDEs. On top panels we show ZTF $r$-band (red points), \swift/UVOT UV W1 band (purple points), ATLAS $o$-band (orange points) and OGLE $I$-band (maroon points), all in magnitudes in the AB system. On bottom panels we show the neutral absorption corrected 0.3-10 keV X-ray flux ($F_X$), \swift/XRT in grey points, \xmm in stars and \srge in diamonds.}
\label{fig:lcs}
\end{figure*}

\subsection{Observations and Data Reduction}

\begin{deluxetable*}{clCCCCCCc}
\label{tab:sources}
\centering
\tabletypesize{\footnotesize}
\tablecaption{Sample Information.}
\tablewidth{0pt}
\tablehead{
\colhead{IAU} &  \colhead{Discovery} & \colhead{Optical/UV peak} & \colhead{Redshift} & \dcolhead{N_{H,G}}  & \dcolhead{\rm{log} \ (M_{*}\rm{/}M_{\odot})^{b}} &  \dcolhead{\sigma_*} &  \dcolhead{\rm{log} (M_{\rm{BH}}\rm{/}M_{\odot})^{c}} &  \colhead{TDE} \\
\colhead{Name} &  \colhead{Name} & \colhead{(MJD)} &  & (10^{20} \ \rm{cm}^{-2})^a& & (\rm{km  \ s^{-1}})&  \dcolhead{} &  \colhead{Classification}
}
\startdata
AT2018zr    & ZTF18aabtxvd  &             58214 &      0.071 &     4.17 & 10.01^{+0.08}_{-0.14} & 49 \pm 5^{(d)}   & 5.83 \pm 0.51 & ATel \#11444 \\
 AT2018hyz   & ASASSN-18zj  &             58422 &      0.045 &     2.59 & 9.96^{+0.09}_{-0.16}  & 57 \pm 1^{(e)}   & 6.12 \pm 0.46 & ATel \#12198\\
 AT2019azh   & ASASSN-19dj  &             58558 &      0.022 &     4.15 & 9.74^{+0.08}_{-0.05}  & 77 \pm 2^{(f)}    & 6.68 \pm 0.46 & ATel \#12568\\
 AT2019dsg   & ZTF19aapreis  &             58600 &      0.051 &     6.46 & 10.55^{+0.09}_{-0.12} & 94 \pm 1^{(g)}    & 7.04 \pm 0.45 & ATel \#12752\\
 AT2019ehz   & Gaia19bpt  &             58611 &      0.074 &     1.42 & 9.65^{+0.09}_{-0.12}  & 47 \pm 11^{(d)}   & 5.75 \pm 0.59 & ATel \#12789\\
 AT2019qiz   & ZTF19abzrhgq  &             58764 &      0.015 &     6.35 & 10.01^{+0.09}_{-0.12} & 70 \pm 2^{(h)}     & 6.49 \pm 0.46 & ATel \#13131\\ 
 AT2019teq   & ZTF19accmaxo  &             58794 &      0.087 &     4.54 & 9.95^{+0.07}_{-0.11}  & \nodata & 6.32 \pm 0.49 & TNSCR \#7482\\
 AT2019vcb   & ZTF19acspeuw  &             58838 &      0.088 &     1.45 & 9.49^{+0.06}_{-0.06}  & \nodata & 5.59 \pm 0.52 & TNSCR \#7078 \\
 AT2020ddv   & ZTF20aamqmfk  &             58915 &      0.16  &     1.35 & 10.30^{+0.13}_{-0.16} & 57 \pm 10^{(d)}   & 6.09 \pm 0.55 & ATel \#13655\\
 AT2020ksf   & Gaia20cjk  &             58976 &      0.092 &     3.61 & 10.12^{+0.13}_{-0.09} & 52 \pm 2^{(i)}    & 5.92 \pm 0.48 & \citet{Gilfanov2020} \\
 AT2020ocn   & ZTF18aakelin  &             58972 &      0.07  &     1.28 & 10.28^{+0.13}_{-0.70} & 81 \pm 8^{(d)}   & 6.77 \pm 0.49 & ATel \#13859\\
 AT2021ehb   & ZTF21aanxhjv  &             59330 &      0.018 &     9.88 & 10.18^{+0.01}_{-0.02} & 93 \pm 5^{(j)}   & 7.04 \pm 0.46 & TNSCR \#10001\\
 AT2021yzv   & ZTF21abxngcz  &             59475 &      0.286 &     8.60  & 10.65^{+0.04}_{-0.06} & \nodata & 7.45 \pm 0.47 &  TNSCR \#11890\\
 \nodata  &        ASASSN-14li       &             56993 &      0.02  &     1.95 & 9.68^{+0.04}_{-0.09}  & 81 \pm 2^{(k)}    & 6.77 \pm 0.46 & ATEL \#6777\\
  \nodata  &        ASASSN-15oi       &             57259 &      0.048 &     4.86 & 10.02^{+0.04}_{-0.03} & \nodata & 6.42 \pm 0.48 & ATEL \#7936\\
 AT2018fyk   & ASASSN-18ul   &             58389 &      0.059 &     1.16 & 10.56^{+0.21}_{-0.12} & 158 \pm 1^{(l)}  & 8.04 \pm 0.44 & TNSCR \#2723\\
  \nodata    &       OGLE16aaa       &             57403 &      0.165 &     2.72 & 10.47^{+0.09}_{-0.11} & \nodata & 7.14 \pm 0.48 &\citet{Wyrzykowski2017} \\
\hline
\hline
AT2018bsi & ZTF18aahqkbtr &             58389 &      0.051 &     4.91 & 10.62^{+0.05}_{-0.07} & 118 \pm 8^{(d)}   & 7.48 \pm 0.46 & ATel \#12035\\
 AT2018hco & ATLAS18way  &             58479 &      0.088 &     4.12 & 10.01^{+0.12}_{-0.16} &\nodata & 6.40 \pm 0.49 & ATel \#12263 \\
 AT2018iih & ATLAS18yzs  &             58558 &      0.212 &     3.19 & 10.81^{+0.11}_{-0.14} &\nodata & 7.70 \pm 0.48 & \citet{vanVelzen2021}\\
 AT2018lna & ZTF19aabbnzo  &             58561 &      0.091 &     6.42 & 9.56^{+0.08}_{-0.14}  & 36 \pm 4^{(d)}   & 5.20 \pm 0.53 & ATel \#12509 \\
 AT2019mha & ZTF19abhejal  &             58705 &      0.148 &     1.71 & 10.01^{+0.14}_{-0.18} &\nodata & 6.41 \pm 0.49 & \citet{vanVelzen2021}\\
 AT2019meg & ZTF19abhhjcc  &             58743 &      0.152 &     5.08 & 9.64^{+0.07}_{-0.08}  &\nodata & 5.81 \pm 0.52 & AN-2019-88\\
 AT2020pj  & ZTF20aabqihu  &             58866 &      0.068 &     2.24 & 10.01^{+0.07}_{-0.08} &\nodata & 6.43 \pm 0.49 & TNSCR \#7481 \\
 AT2020vwl & ZTF20achpcvt  &             59167 &      0.032 &     2.23 & 9.89^{+0.08}_{-0.08}  &\nodata & 6.21 \pm 0.49&  TNSCR  \#8572\\
 AT2020wey & ZTF20acitpfz  &             59156 &      0.027 &     6.19 & 9.67^{+0.09}_{-0.12}  &39 \pm 3^{(d)} & 5.38 \pm 0.51 & TNSCR  \#7769
\enddata

\tablecomments{Top: X-ray detected TDE (main sample). Bottom: X-ray non-detected TDEs with new deeper upper-limits. Abbreviations ATel corresponding to the Astronomer's Telegram (\href{https://astronomerstelegram.org/}{https://astronomerstelegram.org}), AN corresponding to AstroNotes (\href{https://www.wis-tns.org/astronotes}{https://www.wis-tns.org/astronotes}), and TNSCR corresponding to TNS classification reports. a) Galactic absorption column density from \citet{HI4PI2016}; b) Host galaxy stellar mass from SED fitting (see \ref{sec:hosts}); c) Black hole masses. When a $\sigma_*$ measurement is available it is estimated using the \citet{Gultekin2019} $\sigma_*-$\mbh relation, when $\sigma_*$ is not available, this is estimated from the $M_*$-\mbh relation presented in \citet{Yao2023}. d) \citet{Hammerstein2023}; (e) \citet{Short2020}; f) \citet{Liu2021}; g) \citet{Cannizzaro2021}; h) \citet{Nicholl2020}; i) Wevers et al., in prep; j)\citet{Yao2022}; k) \citet{Wevers2019_Mbh}; l) \citet{Wevers2019_18fyk}}
\end{deluxetable*}
\subsubsection{\xmm}

The primary data set underlying this work is based on \xmm observations, these were obtained primarily from a series of Announcement of Opportunity (AO) programs (AO-18 84259, AO-20 88259, P.I.: Gezari, and AO-21 90276 P.I. Yao) aimed on the deep X-ray follow-up of ZTF-discovered TDEs. These observations were taken in Full Frame mode with the thin filter using the European Photon Imaging Camera \citep[EPIC;][]{Struder2001} and are presented here for the first time. We also included publicly available observations from several other AO and Director Discretionary Time (DDT) proposals. The details on the \xmm observations are shown in Table \ref{tab:xmm_obs}.

The observation data files (ODFs) were reduced using the \xmm Standard Analysis Software \citep[SAS;][]{Gabriel_04}.
The raw data files were then processed using the \texttt{epproc} task. 
Since the pn instrument generally has better sensitivity than MOS1 and MOS2, we only analyze the pn data. 
Following the \xmm data analysis guide, to check for background activity and generate ``good time intervals'' (GTIs), we manually inspected the background light curves in the 10--12\,keV band. 
Using the \texttt{evselect} task, we only retained patterns that correspond to single and double events (\texttt{PATTERN<=4}). The source spectra were extracted using a source region of $r_{\rm src} = 35^{\prime\prime}$ around the peak of the emission. 
The background spectra were extracted from a $r_{\rm bkg} =
108^{\prime\prime}$ region located in the same CCD. The ARFs and RMF files
were created using the \texttt{arfgen} and \texttt{rmfgen} tasks,
respectively. Some of the observations for ASASSN-14li and AT2020ocn
presented pile-up effects. Therefore, we followed the SAS
guide\footnote{https://www.cosmos.esa.int/web/xmm-newton/sas-thread-epatplot} by excising the core of the point spread function (PSF), up to a radius where the pile-up fraction becomes negligible following the
\texttt{epatplot} command results.

\subsubsection{\swiftall}\label{sec:obs_xrt}
All the sources were observed by the X-ray Telescope (XRT; \citealt{Burrows2005}) and the Ultra-Violet/Optical Telescope (UVOT; \citealt{Roming2005}) on board \swiftall \citep{Gehrels2004}. The number of observations varies from a few for the more distant sources to a hundred for the most well-sampled sources. 

The 0.3–10 keV X-ray counts light curves were produced using the UK Swift Data center online XRT data products tool, which uses the HEASOFT v6.22 software \citep[][]{Arnaud1996}. We used a fixed aperture at the ZTF coordinate of the transient, generating one count rate point per visit (i.e., per ObsID) for most of the sources; for faint sources (AT2018zr, AT2018hyz and AT2019qiz), in which the count rate of individual visits were close to the XRT detection limit ($\sim 10^{-3}$ count $s^{-1}$) we stack few observations using the `dynamical binning' as described in \citet{Evans2007}, in order to obtain a smooth X-ray light curve. 

The short XRT exposures do not allow for spectral fitting of individual visits, so we stacked consecutive observations using an automated online tool\footnote{\url{https://www.swift.ac.uk/user_objects}} \citep{Evans2009}. We aimed to have at least 100 counts per stacked spectrum, allowing the bins to be at maximum 100 days long, we also ensure there was no large evolution in the hardness ratio within each bin. For AT2018zr, AT2019vcb, AT2020ddv and ASASSN-15oi there were not enough counts to generate a fitable spectrum, even combining all the XRT observations, therefore we restrict our XRT analyzis on their light curves and only perform spectral analysis in their \xmm data, which will be described in  detailed in \S\ref{sec:X-ray_fit}.

We used the \texttt{uvotsource} package to analyze the Swift UVOT photometry, using an aperture of 5\arcsec\xspace for all sources except AT2019azh, AT2019qiz, and AT2019dsg, which required a larger aperture to capture the host galaxy light. We subtracted the host galaxy flux estimated from the population synthesis modeling of archival pre-event photometry described in \S\ref{sec:hosts}. We apply  Galactic extinction correction on all bands using $E(B-V)$ values at the position of each source from \citet{Schlafly2011}.
\subsubsection{Zwicky Transient Facility}
We performed point spread function (PSF) photometry on all publicly available ZTF data using the ZTF forced-photometry service \citep{Masci2019,Masci2023} in $g$- and $r$-band. Similar to UVOT, ZTF photometry was corrected for Galactic extinction.

\begin{deluxetable*}{ccCCCCCc}
\label{tab:xmm_obs}
\tabletypesize{\scriptsize}
\tablecaption{Summary of \xmm Observations}
\tablewidth{0pt}
\tablehead{
\colhead{Source} &  \colhead{Obs ID} & \colhead{Obs Date}  & \colhead{phase ($\Delta t$)$^{a}$}  &\colhead{log $F_{\rm X}^b$} & \colhead{log $L_{\rm X}^b$} & \colhead{$L_{\rm{BB}}$ / $L_{\rm X}$} & \colhead{First} \\
\colhead{} &  \colhead{} & \colhead{MJD} & \colhead{(days)}  &\colhead{(erg cm$^{2}$ s$^{-1}$)} & \colhead{(erg s$^{-1}$)} & \colhead{}  & \colhead{presented in}
}
\startdata
\multirow{3}{*}{AT2018zr}    & 0822040301 & 58220 &       5 & -13.538^{+0.040}_{-0.045} & 41.610^{+0.040}_{-0.045} & 95.46^{+9.26}_{-9.30}    &          \multirow{2}{*}{\citet{vanVelzen2019}} \\
                              & 0822040501 & 58241 &      25 & -13.831^{+0.052}_{-0.059} & 41.317^{+0.052}_{-0.059} & 127.48^{+16.28}_{-16.24} &           \\ \cline{8-8}
                              & 0822040401 & 58569 &     331 & -13.493^{+0.033}_{-0.036} & 41.656^{+0.033}_{-0.036} & 12.38^{+0.99}_{-1.00}    &          this work \\\hline
 \multirow{5}{*}{AT2019azh}   & 0822041101 & 58579 &      20 & -12.872^{+0.016}_{-0.017} & 41.178^{+0.016}_{-0.017} & 889.10^{+34.07}_{-34.09} &          \multirow{4}{*}{this work}\\ 
                              & 0842591001 & 58760 &     197 & -10.924^{+0.002}_{-0.002} & 43.125^{+0.002}_{-0.002} & 0.58^{+0.01}_{-0.01}     &            \\
                              & 0823810401 & 58788 &     225 & -10.881^{+0.001}_{-0.001} & 43.168^{+0.001}_{-0.001} & 0.41^{+0.01}_{-0.01}     &            \\
                              & 0842592601 & 58971 &     404 & -12.248^{+0.013}_{-0.014} & 41.801^{+0.013}_{-0.014} & 2.14^{+0.07}_{-0.07}     &            \\
                              & 0902761101 & 60049 &    1458 & -14.114^{+0.116}_{-0.158} & 39.936^{+0.116}_{-0.158} & \nodata &           \\\hline
 \multirow{2}{*}{AT2019dsg}   & 0842590901 & 58633 &      31 & -11.483^{+0.004}_{-0.004} & 43.316^{+0.004}_{-0.004} & 3.16^{+0.03}_{-0.03}     &          \multirow{2}{*}{\citet{Stein2021}} \\
                              & 0842591901 & 58779 &     170 & < -13.046    & < 41.753     & > 15.99      &           \\\hline
 \multirow{1}{*}{AT2019ehz}   & 0842590801 & 58633 &      20 & -13.138^{+0.022}_{-0.023} & 41.998^{+0.022}_{-0.023} & 50.89^{+2.63}_{-2.63}    &          this work \\\hline
 \multirow{2}{*}{AT2019teq}   & 0842591701 & 58841 &      43 & -12.549^{+0.011}_{-0.012} & 42.736^{+0.011}_{-0.012} & 2.25^{+0.06}_{-0.06}     &          \multirow{2}{*}{this work}  \\
                              & 0842592401 & 58915 &     111 & -12.173^{+0.005}_{-0.006} & 43.111^{+0.005}_{-0.006} & 0.49^{+0.01}_{-0.01}     &          \\\hline
 \multirow{2}{*}{AT2019vcb}   & 0871190301 & 58991 &     140 & -12.967^{+0.020}_{-0.021} & 42.339^{+0.020}_{-0.021} & 1.71^{+0.08}_{-0.08}     &          \citet{Quintin2023}  \\\cline{8-8}
                              & 0882591401 & 59764 &     851 & -14.136^{+0.106}_{-0.150} & 41.169^{+0.106}_{-0.150} & 7.12^{+1.97}_{-2.08}     &          this work \\\hline
 \multirow{1}{*}{AT2020ddv}   & 0842592501 & 58967 &      44 & -12.332^{+0.008}_{-0.008} & 43.523^{+0.008}_{-0.008} & 5.05^{+0.09}_{-0.09}     &          this work \\\hline
 \multirow{1}{*}{AT2020ksf}   & 0882591201 & 59725 &     749 & -12.280^{+0.009}_{-0.010} & 43.057^{+0.009}_{-0.010} & 2.92^{+0.06}_{-0.06}     &          this work \\\hline
 \multirow{3}{*}{AT2020ocn}   & 0863650101 & 59048 &      76 & -12.238^{+0.005}_{-0.006} & 42.848^{+0.005}_{-0.006} & 1.42^{+0.02}_{-0.02}     &          \citet{Pasham2024} \\  \cline{8-8}
                              & 0872392901 & 59349 &     377 & -11.902^{+0.004}_{-0.004} & 43.183^{+0.004}_{-0.004} & 0.06^{+0.01}_{-0.01}     &           \multirow{2}{*}{this work}  \\ 
                              & 0902760701 & 59712 &     685 & -13.094^{+0.020}_{-0.021} & 41.991^{+0.020}_{-0.021} & 2.73^{+0.13}_{-0.13}     &          \\\hline
 \multirow{3}{*}{AT2021ehb}   & 0882590101 & 59430 &     98 & -10.696^{+0.003}_{-0.004} & 43.177^{+0.003}_{-0.004} & 0.49^{+0.01}_{-0.01}     &         \multirow{2}{*}{\citet{Yao2022}}  \\
                              & 0882590901 & 59604 &     269 & -11.203^{+0.003}_{-0.003} & 42.670^{+0.003}_{-0.003} & 0.52^{+0.01}_{-0.01}     &           \\\cline{8-8}
                              & 0902760101 & 59825 &     486 & -11.489^{+0.006}_{-0.006} & 42.384^{+0.006}_{-0.006} & 0.67^{+0.01}_{-0.01}     &          this work \\ \hline
 \multirow{2}{*}{AT2021yzv}   & 0882591001 & 59654 &     139 & -13.844^{+0.054}_{-0.062} & 42.580^{+0.054}_{-0.062} & 99.26^{+13.22}_{-13.19}  &         \multirow{2}{*}{this work} \\
                              & 0882591501 & 59837 &     281 & -14.222^{+0.301}_{-0.176} & 42.204^{+0.301}_{-0.176} & 68.48^{+68.48}_{-22.83}  &           \\\hline
 \multirow{6}{*}{ASASSN-14li} & 0694651201 & 56997 &       3 & -10.549^{+0.002}_{-0.002} & 43.442^{+0.002}_{-0.002} & 2.33^{+0.01}_{-0.01}     &          \citet{Miller2015} \\ \cline{8-8}
                              & 0694651401 & 57023 &      29 & -10.588^{+0.002}_{-0.002} & 43.403^{+0.002}_{-0.002} & 1.31^{+0.01}_{-0.01}     &          \multirow{2}{*}{\citet{Kara2018}} \\
                              & 0694651501 & 57213 &     215 & -11.135^{+0.004}_{-0.004} & 42.856^{+0.004}_{-0.004} & 0.78^{+0.01}_{-0.01}     &        \\ \cline{8-8}
                              & 0770980501 & 57399 &     397 & -11.312^{+0.005}_{-0.005} & 42.679^{+0.005}_{-0.005} & 0.49^{+0.01}_{-0.01}     &          \multirow{3}{*}{\cite{Wen2020}} \\
                              & 0770980701 & 57726 &     718 & -11.855^{+0.007}_{-0.008} & 42.136^{+0.007}_{-0.008} & 2.04^{+0.04}_{-0.04}     &         \\
                              & 0770980901 & 58092 &    1076 & -12.318^{+0.008}_{-0.009} & 41.673^{+0.008}_{-0.009} & 5.13^{+0.10}_{-0.10}     &          \\\hline
 \multirow{2}{*}{ASASSN-15oi} & 0722160501 & 57324 &      62 & -12.907^{+0.027}_{-0.029} & 41.837^{+0.027}_{-0.029} & 77.94^{+5.07}_{-5.06}    &          \multirow{2}{*}{\citet{Gezari2017}} \\
                              & 0722160701 & 57482 &     212 & -12.051^{+0.011}_{-0.011} & 42.693^{+0.011}_{-0.011} & 1.22^{+0.03}_{-0.03}     &           \\\hline
 \multirow{2}{*}{AT2018fyk}   & 0831790201 & 58461 &      67 & -12.056^{+0.005}_{-0.005} & 42.874^{+0.005}_{-0.005} & 29.62^{+0.31}_{-0.31}    &         \multirow{2}{*}{\citet{Wevers2021}} \\
                              & 0853980201 & 58783 &     372 & -11.803^{+0.003}_{-0.003} & 43.127^{+0.003}_{-0.003} & 2.97^{+0.02}_{-0.02}     &          \\\hline
 \multirow{2}{*}{OGLE16aaa}   & 0790181801 & 57548 &     124 & -13.563^{+0.056}_{-0.064} & 42.322^{+0.056}_{-0.064} & 15.06^{+2.06}_{-2.06}    &          \multirow{2}{*}{\citet{Kajava2020}} \\
                              & 0793183201 & 57722 &     273 & -12.543^{+0.012}_{-0.012} & 43.342^{+0.012}_{-0.012} & 0.49^{+0.01}_{-0.01}     &           \\\hline \hline
AT2018bsi & 0822040801 &  58389 &        164 &    < -13.453 & < 41.319 &              > 42 &  this work \\  \hline
 AT2018hco & 0822040901 &  58479 &         71 &   < -13.974 & < 41.312 &              > 291 &  this work\\  \hline
 AT2018iih & 0822040701 &  58558 &        95 &   <  -14.250 & < 41.867 &              > 254 &  this work\\  \hline
 AT2018lna & 0822041001 &  58561 &         48 &  < -14.057 & < 41.260 &              > 309 &  this work\\  \hline
 AT2019mha & 0842592201 &  58705 &         33 &  <  -14.288 & < 41.483 &              > 43 &  this work \\  \hline
 AT2019meg & 0842592101 &  58743 &         40 &  < -13.919 & < 41.878 &              > 82 &  this work\\
 \hline
 AT2020wey & 0902760401 &  59851 &        676 & <-13.0857 & <41.1449 &          \nodata^{c}      &    this work \\ \hline
 AT2020pj  & 0902760801 &  59809 &        882 &  <-13.5102 & <41.5488 &       \nodata^{c}     &  this work   \\ \hline
 AT2020vwl & 0902760301 &  59776 &        590 &  <-13.568  & <40.8932 &       \nodata^{c}           & this work  \\ \hline
\enddata
\tablecomments{a) Relative to the rest-frame UV/optical peak; b) 0.3-10.0 keV band, upper limits are 3$\sigma$. c)  No simultaneous UV/optical 
 detections available.}
\end{deluxetable*}

\subsubsection{Additional Data}
OGLE16aaa was discovered by the fourth phase of the Optical Gravitational Lensing Experiment \citep[OGLE-IV,][]{Udalski2015} survey, we added the I band (7970 \AA) data, which is the best sampled optical light curve of the
source; the data was retrieved from the survey website\footnote{http://ogle.astrouw.edu.pl/ogle4/transients/2017a/transients.html}.

The optical peak of AT2020ksf was missed by ZTF, however the Asteroid Terrestrial-impact Last Alert System \cite[ATLAS,][]{Tonry2018} started observing the 
field containing the source $\sim$60 days before ZTF, hence allowing us to measure the date of its optical peak; we added available ATLAS data to our light curve of AT2020ksf. The source was first detected in the X-rays by the \srge, although the spectral data is not yet publicly available, we added the reported detection flux of $1.7 \times 10^{-12}$ \ergcms ($2.85
\times 10^{-12}$ \ergcms unabsorbed) observed in 20th November 2020 (MJD 59162), as well as the previous X-ray upper limit ($\leq 2.85 \times 10^{-13}$ \ergcms) from a visit taken on 8th May 2020 (MJD 58977) which is fortuitously 
timed at the optical peak of the source, both reported by \citet{Gilfanov2020}.

\begin{figure}
\includegraphics[width=1.0\columnwidth]{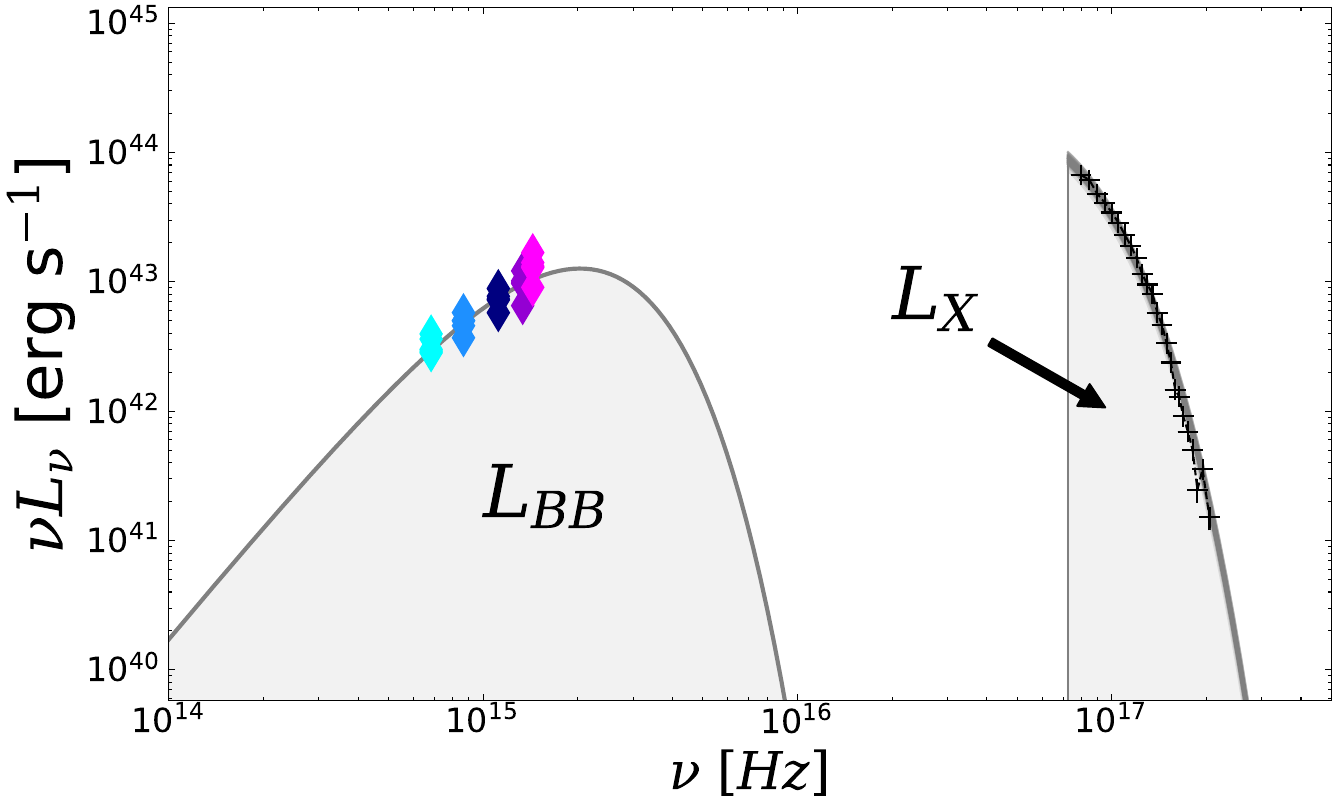}

\caption{Illustration of how the UV/optical blackbody luminosity (\lbb) and the 0.3-10 keV X-ray luminosity (\lx) are measured and what those values represent regarding the full SED of the transient.
Colored points indicate observed UV/optical photometry, underline it the best-fitting blackbody function is shown in dark gray, and the area below the curve is the measured \lbb. The black crosses show the observed X-ray spectra and best-fitted X-ray model, and the measured \lx is also shown in gray. Both UV/optical and X-ray components are corrected for Galactic extinction/absorption.}
\label{fig:Lbb_Lx_measurments}
\end{figure}



\subsection{UV/optical Data Analyses}\label{sec:analyses}
\subsubsection{Host Galaxy SED modeling}\label{sec:hosts}
Before analyzing the transient's UV/optical light curve, the host contamination needs to be subtracted. We followed 
\citet{vanVelzen2021} prescription to fit the host galaxy pre-flare photometry. We compile the host galaxy spectral 
energy distribution (SED) using archival observations in the UV through IR bands. We use the \texttt{Prospector} software 
\citep{Johnson2021} to run a Markov Chain Monte Carlo (MCMC) sampler \citep{Foreman-Mackey2013} to obtain the 
posterior distributions of the Flexible Stellar Population Synthesis models \citep{Conroy2009}. We adopted a simple 
power law star formation history , with the same ranges and priors as in \citet{vanVelzen2021}, for the 5 free 
parameters: stellar mass, \citet{Calzetti2000} dust model optical depth, stellar population age, metallicity, and the
e-folding time of the star formation history. The resulting host stellar mass (\mstar) are presented in Table 
\ref{tab:sources}, the fitting for all host galaxies of our sample are 
presented either in \citet{vanVelzen2021}, \citet{Hammerstein2022} or \citet{Yao2023}, the reader is referred to these papers for the 
full list of best-fitting parameters.  We  subtract the host contribution to the transient's UVOT filter's photometry. The UV\,W1 band 
(2600 \AA, observed frame) host-subtracted light curves are shown in the top panels of Figure \ref{fig:lcs} for 
individual sources.

\subsubsection{UV/optical light curves}\label{sec:UV_opt}

Following the standard approach for optical TDEs \citep[e.g.,][]{vanVelzen2021,Hammerstein2022} we estimate the integrated UV/optical luminosity ($L_{BB}$) by fitting the transient UV/optical SED with an evolving, Gaussian rise and power law decay, blackbody (BB) function. The model can be written as:

\begin{multline}
    L_{BB}(t) = L_{BB,peak} \frac{\pi B_{\nu}(T(t))}{\sigma_{SB}T^{4}(t))} \\ \times\left\{\begin{matrix}
 e^{-(t-t_{peak})^2/2\sigma^2},& t\leq t_{peak}\\ 
 [(t-t_{peak})/t_0]^{p},& t > t_{peak}
\end{matrix}\right.
\label{eq:lbb}
\end{multline}

\noindent
We consider a non-parametric temperature evolution; we fit the temperature at grid points spaced $\pm$ 30 days apart beginning at peak and use a log-normal Gaussian prior at each grid point. 
We use a Gaussian likelihood function to estimate the parameters of the models above; we use the \texttt{emcee} sampler \citep{Foreman-Mackey2013}. Details on the fitting process and resulting UV/optical integrated light curves can be seen in either \citet{vanVelzen2021} or \citet{Hammerstein2022} for all the sources. The time dependent model assumed in Eq. \ref{eq:lbb}, was assumed for the first 350 days of the optical light curve, after that the light curve shape usually deviates from the power-law decay \citep{vanVelzen2021,Hammerstein2022}. For the later epochs, we measured the integrated UV/optical luminosity by fitting again with a blackbody function for the available UV/optical photometry data, epoch by epoch, when \lbb measurements were necessary (see \S\ref{sec:results}).
In Fig. \ref{fig:Lbb_Lx_measurments}, we illustrate what fitting a blackbody to the UV/optical broad-band photometric data means in terms of the full SED of the transient; it also illustrates how misleading it can be to interpret \lbb as a \textit{`bolometric'} luminosity, as is commonly done by some authors; such aspects will be further explored in \S\ref{sec:SED}.

\subsection{Black Hole Masses}\label{sec:M_BH}

The black holes masses (\mbh) of the TDE hosts were estimated from the 
host galaxy scaling relations. If a measurement of the velocity dispersion 
($\sigma_*$) of their nuclear stellar populations were publicly available 
\citep[e.g., ][]{Wevers2019_Mbh,Yao2023,Hammerstein2023} \mbh was estimated from $\sigma_*-$\mbh 
relation by \citet{Gultekin2019}. Alternatively, if $\sigma_{*}$ was not available, \mbh 
was estimated from the host galaxy mass ($M_*$, as measured from 
\ref{sec:hosts}) using the relation presented in \citealt{Yao2023}. The 
$\sigma_{*}$, $M_*$ and \mbh values are shown in Table \ref{tab:sources}. 
Uncertainties in \mbh are the result of the addition in quadrature of the 
statistical uncertainty of $\sigma_*$/$M_*$ and the systematic spread of 
the scaling relations, and are usually $\sim$0.5 \textrm{dex}.

\section{X-ray Spectral Fitting}\label{sec:X-ray_fit}

\begin{figure*}
\centering
\includegraphics[width=0.8\textwidth]{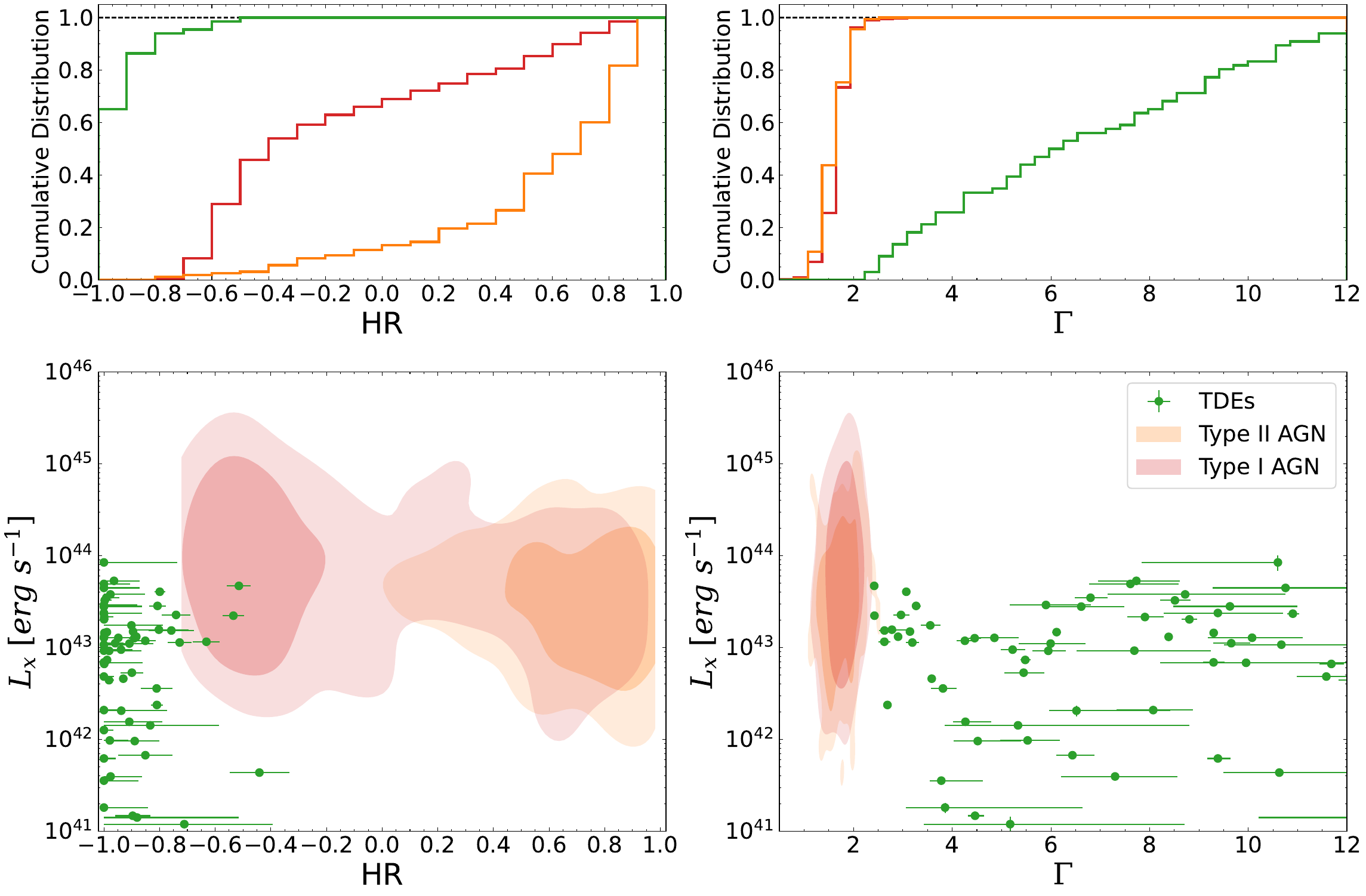}
\caption{Comparison between X-ray spectral properties of TDEs (green) and AGN (Type I in red and Type II in gold) from the BAT AGN Spectroscopic Survey \citep[BASS,][]{Ricci2017}. In the left panels we compare the \swift/XRT or \xmm hardness ratio (\hr, see text for definition), in the right panels we compare the $\Gamma$ power-law index when TDEs are fitted with an absorbed power-law model. The top panels show the cumulative distribution of the parameters, while the bottom panels show the distribution of samples in the \lx vs. \hr and \lx vs. $\Gamma$ parameter space. For AGNs the contours represent 68\% and 90\% of the sample distribution, for TDEs each point is a \xmm or \swift/XRT stacked spectrum.}
\label{fig:TDExAGN}
\end{figure*}

The following procedures were performed using the python version of \texttt{xspec} \citep{Arnaud1996}, \texttt{pyxspec}\footnote{https://heasarc.gsfc.nasa.gov/xanadu/xspec/python}.
For all spectral models described below, we included 
the Galactic absorption using the \texttt{TBabs} model \citep{Wilms2000}, with the hydrogen-equivalent neutral column density $N_{H,G}$ fixed at the values shown in Table \ref{tab:xmm_obs} \citep{HI4PI2016}. We shifted the TDE emission using the convolution model \texttt{zashift}, with the redshift $z$ shown in Table \ref{tab:sources}. \xmm spectra were grouped to have at least 25 counts per bin, but limiting the over-sampling of the instrumental resolution to a factor of 5; we assume a $\chi^2$-statistic. For \swift/XRT data fitting, the stacked spectra were grouped to have at least 1 count per bin; we assume a C-statistic \citep{Cash1979}. To convert the count rate of each visit to flux, we assume the closest in time best-fit model to the stacked spectra. We check for the convergence of the fitting using the \texttt{steppar} command in \texttt{xspec}. \\


\subsection{General Spectral Properties}

The most distinct characteristic property of the X-ray spectra of TDEs is its softness --  a visual inspection of our \xmm observations confirms this trend for our optically detected X-ray bright TDEs sample -- this makes TDE X-ray spectra clearly distinguishable from the dominant sources of extragalactic X-ray emission from AGN. While AGN usually emits from the soft X-ray up to the hard X-rays ($E >> 10$ keV), with a non-thermal (power-law) spectrum, TDEs rarely show emission at energies higher than 2.0 keV. 

To demonstrate this, we compare the X-ray spectral properties of our TDE sample with the X-ray properties of the non-blazar Type I and Type II of the BAT AGN Spectroscopic Survey \citep[BASS,][]{Ricci2017}. Details on how the BASS comparison sample was retrieved is described in appendix \S\ref{app:bass}.
In the left panels of Fig.~\ref{fig:TDExAGN} we compare the model independent -- but instrument dependent -- hardness ratio (\hr) of the samples, which is defined as \hr = $(H-S)/(H+S)$, where $S$ is the 0.3-2.0 keV count rate and $H$ is the 2.0-10.0 keV count. Our entire TDEs sample has \hr $\leq$ -0.5, with 85\% having \hr  $\leq$ -0.80. The Type I AGN sample is concentrated between -0.7 $\leq$ \hr $\leq$ 
0.0, while the increased column density ($N_H$) in obscured type
II AGN makes their \hr to range from 0.0 to 1.0. Such \hr is dependent on the X-ray instrument response, these values are valid for \swift/XRT and/or \xmm, which have similar \textit{relative} soft-to-hard sensibility.


\subsection{Absorbed Power-law Model} 

In AGN the main X-ray emission originates from the ubiquitous corona, and it is usually fitted with a phenomenological \texttt{powerlaw} model. Fitting TDE's extremely soft spectra with an absorbed power-law (i.e. \texttt{TBabs*zashift*(TBabs*powerlaw)} in \texttt{xspec}) like in AGN, usually results in best-fitted $\Gamma$ parameter higher than 5; such high values are nonphysical in the case of inverse up-scattering of seed photons by a hot corona \citep{Titarchuk_95} and hence do not represent any meaningful physical measurement. 

Nevertheless, fitting with \texttt{powerlaw} may be useful to differentiate X-ray TDE spectra from AGN ones. In the right panels of Fig.~\ref{fig:TDExAGN}, we compare the photon index $\Gamma$ 
with those of AGN from the BASS comparison sample. In AGN, both Type~I and Type~II, $\Gamma$ only varies between $\sim$ 1.0 and 2.5, while in TDEs these are much stepper, with $\Gamma \in (2,12)$. The large uncertainties in the spectra with $\Gamma \geq 4$ is not due to low S/N, instead by the inadequacy of the absorbed power-law to described TDE spectra, and the large degeneracy between the intrinsic $N_H$ and $\Gamma$, when a power-law model is fitted in a underlying thermal/soft spectra, epochs/spectra with $\Gamma \leq 4$ are those in which corona formation is observed (see \S\ref{sec:soft_hard}). 

In summary, an absorbed powerlaw model is an inappropriate model for the emission of non-jetted TDEs, usually no physical interpretation can be derived from such fit, however, it still may be a good tool to differentiate (when more information is lack) TDEs from AGN, as clearly demonstrate by Fig. \ref{fig:TDExAGN}.

\subsection{Blackbody Model}
TDE soft spectra can be fitted with thermal models, in their simplest form, a single temperature blackbody (\texttt{blackbody} or  \texttt{bbodyrad}, in \texttt{xspec}), has been used for some of the first X-ray discovered TDEs \citep[see review by][]{Saxton2020}. We employed such model in our data (\texttt{TBabs*zashift*bbodyrad} in \texttt{xspec}), the model fits well spectra with low signal-to-noise ratio (S/N). However, the model seems to be insufficient to fit spectra of high count rate observations, in which the model results in systematically worse fit (in terms of reduced $\chi^2$) than multi-temperature thermal models, which are usually associated with a standard accretion disk \citep[][see below]{Shakura1973,Mitsuda1984,Mummery2021}. Furthermore, as shown by \citet{Mummery2021}, \texttt{bbodyrad} can led to unphysical emitting regions sizes.

\begin{figure}
\includegraphics[width=1.0\columnwidth]{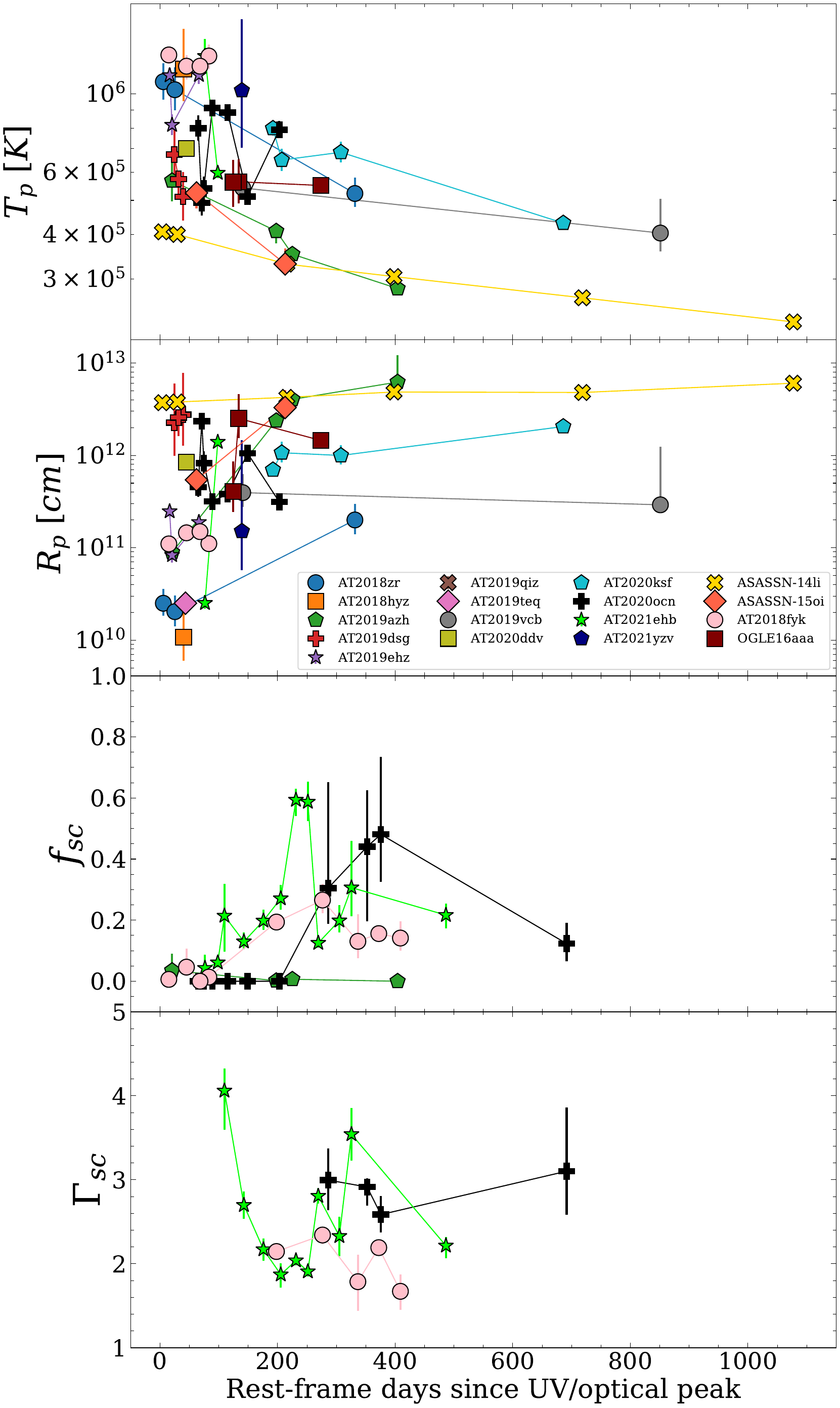}
\caption{Best-fitting parameters from X-ray spectral modeling as function of time from the UV/optical peak. From top to bottom: Peak temperature ($T_p$) and Radius ($R_p$) from \texttt{tdediscspec}; fraction of comptonized photons ($f_{sc}$) and power-law photon index ($\Gamma_{sc}$) of \texttt{simPL}.}
\label{fig:fit_pars}
\end{figure}

\subsection{Accretion Disk + Comptonization Model}

In terms of multi-temperature thermal models, the \texttt{diskbb} model \citep{Shakura1973,Mitsuda1984} developed for stellar black holes in X-ray binaries (XRB) is usually employed in TDE spectra. The model, however, assumes quasi-static state condition which are not necessarily present in the newly form accretion disk of TDEs.

\begin{figure*}
\centering
\includegraphics[width=1.0\textwidth]{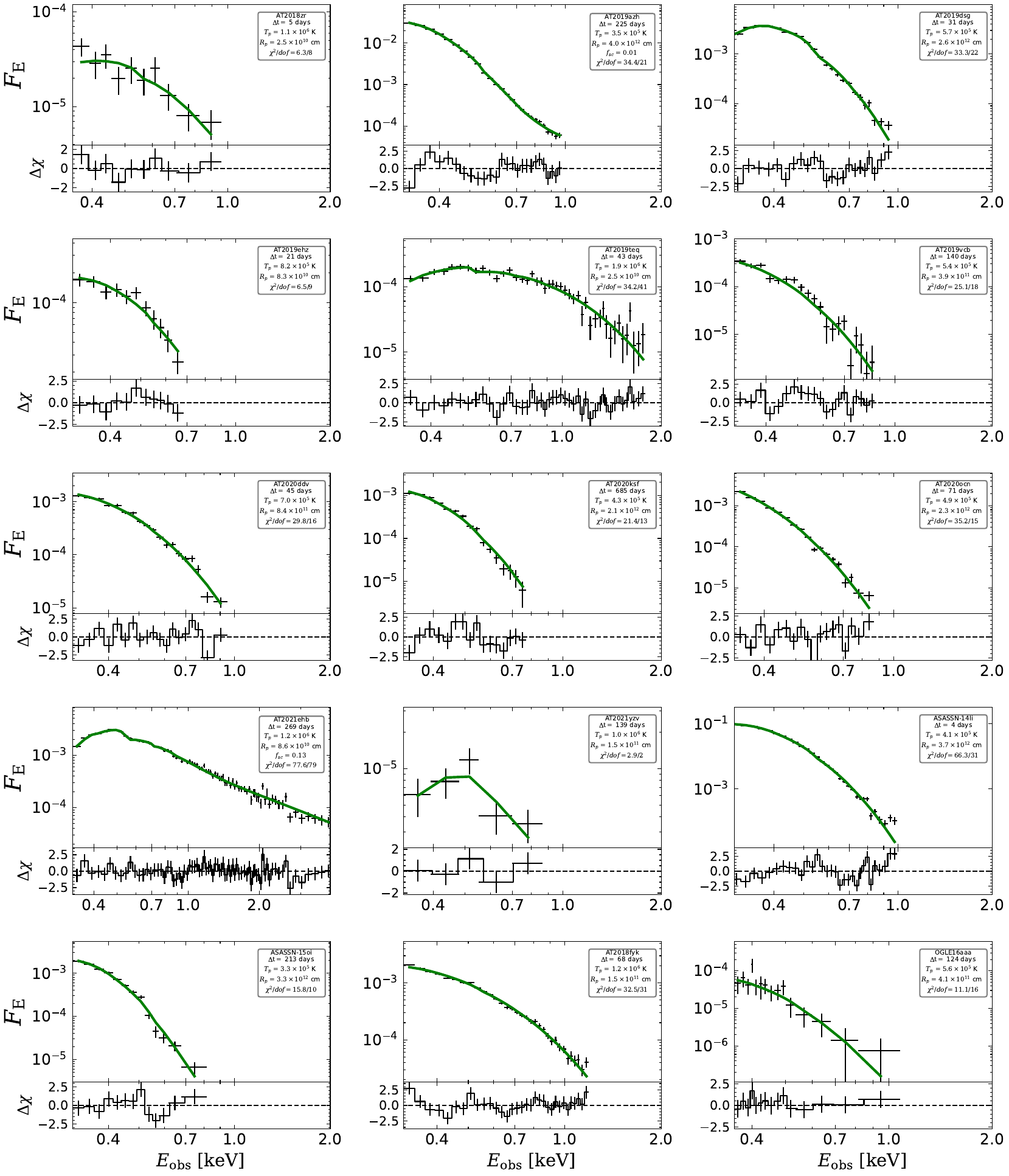}
\caption{Representative unfolded spectra, best-fit model and residuals for all the source with available \xmm data in our sample. Top panels shows the unfolded spectra in black and the best-fitting model in green (either \texttt{tdediscspec} or \texttt{simPL$\otimes$tdediscspec}) in units of $\rm{keV \ cm^{-2} \ s^{-1} \ keV^{-1}}$. Bottom panels show the residuals in $\Delta \chi$.}
\label{fig:xmm_spec}
\end{figure*}

Recently, \citet{Mummery2021} developed a model specifically tailored for TDE accretion disks; the author based their modeling on the convenient property of TDE disks being relatively cool, with their spectra peaking below the low bandpass of X-ray telescopes, $kT \leq 0.3$ keV. This means that X-ray observations of TDE disks probe the quasi-Wien tail of the disk spectrum; hence no assumption about the disk temperature profile needs to be made; instead, the only assumption inherent to the model is that each disk radius emits like a color-corrected blackbody and that there
exists some disk radius where the disk temperature peaks. The \texttt{xspec} model, called \texttt{tdediscspec}, fits the following expression to the observed X-ray spectra \citep{Mummery2021,Mummery_Balbus2021}:

\begin{multline}
     F_\nu(R_p, \tilde{T_p}, \gamma)  \simeq \\ \frac{4\pi \xi_1 h\nu^3}{c^2f_{\rm col}^4}\left( \frac{R_p}{D} \right)^2 \left(\frac{k \tilde{T_p}}{h \nu} \right)^\gamma \exp\left(- \frac{h\nu}{k \tilde{T_p}} \right)  
\end{multline}
 
\noindent where $\tilde{T_p} \equiv f_{\rm col} f_\gamma T_p$, and $T_p$ is the parameter of interest, i.e., the hottest temperature in 
the accretion disc. The factor $f_{\gamma}$ is the photon energy-shift factor, defined as the ratio of the observed photon frequency $\nu$
to the emitted photon frequency $\nu_{emit}$, $f_{\gamma} = \nu/\nu_{emit} \approx 1/\sqrt{2}$ \citep[see][and reference therein for details]{Mummery_Balbus2021}, while $f_{\rm col}$ is the `color-correction' factor, which is included to model disk opacity effects. This correction factor generally takes a value  $f_{\rm col}$  $\sim$ 2.3 for typical TDE disk temperatures \citep{Done2012,Mummery2021,Mummery2023}, which is the value assumed in the \texttt{tdediscspec} model.
The radius $R_p$ is a normalization parameter that corresponds to radius of the hottest region. The constant $\gamma$ depends on assumptions about both the inclination angle of the disk and the disk's inner boundary condition and is limited to the range $1/2 \leq \gamma \leq  3/2$. The properties of $\gamma$ are discussed in more detail in \citet{Mummery2021} and references therein. In practice, the observed accretion disk spectrum is only weakly dependent on $\gamma$, which cannot be strongly constrained from observations \citep{Mummery2021}. In this model, $\gamma$ is therefore treated as a nuisance parameter, letting it vary between its allowed limits for each source. In fact, the 1-$\sigma$ uncertainties on $\gamma$ typically fill the entire permitted range, and as such, the $\gamma$ parameter merely extends the uncertainty range of the parameters $T_p$ and $R_p$. 


We fit our spectra with \texttt{TBabs$\times$zashift} \texttt{$\times$tdediscspec}, the model shows both better fitting quality in terms of fitting statistics (see Table \ref{tab:best_fit_ta} and Fig.~\ref{fig:model_stat}) as well as physical interpretation of the derived parameters, as compared to the aforementioned models, we hence adopted it as our main model for the soft component in TDEs.

\begin{deluxetable*}{c|C|C|l}
\label{tab:models}
\tabletypesize{\footnotesize}
\tablecaption{Summary of X-ray spectral modeling described in \S\ref{sec:X-ray_fit}.}
\tablewidth{10pt}
\tablehead{
\colhead{Model} &  \colhead{\texttt{XSPEC$^{a}$}} & \colhead{Free Parameters}  & \colhead{Notes}
}
\startdata
\hline
Absorbed power-law              & {\tt TBabs$\times$powerlaw}   & N_H, \Gamma, {\rm norm}                                 & \begin{tabular}[c]{@{}l@{}}$\bullet$ Inappropriate model for X-ray emission \\ of  non-jetted TDEs\\ $\bullet$ Overestimate intrinsic $N_H$ \\ $\bullet$ It can be used to differentiate the emission \\ of AGN and TDEs (See Fig.~\ref{fig:TDExAGN})\end{tabular}    \\ \hline
Single Temperature Blackbody    & {\tt bbodyrad}                 & $T, R$                                          & \begin{tabular}[c]{@{}l@{}} $\bullet$  Better fit than absorbed power-law\\ $\bullet$  Good fit for low S/N spectra\\ $\bullet$  Leave significant residuals for high S/N spectra\end{tabular}                                                                                                                                                                                                                   \\\hline
Accretion Disk                  & {\tt tdediscspec}^{b}              & $T_{p}$, $R_{p}$                 &                                                      \begin{tabular}[c]{@{}l@{}} $\bullet$ Good fit for most sources\\ $\bullet$ {\bf Final model$^{c}$ for source with no `hard excess'}\\ $\bullet$ Leaves residuals at high energies for source \\ with some hard emission\end{tabular}                                                                                                                                                                              \\\hline
Accretion Disk + Comptonization & {\tt simPL$\otimes$tdediscspec}$^{c}$ & $T_{p}$, $R_{p}$,  $f_{\rm sc}$, $\Gamma_{\rm sc}$ &    \begin{tabular}[c]{@{}l@{}} $\bullet$ Accounts for comptonization of a fraction of disk's photons\\ $\bullet$ {\bf Final model$^{c}$ for source with `hard excess'}\end{tabular}  
\enddata
\tablecomments{a) All models were preceded by \texttt{TBabs$\times$zashift$\times$}. b) \texttt{tdediscspec} has an additional nuance parameter $\gamma$, see text for discussion. c)  We also check whether an additional intrinsic neutral absorption was necessary by adding a \texttt{zTBabs} component to the model. Only for ASASSN-14li and AT2019dsg it significantly improve the fit, hence added to the final model. For all other sources, the intrinsic neutral absorption component was negligible.}
\end{deluxetable*}

For several sources/epochs, when fitted only with a soft component, a large residual at higher energies is present. 
Such a `hard excess' is well described by a power-law function and was already reported for both X-ray \citep[e.g.,][]{Saxton2017} and optically selected TDEs \citep[e.g.,][]{Wevers2021,Yao2022}, and is associated with Compton up-scattering of the soft photons by a corona of hot electrons near the accretion disk \citep{Titarchuk_95}. 

In AGN, this component is the dominant component of their X-ray spectra and is modeled with a phenomenological model ($F(E) \propto E^{-\Gamma}$, \texttt{powerlaw} in \texttt{xspec}). 
However, for soft and intermediate state XRBs, which in contrast to AGN and similar to TDEs, have both their main continuum and the power-law in the X-ray band, a more physically motivated and self-consistent model was developed by \citet[]{Steiner2009} to describe the Comptonization process, called \texttt{simPL} in \texttt{xspec}. The model has two free parameters, $f_{sc}$\footnote{This should not be confused with the fractional flux of the power-law with respect to the total flux, at $f_{sc} \sim 0.3$ the total flux is already dominated by the powerlaw component as shown in \S\ref{app:simpl}.} and $\Gamma_{sc}$\footnote{We will use $\Gamma_{sc}$ for the power-law photon index when the TDE X-ray spectra are fitted with  \texttt{simPL$\otimes$tdediscspec}, and $\Gamma$ when the spectra are fitted with \texttt{powerlaw} model.}; the first is the fraction of photons from the soft component that are up-scattered to create the power-law, and the second is the photon index of the resulting power-law. A more detailed discussion on using \texttt{simPL} on X-ray TDE spectra is presented in Appendix \S\ref{app:simpl}.

We employ the models described above, either \texttt{tdediscspec} or \texttt{simPL$\otimes$tdediscspec}, depending on the need for the `hard tail', based on a simple $\chi^2/d.o.f$ analysis. Therefore, our X-ray modeling has either two free parameters ($T_p$ and $R_p$), if fitted only with \texttt{tdediscspec}, or four free parameters ($T_p$, $R_p$, $f_{sc}$ and $\Gamma_{sc}$) if fitted with \texttt{simPL$\otimes$tdediscspec}. We also check whether an additional intrinsic neutral absorption was necessary by adding a \texttt{zTBabs} component to the model. Only for ASASSN-14li and AT2019dsg, $N_H$'s of respectively $\sim5\times10^{20}$ cm$^{-2}$ and  $\sim 3\times10^{20}$ cm$^{-2}$, significantly improve the residuals at the most soft energies, and were added to the final model. For all other sources the intrinsic absorption component was negligible to the fit quality which constrained their intrinsic $N_H$ to be  
$\ll 10^{20}$ cm$^{-2}$.

The stacked \swift/XRT spectrum of AT2019qiz and the second \xmm epoch of AT2019teq show a divergent degeneracy between some of the model parameters, hence no uncertainties on the parameters could be determined and therefore no interpretation of the derived best-fitting parameters will be done, although we use the best-fitted parameters to scale the count-rates to flux/luminosity. The measured temperature of second epoch of AT2019teq is hotter than the $T_p$ maximum limit on quasi-Wien approximation of \texttt{tdediscspec}, similarly no physical interpretation is attributed from the derived physical parameter of this epoch.

In Fig.~\ref{fig:fit_pars}, we show the best-fit value and its evolution for all four parameters in our modeling.  A representative \xmm observation and modeled spectra, as well as residuals for each source, are shown in Fig. \ref{fig:xmm_spec}.  In Table \ref{tab:models} we summarize the spectral fitting procedures and models describe in this section.

Our final continuum models fit very well the continuum of most observations ($\chi^2/d.o.f \sim 1.0 $). However, some sources present features that resemble absorption lines, usually associated with O\,VII ($\sim$ 0.54 keV) and/or O\,VIII ($\sim$ 0.65 keV); these can be highly blueshifted (up to 0.2$c$) and interpreted as ultra fast outflows \citep[UFO, e.g.,][]{Kara2018}. We will explore these absorption lines' detection, physical interpretation, and modeling in separate publications for those sources in which the statistical significance of the absorption detection can be assured.

\section{Results}\label{sec:results}
\begin{figure*}

\centering
\includegraphics[width=0.74\textwidth]{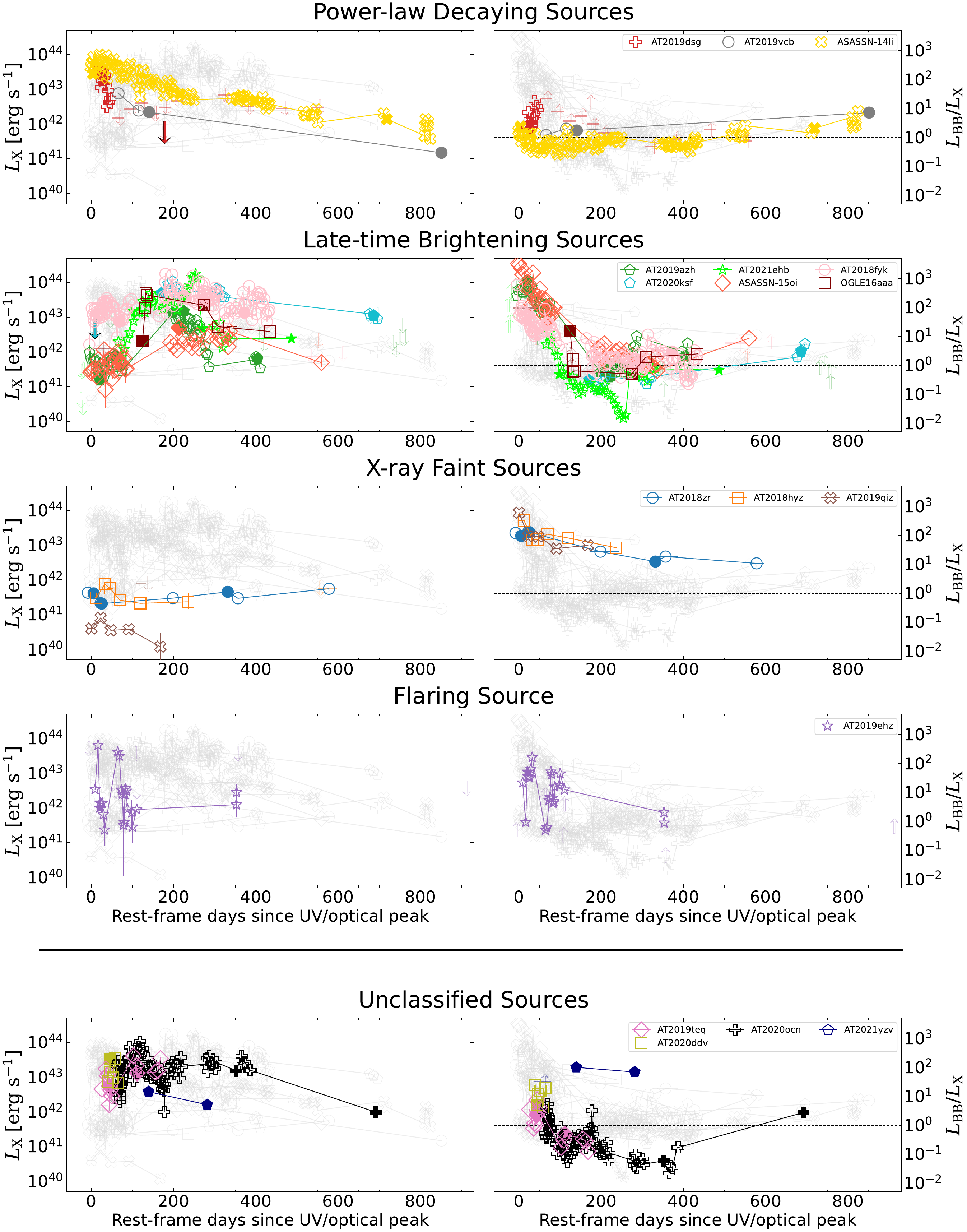}
\caption{Left: The 0.3-10 X-ray luminosity (\lx). Right: ratio of UV/optical blackbody luminosity to X-ray luminosity (\lbblx), grey dotted line shows $L_{BB}=L_X$. Top to Bottom panels show the different classes of X-ray evolution as described in \S\ref{sec:lc_div}.}
\label{fig:lc_diversity}
\end{figure*}


In Fig. \ref{fig:fit_pars}, we show the best-fitting parameters from the X-ray spectral fitting, first the main parameters of \texttt{tdediscspec}, the peak temperature ($T_p$) and apparent radius in which $T_p$ occurs ($R_p$), and for the power-law component (\texttt{simPL}), the fraction of the comptonized photons ($f_{sc}$) and the photon index of the power-law ($\Gamma_{sc}$), the later is only shown for three sources (AT2018fyk, AT2020ocn, AT2021ehb) in which a strong enough power-law is detected at some epoch (i.e., $f_{sc} \geq 0.1$), given the high uncertainty in $\Gamma_{sc}$ when the component is only marginally detected. We also do not show the measurements of $T_p$ and $R_p$ for the epochs in which  $f_{sc} \geq 0.2$ given that the power-law emission completely dominates the spectra; hence no trustworthy information on the underline thermal emission can be recovered, as discussed in \S\ref{app:simpl}.

In Fig.~\ref{fig:lc_diversity}, we show in left panels the neutral absorption corrected 0.3-10 keV luminosity (\lx) and in the right panels the ratio (\lbblx) between the UV/optical integrated luminosity ($L_{BB}$, Eq. \ref{eq:lbb}) and \lx, for the light curves of all the sources, as classified below.

\subsection{Diversity of X-ray Light Curves}\label{sec:lc_div}
A surprising characteristic of optically discovered X-ray bright TDEs -- first observed in ASASSN-15oi by \citet{Gezari2017} -- is that their X-ray light curve does not necessarily follow the theoretically expected fallback rate 
\citep[$\propto  t^{-5/3}$,][]{Rees1988}, not even a more general lower-law decay ($\propto t^{-\alpha}$), which is now established as the general evolution of their optical component \citep{Hammerstein2022}.  Instead, a wide diversity 
of time evolution and luminosity ranges are observed, in extreme contrast with some predictions, for example those by \citet{Lodato2011}.
This diversity also means that the time evolution of the ratio between the UV/optical 
and X-ray components also evolves in distinct ways in different sources. To search for a general picture of such diversity, we 
classified the sources in groups with similar evolution in terms of \lx and \lbblx; in Fig. \ref{fig:lc_diversity}, we show from top to bottom panels the following classes:

\textit{Power-law Decaying} -- ASASSN-14li, AT2019dsg and AT2019cvb are the only three source to show prompt bright (\lx $\geq 10^{43}$ \ergs) X-ray emission during the optical peak. The three sources show a power-law-like decay X-ray light curve, although the decay rate of the X-ray light curve in AT2019dsg is much higher than ASASSSN-14li and AT2019cvb. Given the X-ray behavior, their \lbblx do not show much variability staying in the $0.5 \leq$ \lbblx $\leq 10$ range during the entire evolution of the sources.

\textit{Late-time Brightening} -- Most sources in our sample show a \lbblx $\gg 100$ near the optical peak, usually resulting from the faint X-ray emission (\lx $\leq 10^{42} $ \ergs) at early times. However, between 100-200 days after the optical peak, they show a significant brightening (more than an order of magnitude, \lx $\geq 10^{43}$ \ergs) in the X-rays, simultaneously to the UV/optical dimming and plateau, consequently \lbblx tends to approach $\sim$ 1.

\textit{X-ray Faint} -- AT2018zr, AT2018hyz, and AT2019qiz also show \lbblx $\gg 100$ near the optical peak, again resulting from the faint X-ray emission (\lx $\leq 10^{42} $ \ergs) at early times. However, the three sources never show a bright X-ray phase, although similar with previous classes their \lbblx values also decrease with time.

\textit{Flaring} -- AT2019ehz show a unique behavior, the X-ray light curve show flares of almost two orders of magnitudes, from \lx $\sim few \times 10^{41}$ \ergs up to \lx $\sim few \times 10^{43}$ \ergs in a short scale of tens of days, while the \lbb show a standard smoothly evolution.
      
We could fit 13 of our 17 sources in these four classes; however, for the remaining sources, we do not have enough time coverage  to assign them to one of these classes of X-ray evolution, either lack of sampling (AT2021yzv), lack of long-term follow-up (AT2019vcb and AT2020ddv), or lack of observations within the first 50 days from the optical peak (AT20220ocn). The later being fundamental to access whether the prompt bright X-ray emission is present or not.

\subsection{Temperature and Radius Evolution}\label{sec:T_x_L}

In this section, we explore the evolution of the temperature ($T_p$) and radius ($R_p$) derived from the X-ray continuum fitting (\texttt{tdediscspec}); we focus our analyses in the sources with higher temporal coverage in which some assessments on the long-term evolution of the derived parameters can be made. Due to the degeneracy between $f_{sc}$ and $T_p$, at high values of  $f_{sc}$, and the underline thermal can not be uniquely recovered (see \S\ref{app:simpl}), we exclude the sources with strong corona formation (i.e., $f_{sc} \geq 0.1$) from this analyses, these sources will be separately discussed in \S\ref{sec:soft_hard}.

\begin{figure}

\includegraphics[width=1.0\columnwidth]{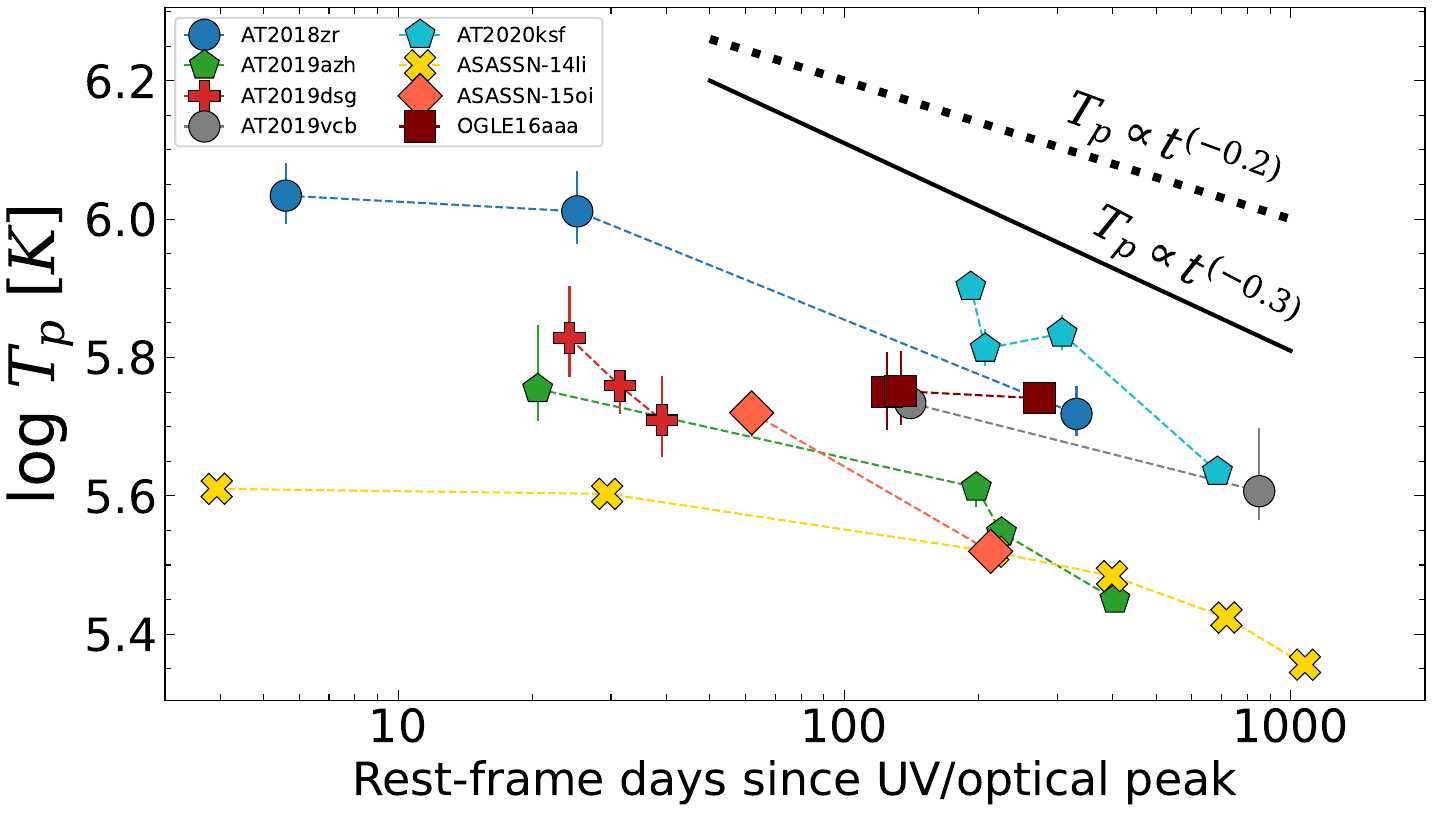}
\caption{Peak temperature ($T_p$) of the accretion disk model as function of days since the UV/optical peak. Only sources with at least two fittable spectra and with no corona emission (see \S\ref{sec:soft_hard}) are shown. The solid black line show the expected theoretical evolution for a finite stress at the innermost stable circular orbit.}
\label{fig:T_x_t}
\end{figure}

The temporal evolution of $T_p$ is shown in Fig. \ref{fig:T_x_t}; the cooling of the X-ray continuum is clearly observed: most sources show a peak X-ray temperature at their first available spectra, i.e., the closest to the peak of the UV/optical emission, with a decreasing $T_p$ with time. This behavior is observed even for the sources with faint X-ray emission at early times.

In a Newtonian time evolving standard disk \citep{Shakura1973,Cannizzo1990}, the peak temperature 
follows a power law in time for a power-law declining accretion rate, $T_p \propto t^{(-n/4)}$, where $n$ depends on the boundary conditions of the accretion disk: for finite
stress at the innermost stable circular orbit (ISCO), $n \approx 0.8$, for a vanishing ISCO stress $n \approx  1.2$ \citep{Mummery2020}. These solutions seem to qualitatively agree with the observed behavior shown by the filled ($T_p \propto t^{(-1.2/4)}$) and dotted ($T_p \propto t^{(-0.8/4)}$) black lines in  Fig. \ref{fig:T_x_t}. The current quality/cadence of available data on X-ray TDEs does not allow for a more detailed assessment of the temperature evolution.

For a standard disk emission, $T_p$ should also correlate with the bolometric luminosity, in the form $L_{disk,bol} \propto 
T^{4}$, while the relation with the observed X-ray luminosity should have a more general form of \lx $\propto T^{\alpha}$, where 
$\alpha$ is related to both the temperature evolution (i.e., $n$) and the measured temperature itself, given that the latter dictates the fraction of the bolometric luminosity emitted in the X-ray band.

\begin{figure}
    
    \includegraphics[width=1.0\columnwidth]{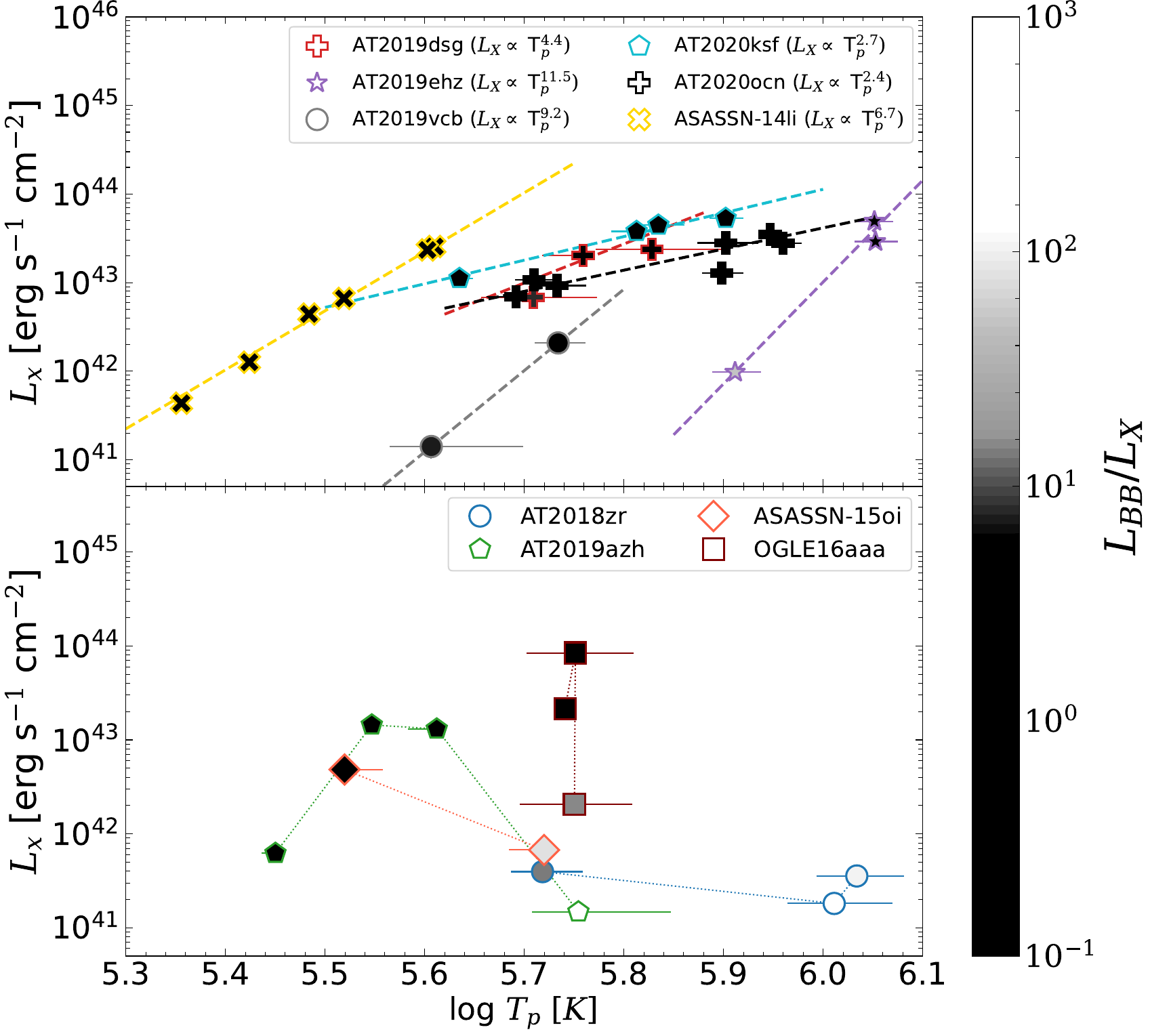}
    \caption{X-ray luminosity (\lx) as a function of peak temperature ($T_p$). The upper panel show sources where a \lx $\propto$ $T_p^{\alpha}$ relation is observed, with the best-fitted $\alpha$ for each source shown in the legend. The bottom panel show the source where a clear decoupling between \lx and $T_p$ is present. The color in which the markers are filled maps the \lbblx ratio between the UV/optical luminosity and the \lx following the color-bar in the right side if the figure.}
    \label{fig:T_x_L}
\end{figure}

In the top panel of Fig. \ref{fig:T_x_L}, we show that such correlation is observed for several sources. However, some sources, namely AT2018zr, AT2019azh, ASASSN-15oi, and OGLE16aaa, show a decoupling between \lx and $T_p$. Such decoupling is a result of their maximum $T_p$ occurring at early times (like all other sources), while their X-ray luminosity is at the faintest levels at early times (\lx $\leq10^{42}$), which runs contrary to the expectation of higher luminosity at higher temperatures. All these sources are either \textit{Late-time brightening} or \textit{X-ray faint} sources. The resulting decoupling, driven by the faint early-time X-ray emission, can also be seen by the color of the points in Fig. \ref{fig:T_x_L} representing their \lbblx values: all sources that show \lx $\propto T^{\alpha}$ (top panel) have \lbblx $< 10$ in all epochs, sources with decoupling between \lx and $T_p$ have epochs with \lbblx $\gg 10$, and it is in those specific epochs that the decoupling is observed. These results points towards a suppression of the X-ray flux/luminosity in this sources/epochs, while the observed $T_p$ seems be following its expected behavior. An X-ray spectrum of AT2020ksf (another \textit{Late-time brightening} source) at its early-time X-ray faint phase is not available to confirm whether such decoupling from the expected relationship is also present.

A similar analysis can be done regarding the apparent radius of the peak disk temperature ($R_p$); for a standard disk (with no 
ionizing/neutral absorption or/and reprocessing), the temperature should peak near the ISCO, i.e., $R_p \approx R_{ISCO} = (1-6) \times R_g$ 
(depending on the MBH spin), where $R_g = GM_{BH}/c^2$ is the gravitational radius. Given that the (systematic plus statistical) uncertainties in \mbh derived from scaling relations (see \S\ref{sec:M_BH}) are in the order of 0.5 \textrm{dex}, any measured  radius in the $0.3 \leq R_p/R_g \leq 20$ range is still statically consistent with being emitted near the $R_{ISCO}$.

From  Fig. \ref{fig:R_x_L}, it is clear that for most sources/epochs, the recovered $R_p$ are within the physically valid interval (i.e., $0.3 \leq R_p/R_g \leq 20$) hence are consistent with $R_{ISCO}$, in agreement with \citet{Mummery2023} findings. However, some epochs/sources show unphysically low values, i.e., $R_p/R_g \ll 0.3$, interestingly these sources show a large apparent $R_p$ evolution, from unphysical values at early times to valid reasonable values at late times;
a consequence of this is that the $R_p/R_g$ seems closely connected with \lbblx (and hence the distinct light curve evolution classes): the epochs with $R_p/R_g \ll 0.3$ are the epochs with higher \lbblx (this can also be seen in Fig. \ref{fig:SED_plot}). This correlation is not only observed for the general sample but also in distinct epochs of the same source (see, e.g., AT2019azh, OGLE16aa, AT2018zr in Fig. \ref{fig:R_x_L}). 

The \lbblx (hence the shape of the SED) for a standard accretion disk also has a limiting range of values; based on our simulation shown in \S\ref{app:disk_lbb_lx}, for $T_p$ range measured in our sample (i.e., $5.5 \leq \rm{log} ~ T_p \leq 
6.1$ K), the ratio between the observe \lbb and \lx, can only be  $5\times10^{-2} \leq $ \lbblx $\leq 70$. Combing the valid ranges for \lbblx and for $R_p/R_g$, in Fig. \ref{fig:R_x_L} we show in a gray shaded region the space of 
parameters in which derived properties \textit{could}, in principle, be explained by a \textit{bare/unreprocessed} accretion disk. For observations that fall outside this region, however, the derived parameters are inconsistent with the
ones of a \textit{bare/unreprocessed}  disk; hence additional radiative processes must be present. Interestingly, the epochs/sources in which the measured \lbblx and $R_p/R_g$ fall outside the disk region are the same epochs/sources that are decoupled from the \lx $\propto T^{\alpha}$ relation. i.e., all sources showing faint X-ray emission at early times. 

In the following paragraphs we aim to demonstrate that the apparent unphysical $R_p/R_g$ values derived from \textit{X-ray faint} sources and the early-time observations of \textit{Late-time brightening} are evidence for the suppression of the emitted X-ray in these sources/epochs and can be explained by the way that $R_p$ is `measured' by the X-ray continuum fitting.

In the color-corrected quasi-Wien approximation of \texttt{tdediscspec}\footnote{The specific equations are for \texttt{tdediscspec} but the same arguments holds for any thermal model.}, the X-ray spectrum ($F_X (\nu)$) is related to $R_p$ and $T_p$ as follows:

\begin{equation}
    F_X (\nu) \propto \left ( \frac{R_{p}}{D}\right )^2 \ \tilde{T}_{p}^{\gamma} \  exp\left ( -\frac{h\nu}{k\tilde{T}_{p}} \right )
\end{equation}

where $\tilde{T}_{p} = f_{col}f_{\gamma}T_p$ (see definitions in \S\ref{sec:X-ray_fit}), and $D$ is the source luminosity distance. This means that the shape of the X-ray spectra shape depends exclusively on $T_p$, but not on $R_p$ which is only a \textit{`physically scaled'} normalization factor that translates the observed count rate per energy bin to a flux per energy bin. Similarly,

\begin{equation}\label{eq:Lx_disk}
  L_X \propto R_{p}^{2} \ \tilde{T}_{p}^{\gamma} \  \int_{\nu_{i}}^{\nu_f} exp\left ( -\frac{h\nu}{k\tilde{T}_{p}} \right ) d\nu
\end{equation}

where $\nu_i=\frac{0.3~keV}{h}$  and $\nu_f=\frac{10~keV}{h}$. Therefore for a constant $T_p$,  $L_{X} \propto R_p^2$, while $L_X$ decreases with decreasing $T_p$ for constant $R_p$. 

As we have shown, all sources -- including those with faint X-ray emission at early times -- show a decreasing or constant $T_p$ with time; this means that their faint X-ray emission and late time X-ray brightening (or constant $L_X$, in AT2018zr) will translate into to an extremely low $R_p$ at early times and an order of magnitude increase in $R_p$ at late times. Such behavior is a consequence of eq. \ref{eq:Lx_disk} in which $R_p$ can not be constant with time if $T_p$ is decreasing (or held constant) and \lx is not decreasing -- as is the case for all sources in the bottom panel of Fig. \ref{fig:T_x_L}.

One can take the earliest ($\Delta t =21$ days) spectrum of AT2019azh as an example: the measured temperate and apparent radius were $\rm{log} \ T_p \approx 5.75$ K and $R_p/R_g \approx 6 \times 10^{-2}$, while \lx $\approx 2\times 10^{41}$ \ergs. In order for the $R_p/R_g$ to be within the physically valid range (i.e., to be at least 0.3), based on eq. \ref{eq:Lx_disk}, the observed luminosity would have to be higher by a factor of $\left ( \frac{0.3}{0.06} \right )^{2}  \approx 25$, similarly, for the early time  $R_p/R_g$ to be at the same value as the late-time $R_p/R_g \approx 3$ (hence a physically valid and approximately constant value during its entire evolution) the observed early time \lx would have to be higher by a factor of $\left ( \frac{3}{0.06} \right )^{2}  \approx 2 \times 10^{3}$. Such an increase would make the early time \lx to be coupled to the $T_p$ during the entire evolution of the source, as can be seen by the bottom panel of Fig. \ref{fig:T_x_L}. 

A similar analysis can be done for the early time spectra of all the \textit{Late-time Brightening} and \textit{X-ray faint} sources, that show this unphysical value/evolution of the apparent $R_p$, for all of them an increase in the \lx by a factor $\gg 10$ is necessary for a physical value of $R_p/R_g$, given their measured temperature evolution. 

In summary, the unphysical values of $R_p/R_g$, accompanied by decoupling between $L_X$ and $T_p$, and a SED shape (\lbblx) inconsistent with an accretion disk, indicates suppression of the emitted X-rays in these epochs/sources. However, such suppression of the X-ray emission seems to have a small effect in the measured $T_p$ -- given that the cooling of the accretion disk is still observed and is independent of the observed flux level -- while suppression of the total observed X-ray flux is high. We will discuss possible mechanisms responsible for such suppression in \S\ref{sec:disc_lc}. In \citet{Mummery2023}, when the authors average all $R_p$ values obtained in each spectrum to find a $\left \langle R_p \right \rangle$ for each source, the information on the unphysical nature of the $R_p/R_g$ obtained at early times was missed. Furthermore important data such as high S/N X-ray spectra of, for example, AT2019azh at early times and AT2018zr at late times, as well as a \mbh-$\sigma_*$ measurement of AT2018zr's black hole mass were not available to the authors, but are presented here. Although, we agree on the authors main claim, i.e. $R_p$ is tracing $R_{ISCO}$, we show that this is only valid in the cases where the X-rays are not suppressed and the SED is consistent with a \textit{bare/unreprocessed} disk, which is usually not the case for the very early times of optically discovered TDEs (as we will discuss in \S\ref{sec:SED}).

\begin{figure}
\includegraphics[width=1.0\columnwidth]{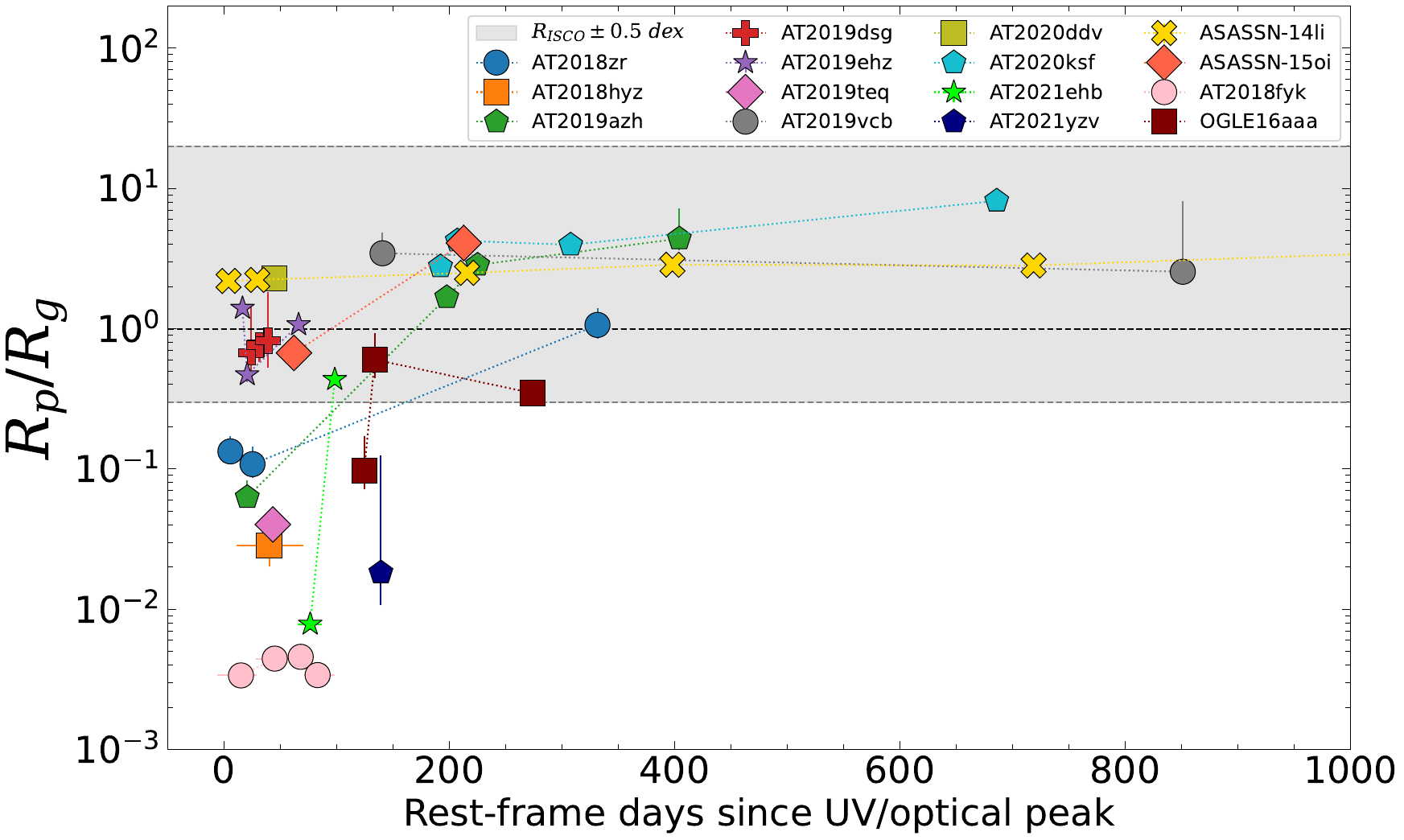}
\caption{Apparent radius ($R_p$) of the peak temperature normalized by the gravitational radius ($R_g$) as function of days since the UV/optical peak. Only sources with no corona emission (see \S\ref{sec:soft_hard}) are shown. Gray shaded region shown the  $0.3 \leq R_p/R_g \leq 20$ interval in which the measured $R_p/R_g$ is statistically consistent with $R_{ISCO}$.}
\label{fig:R_x_t}
\end{figure}

\begin{figure}
    \centering
    \includegraphics[width=1.0\columnwidth]{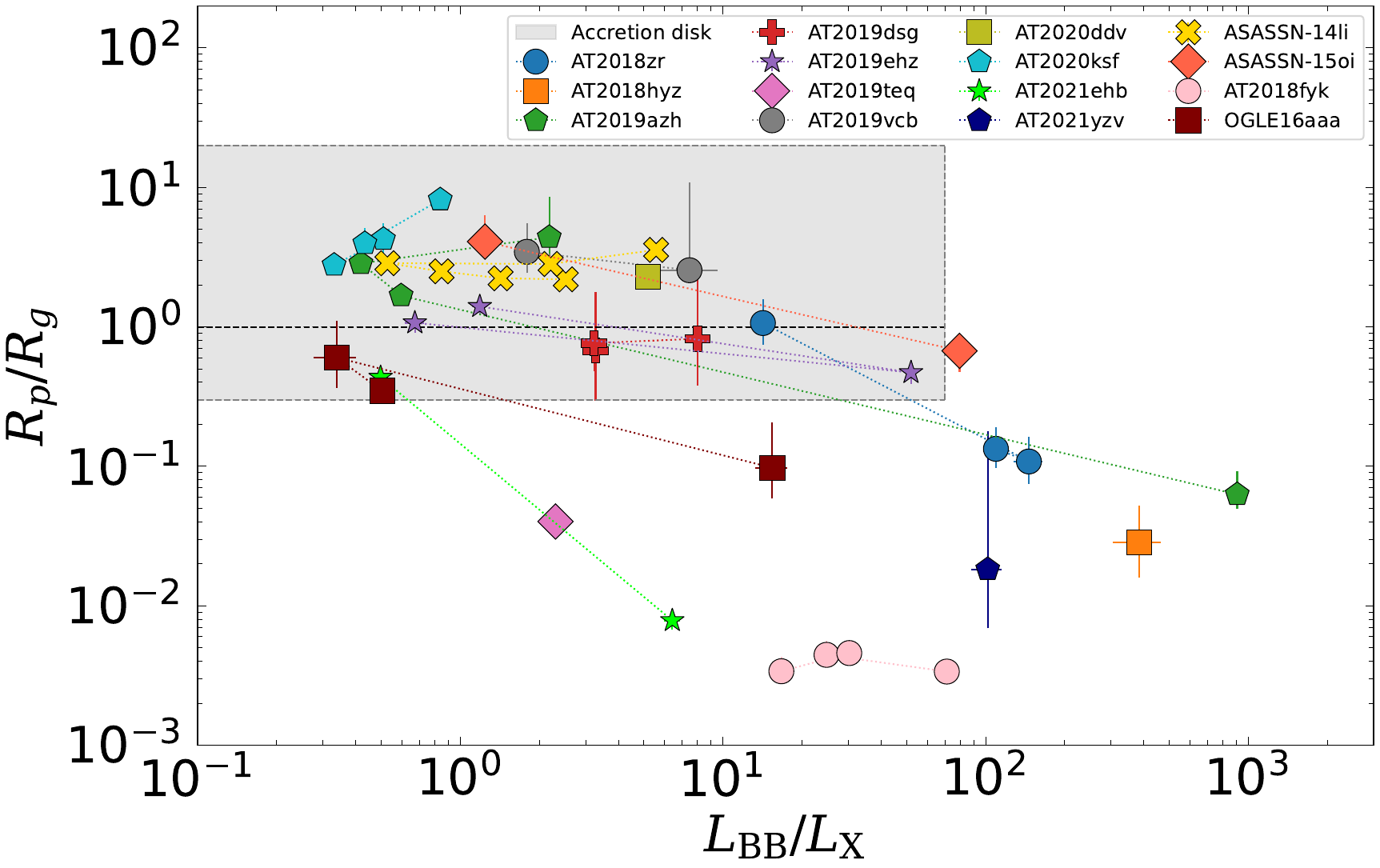}
    \caption{Distribution of apparent disk radius ($R_p$) normalized by the gravitational radii ($R_g$), and the \lbblx ratio between the UV/optical luminosity and the X-ray luminosity. The shaded gray region delimits the parameters space in which the emission can be explained by a \textit{bare/unreprocessed} standard accretion disk, see text for details. Several sources/epochs fall outside this region having unphysically low $R_p/R_g$ values while also having SED shapes (\lbblx) that, deviate form the allowing SED shape of disk, and require additional radiative processes.}
    \label{fig:R_x_L}
\end{figure}

\subsection{Soft $\rightarrow$ Hard Transition: Real-Time Corona Formation}\label{sec:soft_hard}

\begin{figure*}
    
    \centering
    \includegraphics[width=0.7\textwidth]{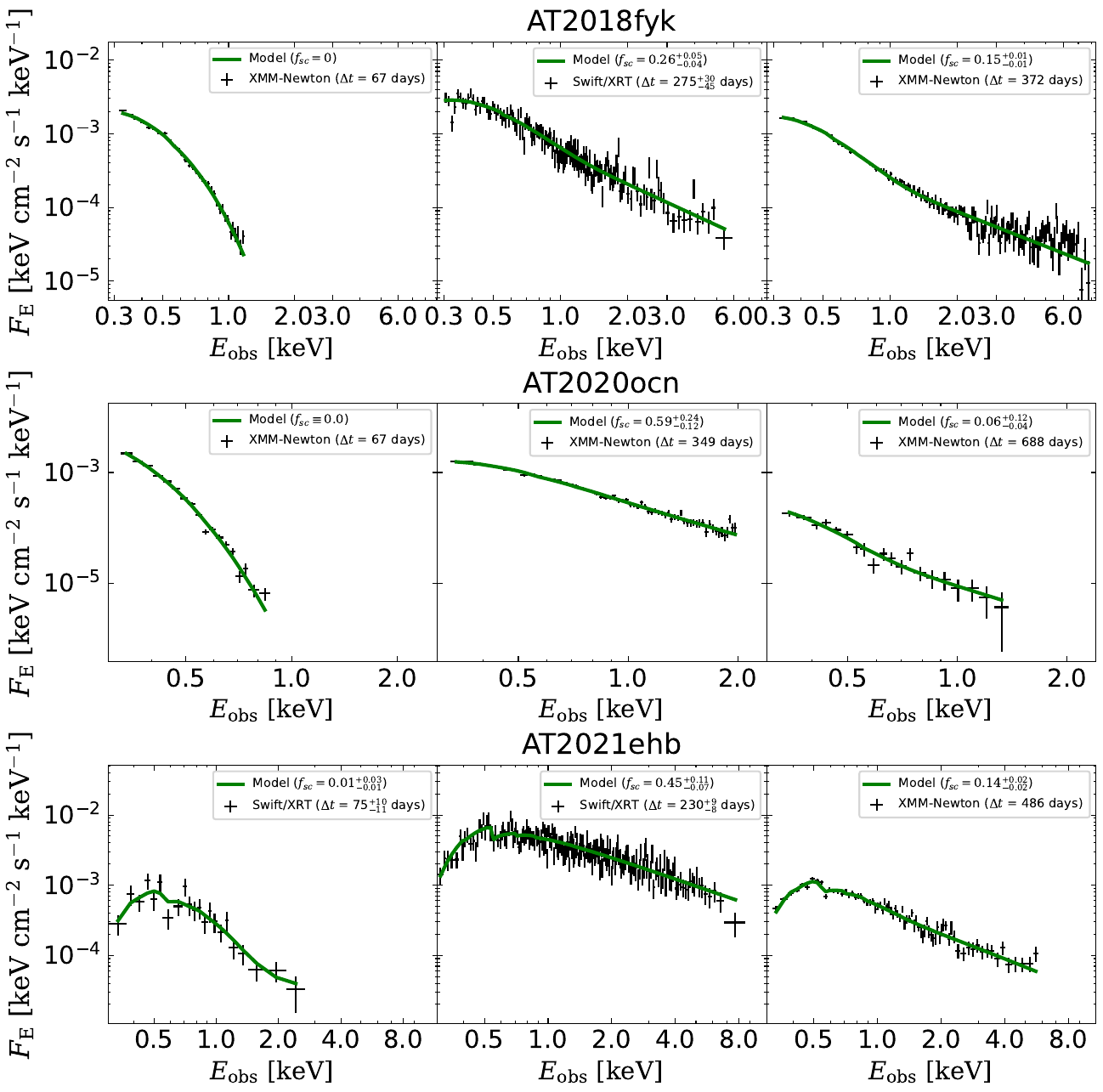}
    \caption{Spectral evolution of the three sources that show soft $\rightarrow$ hard state transition, i.e., corona formation. From top to bottom: AT2018fyk, AT2020ocn, and AT2021ehb. From left to  right: early time soft state spectra, hardest state spectra, and late-time intermediate state spectra. Unfolded spectra are show in black crosses, best fitted model are show in green.}
    \label{fig:hard_state}
\end{figure*}

The X-ray spectra of AT2018fyk, AT2020ocn, and AT2021ehb exhibit extreme softness at early times ($\Delta t \leq 100$ days), similar to other spectra in our sample. However, these three sources undergo a transition to a hard power-law-like state with a timescale of $\sim$200 days after the UV/optical peak. The power-law component dominates the disk emission in this state and extends to much higher energies than in the soft state. As an extreme example, the \nustar spectra of AT2021ehb presented by \citet{Yao2022} show power-law emission detected above background up to 30 keV.

In the case of AT2018fyk, a faint hard excess (\(f_{sc} \leq 0.05\)) emerges around \(\Delta t \approx 50\) days, followed by a rapid formation of a strong coronal component between \(50 \leq \Delta t \leq 200\) days. By \(\Delta t = 209^{+34}_{-35}\) days, the \(\text{\swift/XRT}\) spectrum already exhibits \(f_{sc} \approx 0.2\), and the corona emission peaks at \(\Delta t = 209^{+45}_{-30}\) days with \(f_{sc} \approx 0.3\). During this phase, the power-law component dominates the X-ray flux over the thermal component, with the fraction of up-scattered photons decreasing to \(f_{sc} \approx 0.15\) at \(\Delta t \approx 350\) days, remaining relatively constant until the source becomes undetectable in the X-rays at \(\Delta t = 500\) days. 
In the case of AT2020ocn, the state transition is more extreme, remaining completely soft and disk-dominated (\(f_{sc} = 0.00\)) up to \(\Delta t \approx 200\) days. However, a spectrum taken after a seasonal gap at \(\Delta t \approx 290\) days reveals the corona dominating the X-ray emission (\(f_{sc} \approx 0.5\)). The source persists in this hard state for at least 100 days, transitioning to an intermediate state again at \(\Delta t \approx 700\) days. 
For AT2021ehb, a hard excess (\(f_{sc} \approx 0.02-0.05\)) is present from the first X-ray detections at \(\Delta t \approx 70-100\) days. \(f_{sc}\) then gradually increases to \(\sim 0.15\) at \(\Delta t \approx 200\) days, followed by an abrupt transition to a corona-dominated state (\(f_{sc} \approx 0.5\)) at \(\Delta t \approx 250\) days, resembling typical AGN X-ray spectra. The source subsequently transitions back to an intermediate state around \(\Delta t \approx 300\) days. Figure \ref{fig:hard_state} illustrates the evolution of the softest early-time spectrum (left panels), the hardest spectrum (middle panels), and the intermediate state at very late times (right panels) for the three sources, depicting the formation of the corona.

Transitions from a soft disk-dominated state to a hard corona-dominated state are commonly observed in stellar-mass black holes in X-ray binary systems. They follow a standard \textit{q-shape} evolution in the hardness-intensity diagram \citep[HID, e.g.,][]{Remillard2006,Wang2022}. However, in the case of MBHs, the corona is a ubiquitous and dominant component of the X-ray spectra of AGN. Dramatic state transitions, such as the appearance or disappearance of X-ray power-law emission, are usually not observed in AGN, except in the case of the Changing-look AGN 1ES 1927+654 \citep{Trakhtenbrot2019,Ricci2020}, where the corona was destroyed and later reformed. Therefore, X-ray bright TDEs provide a new window for studying the emission and formation of this poorly understood component.

A study by \citet{Wevers2021} suggested that AT2018fyk exhibits a ``Fainter harder brighter softer" behavior similar to X-ray binary outbursts. Additionally, \citet{Wevers2020} argued that this behavior could be ubiquitous in X-ray bright TDEs. The HID analysis conducted by \citet{Wevers2021} involved using the $\alpha_{OX}$ parameter, defined as the logarithmic ratio between the UV flux (representing the disk emission) and the 2 keV flux (representing the corona emission), along with the bolometric luminosity ($L_{\rm Bol}$) normalized by the Eddington luminosity ($L_{\rm Edd}$).

\begin{figure*}
    \centering
    \includegraphics[width=\textwidth]{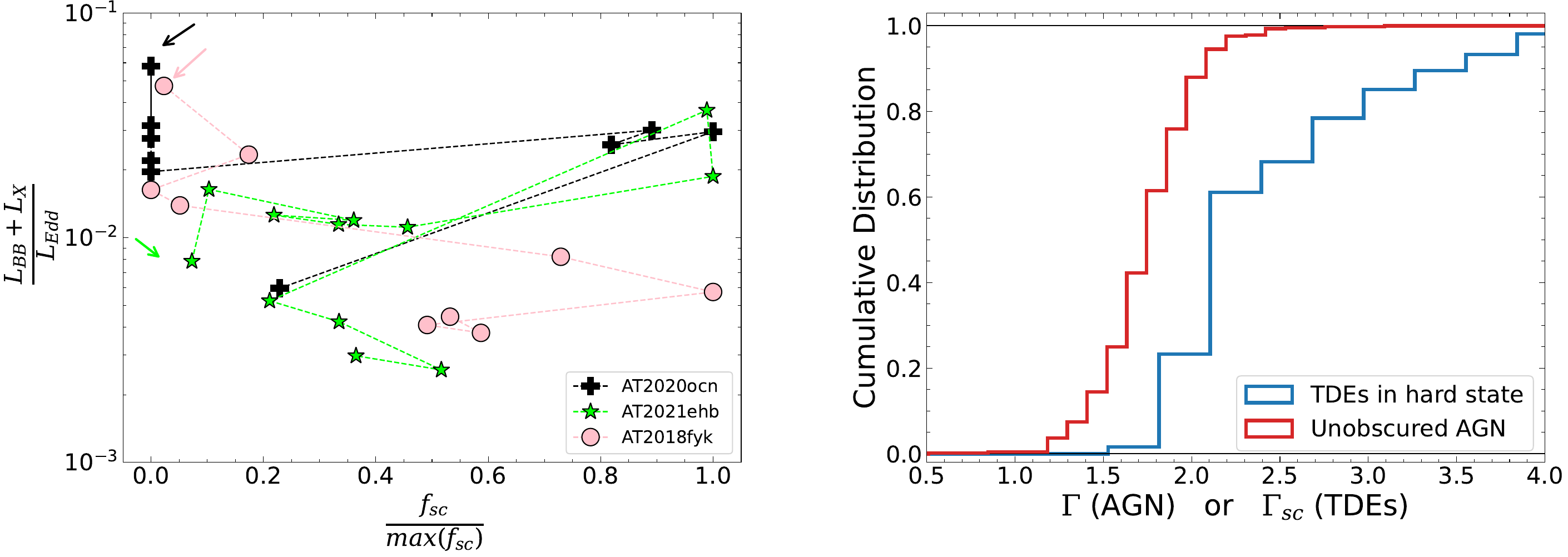}
    \caption{Properties and evolution of the corona emission in TDEs. Left: Hardness intensity diagram (HID), the total observed luminosity (\lbb + \lx) in units of the Eddington luminosity, as a function of the hardness as traced by the normalized fraction ($f_{sc}$) of the photons upper-scatter by the corona. The arrows indicate the first available spectra for each source, and the points are connected by increasing $\Delta t$. Right: comparison between the power-law index of the corona emission spectra in local AGN ($\Gamma$, blue) from the BAT AGN Spectroscopic Survey (BASS), and TDEs ($\Gamma_{sc}$, red) that show corona formation (AT2020ocn, AT2021ehb, AT2018fyk).}
    \label{fig:corona_prop}
\end{figure*}

The HID in the left panel of Fig. \ref{fig:corona_prop} illustrates corona formation for the three sources using $\frac{f_{sc}}{\rm{max}(f_{sc})}$ for hardness and $\frac{L_{BB}+L_X}{L_{Edd}}$ for intensity. Representing the relative corona strength and intensity, respectively. Given the uncertainty in the spectral energy distribution (SED) shape between the UV and X-ray regions, we use ${L_{BB}+L_X}$ as a proxy for $L_{\rm Bol}$\footnote{A significant portion of the bolometric luminosity should be emitted in the unobserved extreme UV (EUV) range, the shape of the SED between the UV and X-ray regions is not fully understood and highly dependent on models, especially when the total SED deviates from a standard accretion disk SED (e.g., when \lbblx $\gg$ 10, see \S\ref{sec:SED}). The actual value of $L_{\rm Bol}$ would be much higher if the EUV emission were taken into account.} and $\frac{L_{BB}+L_X}{L_{Edd}}$ as a proxy for the Eddington ratio. From the left panel of Fig. \ref{fig:corona_prop}, it can be observed that all three sources start in a soft state and then quickly transition to a hard state. The transition occurs at $\frac{L_{BB}+L_X}{L_{Edd}}$ values between $5\times10^{-3}$ and $5\times10^{-2}$ for the three sources. However, as time passes and the luminosity decreases, all sources undergo further transitions, either returning to a completely soft state or transitioning to an intermediate state ($0.05 \leq f_{sc} \leq 0.15$); this behavior contrasts with the \textit{Fainter harder brighter softer} pattern.

In the right panel of Fig. \ref{fig:corona_prop}, we compare the power-law index ($\Gamma_{sc}$) of the emerging corona in the three TDE showing a state transition ($f_{sc} \geq 0.1$) to the corona power-law index ($\Gamma$) measured from the population of unobscured ($N_H \leq 10^{22}$ cm$^{2}$) AGN in the BAT AGN Spectroscopic Survey \citep[BASS,][]{Ricci2017}. In the AGN population, $\Gamma$ exhibits values in a narrow range between 1.2 to 2.5 with a mean value of $\Gamma \approx 1.7$, TDEs however, even at their harder state are rarely as hard as AGN, instead their $\Gamma_{sc}$ can exhibit a broader range of values between 1.7 and 4.0. 

The characteristic power-law spectrum, is thought to be produced by the 
Comptonization of lower energies photons emitted by the accretion disks by a corona of hot electrons 
\citep{Haardt_91,Titarchuk_95} located within a light-hour from the accreting massive black hole \citep[][]{Fabian2009}.
The process depends on the Compton parameter $y$, given by:

\begin{equation}\label{eq:y}
    y = \frac{k \,T_e }{m_e \,c^2} \ \rm{max}(\tau_e, \tau_e^2)
\end{equation}

\noindent
where $\tau_e$ and $T_e$ are, respectively, the corona optical depth and electron temperature of the corona; the resulting power-law spectrum has a photon index has a inverse dependence on $y$.

Although the exact mechanism responsible for the corona formation is still to be fully understood, the need for a strong magnetic field is common feature of the different models. In all three sources, the spectrum is soft at early times, and the complete hardening of the X-ray spectrum is only observed at $\Delta t \geq 200$ days. This gradual hardening process suggests that it takes approximately $10^2$ days for the magnetically dominated hot corona region to develop. The initial weak magnetic fields present in the bound debris could undergo amplification through the combined effects of the disk's differential rotation and the magnetorotational instability \citep{Balbus1991,Miller2000,Yao2022}.

Although the corona forms in these sources, their hard state is short-lived, transitioning back to a soft/intermediate state at later times ($\Delta t \gg$ 400 days). This suggests that the high optical depth corona cannot be sustained as the accretion rate onto the black hole decreases. Another indicator of inefficient corona formation is the measured $\Gamma_{sc}$ (see the right panel of Fig. \ref{fig:corona_prop}). TDE spectra exhibit higher power-law indices compared to typical AGN spectra, translating to lower values of $T_e \times \max(\tau_e, \tau_e^2)$ -- based on the equation \ref{eq:y}, and the inverse relation between $\Gamma$ and $y$ -- when compared with AGN corona. Distinguishing between $\tau_e$ and $T_e$ effects requires detecting the cut-off energy ($E_{cut}$) of the power-law spectra, feasible with instruments like \textit{NuSTAR} for bright AGN but currently challenging for TDEs. The proposed High Energy X-ray Probe \citep[HEX-P]{Madsen2019} may enable $E_{cut}$ measurements in bright TDEs. The reasons why only these three sources exhibit corona formation remains unclear; while all three have \mbh $\geq 5\times 10^{6}$\msun -- at the high-mass end of the TDE \mbh function and consistent with sub-Eddington accretion requirements -- a high \mbh alone does not seem sufficient, as other TDEs with similar \mbh did not show such state transitions.

\subsection{SED Evolution}\label{sec:SED}
\begin{figure}
    \centering
   
    \includegraphics[width=\columnwidth]{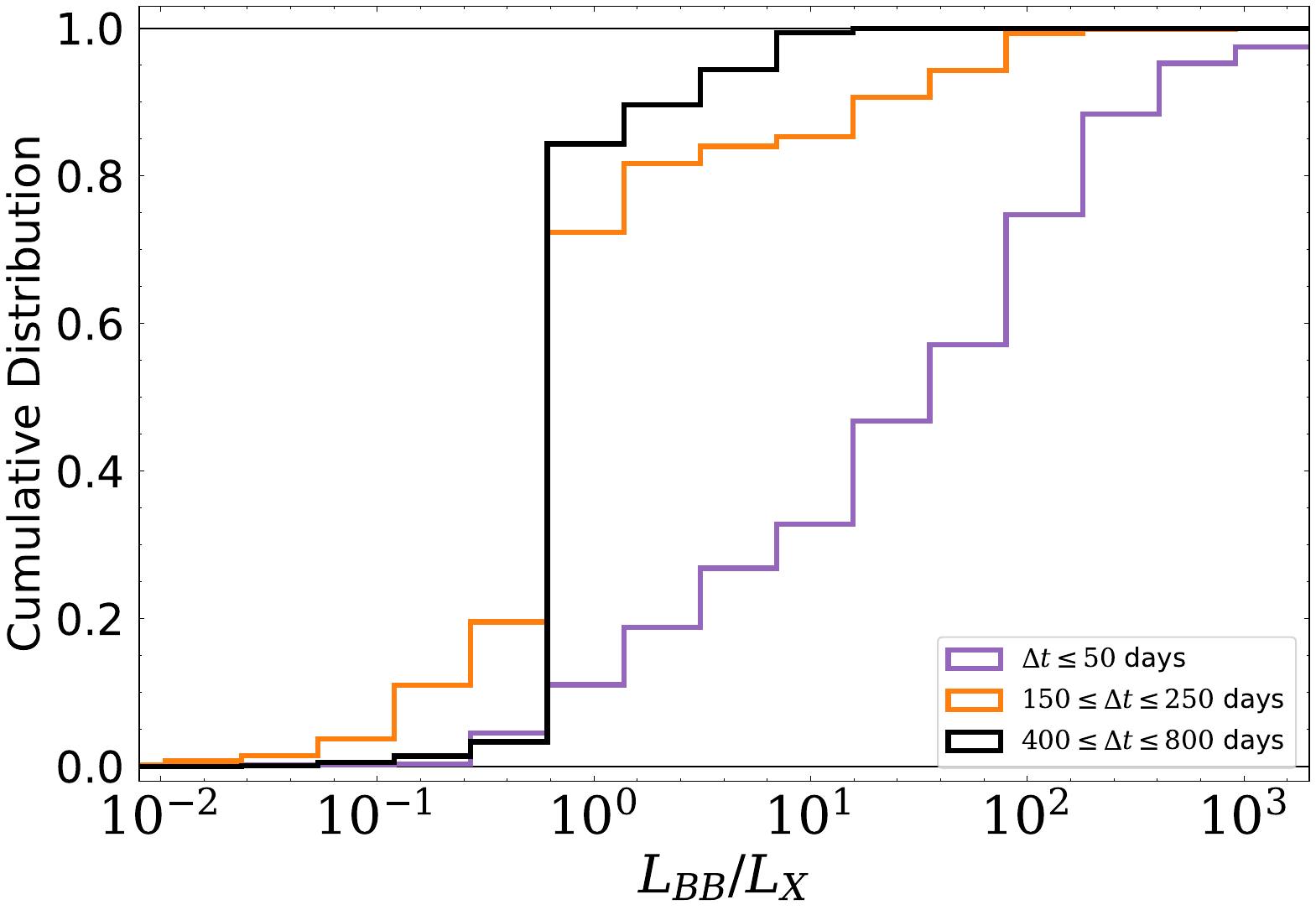}
    \caption{Evolution of the distribution of SED shapes, as trace by the ratio (\lbblx) between the UV/optical luminosity (\lbb) and the 0.3-10 keV luminosity (\lx). Each color show the distribution in a different time bin.
    The contribution of each source for the total distribution was weighed by the number of spectral observation available in each $\Delta t$ interval.}
    \label{fig:SED_hist}
\end{figure}

A natural consequence of the diversity of X-ray light 
curves and the uniformity of the UV/optical light curves 
is that the spectral energy distribution (SED) shows very 
distinct shapes and evolution. The shape of the broad-band SED can 
be probed by the \lbblx ratio. As shown in the right panels of Fig. \ref{fig:lc_diversity}; these ratios can vary 
between $\rm{few} \times 10^{3}$ and $\rm{few} \times 10^{-2}$. In Fig. \ref{fig:SED_hist}, we show the cumulative distribution of \lbblx at  three time bins, early times ($\Delta t \leq 50$ days) in purple, and at late times ($150 \leq\Delta t \leq 250$ days) in orange, and very late times ($400 \leq\Delta t \leq 800$ days) in black. The SED also show a noticeable trend: at early times they have \lbblx as large 
as 3000 and as low as 0.5, but with most sources showing  \lbblx $\geq$ 10; with increase time from optical peak this range of \lbblx shrinks, and at very late times all sources show $0.5 \leq$  \lbblx$ \leq 10$.

As discussed in \S\ref{sec:T_x_L} and demonstrated in Appendix \S\ref{app:disk_lbb_lx}, the SED produced by a \textit{bare/unreprocessed} standard accretion disk with $T_p$ in the range of values find in TDEs (i.e, $5.5 \leq \rm{log} \  T_{p} \leq 6.1$ K) can only produce $5\times10^{-2} \leq $ \lbblx $\leq 70$. Therefore the values of \lbblx in the range 100-3000 found in the early times of a large fraction of our sources indicate that an additional emission mechanism that deviates from a standard accretion disk-corona is operating. The deviation from a standard disk is stronger at early phases, given that the \lbblx ratios converge towards the expected disk values at late times. 

\begin{figure*}
    \centering
    \includegraphics[width=0.99\textwidth]{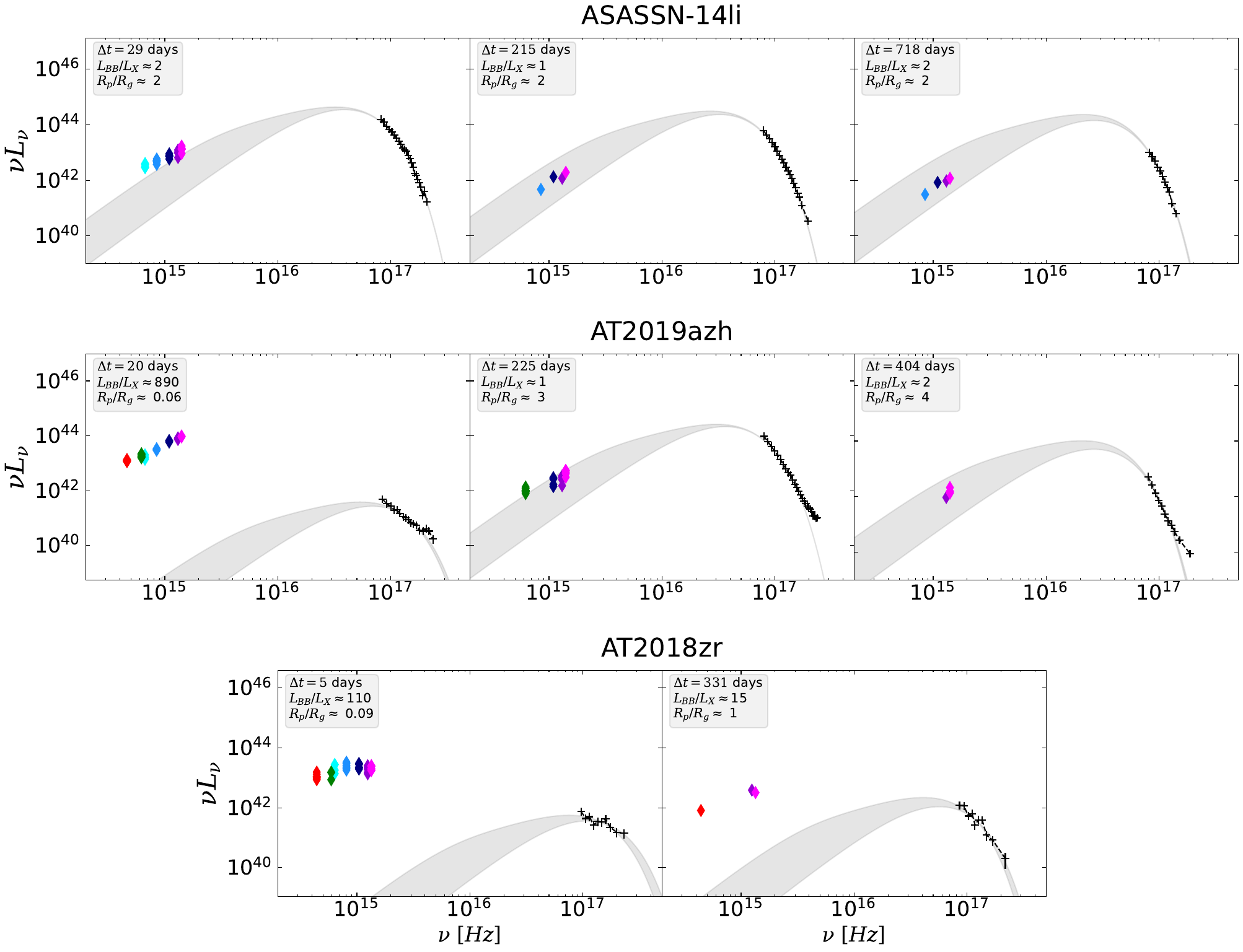}
    \caption{Evolution of the SED for three sources, one \textit{Power-law decaying} (Top), one \textit{Late-time Brightening} (middle), and one \textit{X-ray faint} (bottom). Left panels show early time ($\Delta t \leq$ 30 days) SED, middle panel show late time SED ($\Delta t \approx$ 200 days) and right panel show very-late time SED ($\Delta t \approx$ 400 days). No X-ray spectrum of AT2018zr around $\Delta t \approx$ 200 days is available.
    The colored point show the observed UV/optical photometry (ZTF + \swift/UVOT), the black crosses show the unfolded X-ray spectrum, the gray line show the best-fitted disk model (and uncertainty) for the X-ray spectra, and it is extrapolation to the UV/optical band is shown in shaded region (see text for details). Main parameters of interest are shown in the legend for each epoch.}
    \label{fig:SED_plot}
\end{figure*}

In Fig. \ref{fig:SED_plot}, we explore how a standard disk SED compares with the observed SED for three distinct TDEs: one \textit{Power-law Decaying}, one \textit{Late-time brightening} and one \textit{X-ray faint}. We assume a color-corrected disk solution, where the SED can be obtained numerically, as approximately:

\begin{equation}\label{eq:disk}
    L(\nu) \approx \frac{8\pi^2}{f_{col}^4}   \int_{R_{p}}^{R_{out}}  \ B_{\nu}(\nu, f_{col}T(r))~r~dr
\end{equation}

\noindent  where $B_{\nu}(\nu, T)$ is a Planck function, $T(r)$ is the temperature radial profile of the disk, $T(R_p$) = $T_p$, and $R_{out}$ is the outer radius of the disk. For a given values of the inner temperature ($T_p \pm \delta T_p$) and radius ($R_p \pm \delta R_p$) -- obtained from the the X-ray fitting -- the expected UV/optical emission will depend on the extended disk structure, its size, i.e., $R_{out}/R_p$ ratio, and its temperature profile $T(r)$. At early times $R_{out}$ should be limited to the circularization radii ($R_{circ}$), which for a solar-like disrupted star is:

\begin{equation}
    R_{out} = R_{circ} = 2R_{T} = 94 R_g (M_{BH} / 10^6M_{\odot})^{-2/3}
\end{equation}

\noindent where $R_{T}$ is tidal disruption radius. However, at late times such requirement is lifted due to viscous spread of the disk. To emulate our ignorance on the extended properties of the disk, we generated a series of solutions to equation \ref{eq:disk}, assuming disk sizes between $R_{out}/R_p \in (5, 50)$, temperatures profiles as $T(r) \propto r^{-3/4}$ \citep[for a vanishing ISCO stress,][]{Shakura1973,Cannizzo1990} and $T(r) \propto r^{-7/8}$ \citep[for a finite ISCO stress,][]{Agol2000,Schnittman2016}, and the range of the $1\sigma$ uncertainty for $T_p$ and $R_p$. The shaded gray region in Fig.~\ref{fig:SED_plot} represents the region of possible solutions resulting from the assumed parameters. Our goal is to visualize how much the full SED deviates from a standard disk SED, and how much UV/optical `excess' is present.

From the top panel of Fig.~\ref{fig:SED_plot}, we can see that the SED of ASASSN-14li is not far from an accretion disk SED; very little UV/optical excess is present even at early times; at late time the SED is consistent with an accretion disk, furthermore, the evolution in SED shape with time is very small and compatible with the cooling of the disk. The same holds for the other \textit{Power-law decaying} sources, although for AT2019dsg the rapid decay of X-ray emission indicates a quicker cooling of the disk \citep{Cannizzaro2021}. 

For AT2019azh, however, the disk emission extrapolated from the X-rays under-predicts the UV/optical emission by more than two orders of magnitudes, as can be seen from the left middle panel of the Fig. \ref{fig:SED_plot}. This is mainly a consequence of the low X-ray luminosity \lx $\approx 5 \times 10^{41}$ \ergs and high $T_p$, that result in both an unphysical $R_p/R_g$ and a high \lbblx $\approx 890$. At later times however, \lbb decays and \lx brights (\lbblx $\approx 1$), and the SED is consistent with a standard accretion disk. Such behavior is observed in all \textit{Late-time brightening} sources. One should notice that in AT2021ehb, the \lbblx reaches extremely low value $\sim 10^{-2}$ given the strong corona formation that increases \lx, consequently decreasing \lbblx.

In the case of AT2018zr, the early time is very similar to AT2019azh; however, even at $\Delta t \approx 331$ days, the SED is yet not consistent with a standard disk, although the UV/optical excess does decrease, while \lx is approximately constant during the entire evolution. Similar behavior is found in the other \textit{X-ray faint sources}. The physical interpretation of the diversity of SED shapes at early times and the convergence to a disk-like SED at late times, will be discussed and compared with the theoretical expectations in  \S\ref{sec:disc_lc}.

As discussed in \S\ref{sec:T_x_L}, there is a clear anti-correlation between \lbblx and the radius derived from the X-ray spectral fitting, $R_p/R_g$ assumes nonphysical values at the highest \lbblx (usually with \lbblx $\gg 10$), this can also be seen in the legend of Fig. \ref{fig:SED_plot}. In light of the discussion of this current section and \S\ref{sec:T_x_L}, we can conclude that $R_p/R_g$ values are nonphysical when the SED strongly deviates from the SED of a standard disk. Both lines of evidence point towards the suppression of the observed X-ray emission as discussed in \S\ref{sec:T_x_L}.

\subsection{The ratio of X-ray loud TDEs in optical surveys and its (in)dependence on \mbh}\label{sec:x_ratio}
A surprising characteristic of the population of TDEs discovered by optical surveys is the lack of detectable X-ray emission in most sources, contrary to that first theoretical expectation \citep{Rees1988} and the fact that first TDE were discovered in the X-rays \citep[e.g.,][]{Bade1996}. We would like to determine the fraction ($N_{opt,x}/N_{opt}$) of optically discovered TDEs that show X-ray emission. Therefore, a controlled sample of total discovered TDE and X-ray bright TDEs needs to be compared. 

We will use the first 3 years of the ZTF (October 2018 to August 2020 for ZTF-I and September 
2020 to August 2021 for ZTF-II) survey to determine this ratio. After a ZTF TDE candidate 
\citep[see, ][for selection criteria]{vanVelzen2021,Yao2023} is spectrally classified, a few \swift
observations are performed to confirm the UV brightness characteristic of TDEs and check whether 
the source is detected in the X-rays. The number of visits and the cadence varies from source to
source, but at a minimal a few 1-2~ks long observations (usually with a total of $\sim$ 10~ks) are performed, leading to a detection limit of $\sim 5 \times 10^{-14}$ \ergcms with XRT, which means any TDE with \lx $ > 10^{42}$ \ergs can be detected up to $z \approx 0.09$,
while \lx $\sim 10^{41}$ \ergs can only be detected if extremely nearby ($z < 0.04$). Therefore, we will use \lx $ \geq 10^{42}$ \ergs to define the sub sample X-ray bright TDEs. During the first three years of ZTF 10 sources showed, at some point in time, \lx $\geq 10^{42}$ \ergs, therefore $N_{opt,x} = 10$.

To construct the control sample, we start with all TDEs discovered by ZTF during the same time period \citep[from][]{vanVelzen2021,Hammerstein2022,Yao2023}. For sources at $z < 0.09$ (24 sources), we select those with \swift/XRT observations, for sources with no XRT detection we stacked all their \swift/XRT observations and check whether \lx upper-limit (after correcting for Galactic absorption) was deeper than $10^{42}$ \ergs. Only 3 sources (AT2021mhg, AT2021sdu, AT2021yte) the upper limit were not deep enough to constraint the presence \lx $\geq 10^{42}$ \ergs, particularly because of their higher Galactic absorption $N_{H,G} \geq 10^{21}$ cm$^{-2}$, than the typical $N_{H,G} \approx \rm{few} \times 10^{20} \ \rm{cm}^{-2}$ of the other sources.
For our typical exposure times, \xmm detection limit is around $5\times10^{-15}$ \ergcms, this means that we can detected \lx  $ > 10^{42}$ \ergs up to $z \approx 0.25$. Therefore, we include all sources at $ 0.09 \leq z \leq 0.25$ that had at least one \xmm visit to our control sample (adding 4 more sources). 
This results in 25 sources in which a \lx  $ \geq 10^{42}$ \ergs could be (or was) detected if such level of emission was present, hence $N_{opt} = 25$. Therefore for ZTF, we obtain $N_{opt,x}/N_{opt} = 10/25$, meaning that in ZTF, $\sim 40\%$ of the discovered TDEs had (eventually) some X-ray bright emission (\lx $\geq 10^{42}$ \ergs). The list of the 25 TDEs and their redshift are present in Table \ref{tab:appendix_ratio}. Given the non-uniformity of the X-ray follow-up, particularly at late-time for those at higher $z$, the ratio should be seen as a lower-limit.

\begin{figure}[!ht]
    \centering
    \includegraphics[width=\columnwidth]{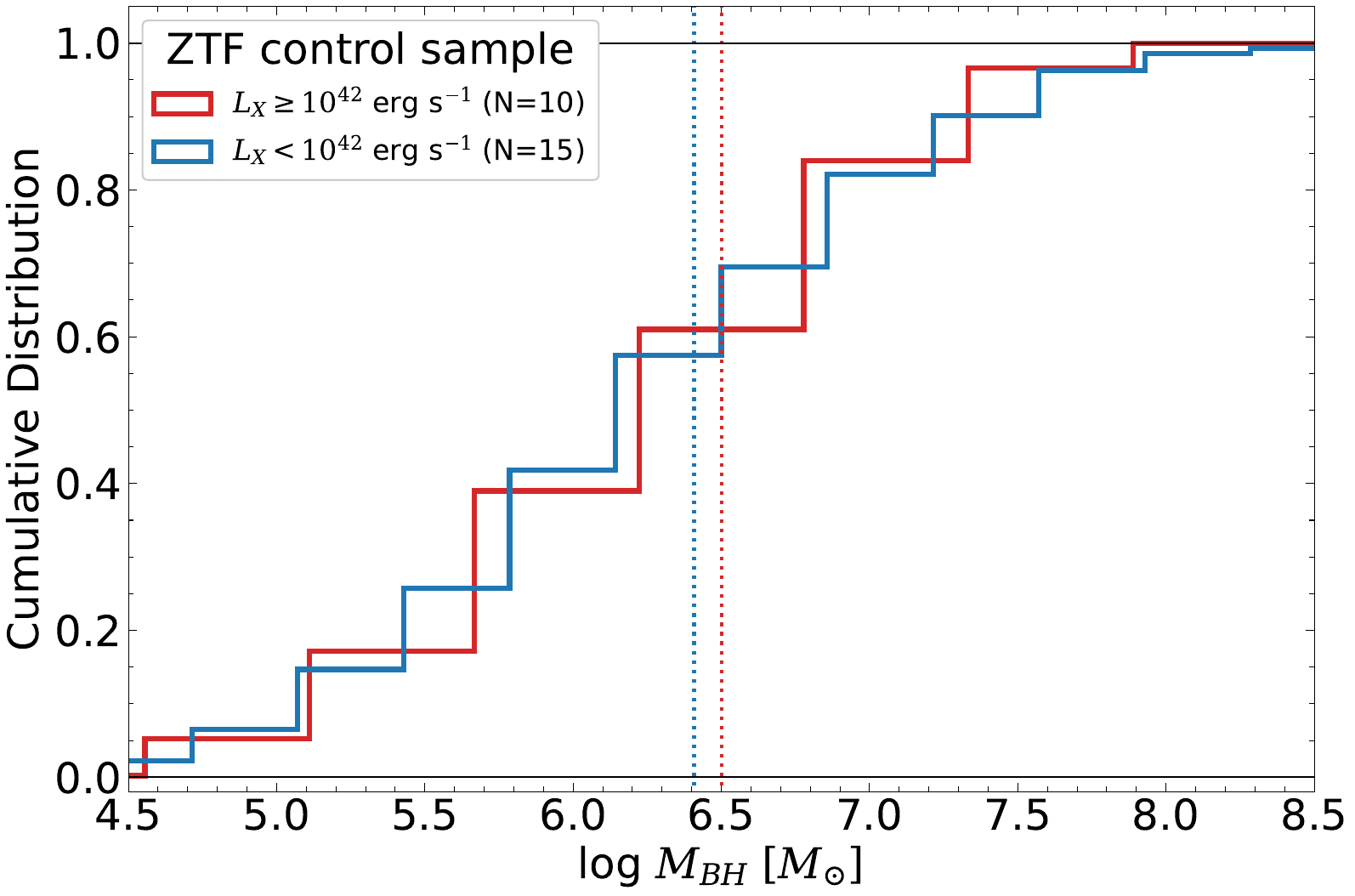}
    \caption{Cumulative distribution of black hole mass for the \textit{ZTF control sample} (see \S\ref{sec:x_ratio} for definition) with \lx higher or lower than $10^{42}$ \ergs. The underlying distribution was constructed by adding the normalized PDF of individual sources \mbh based on the estimated values and its uncertainties. No statistically significant difference in the distributions is observed.}
    \label{fig:mbh_x}
\end{figure}

We also use this above-defined sub samples to investigate whether the presence of \lx $\geq 10^{42}$ is dependent on \mbh. In Fig.~\ref{fig:mbh_x} we show the cumulative distribution of \mbh for the two sub samples (\lx higher and lower than $10^{42}$ \ergs), where the underlying distribution was constructed by adding a normalized probability distribution function (PDF) based on estimated \mbh and their uncertainties (see \S\ref{sec:M_BH}).

From Fig.~\ref{fig:mbh_x} there seems to be no distinction between the \mbh distribution of TDE \lx higher/lower than $10^{42}$ \ergs: the difference in the median of the distributions (dotted vertical lines) is only $\sim 0.1 \rm{dex}$ which is much smaller than the typical uncertainty in \mbh. In order to statistically investigate the (lack of) difference between the distributions we apply a Kolmogorov-Smirnov test, assuming the null-hypothesis that the two samples are drawn from the same distribution, we recover a $p-$value of $\sim 0.2$, which means the null-hypothesis can not be excluded, hence there is no statistically significant difference between the distributions. This result is in agreement with previous ones, that although use distinct samples or selection criteria, also did not find any correlation between X-ray loudness and \mbh \citep{Wevers2019_Mbh,French2020,Hammerstein2022,Hammerstein2023},

\subsection{X-ray Luminosity Function}\label{sec:LF}

When examining sources with constant flux, such as quasars, the estimation of their luminosity function (LF) involves assigning weights to each source based on its maximum detectable volume,  $V_{\rm max}$, as introduced by \citet{Schmidt1968}. This approach enables plotting of the number of sources per unit volume as a function of their luminosity. Shifting to transient events, like supernovae (SNe) and TDEs, our interest lies in understanding their volumetric rate, i.e. the number of events per unit volume per unit time relative to their peak luminosity. Extracting this rate from survey data involves applying the  ``1/$V_{\rm max}$" method, with a modification accounting for both survey duration and area. This adaptation leads to the definition of a modified parameter,  $\mathcal{V}_{\rm max}$:

\begin{align}\label{eq:v_max}
    \mathcal{V}_{\rm max} = V (z_{\rm max}) \ A_{\rm survey} \times \tau_{\rm survey} 
\end{align}

\noindent where  $A_{\rm survey} \times \tau_{\rm survey}$
denotes the product of the effective survey duration and survey area, and $V(z_{max})$ is the volume (per unit solid angle) corresponding to the maximum redshift observable with the survey, given its limiting flux.

The luminosity functions of tidal disruption events (TDEs) across different bands have been investigated by a few studies. \citet{vanVelzen2018} initially presented a \textit{relative} luminosity function in single-band optical and integrated UV/optical for TDEs, utilizing 13 events discovered before 2018 through UV/optical surveys. Subsequently, \citet{Yao2023} provided an \textit{absolute} UV/optical luminosity function derived from a sample of 33 homogeneously selected TDEs identified by ZTF. In the X-ray domain, \citet{Sazonov2021} constructed the first luminosity function based on 13 events selected from X-ray transients detected in the $0^o < l < 180^o$ hemisphere during \srge's second sky survey. The X-ray luminosity function spanned from $10^{42.5}$ to $\leq 10^{45}$ \ergs and was best-fitted with a power-law slope of $0.6 \pm 0.2$. However, as the X-ray bright TDEs in our study were not discovered by an X-ray survey but through inhomogeneous follow-up X-ray observations of optically discovered TDEs, and given the non-uniform criteria and cadence for X-ray follow-up, we refrain from obtaining an \textit{absolute} X-ray luminosity function akin to \citet{Sazonov2021}. Instead, we propose combining TDEs detected by X-ray surveys with our sample of optically selected X-ray detected TDEs to construct a \textit{relative} luminosity function. This approach aims to provide a broader luminosity range and reduced uncertainty per luminosity bin compared to the one derived from \srge, thereby enhancing our ability to constrain its shape.

To construct a large sample of X-ray TDEs, we combine our 17 sources, the 13 discovered by \srge \citep{Sazonov2021}, with 6 discovered by \rosat/RASS and 8 discovered by the \xmm Slew Survey as presented in the recent review by \citet{Saxton2020}, resulting in 44 X-ray detected TDE ranging from $10^{41}$  $\leq$ \lx $\leq 10^{45}$ \ergs.

We follow the procedures detailed in \citet{vanVelzen2018} to construct a \textit{relative} luminosity function from a combined sample discovered by distinct surveys with distinct selection functions and detection efficiency, assuming that each survey discovers events from the same parent distribution so that we can use the discovered/detected number of TDE in each survey to compare the selection efficiencies and thus obtain the \textit{relative} LF. The \textit{effective} ($A_{\rm survey} \times \tau_{\rm survey})^*$ (see Eq. \ref{eq:v_max}) for each survey can be estimated from:

\begin{align}\label{eq:N_TDE}
   (A_{\rm survey} \times \tau_{\rm survey})^* \approx  \frac{N_{\rm TDE, detected}}{ \dot{N} \  V(z_{max*}) }
\end{align}

\noindent where $N_{\rm TDE, detected}$ is the number of TDEs detected by the survey, and $V(z_{max*})$ denotes the comoving volume (per solid angle) corresponding to maximum redshift ($z_{max*}$) the survey can detect a `typical' X-ray TDE. We define a `typical' X-ray TDE to have $L_{X}^{*} = 10^{43}$ \ergs and $T_{p}^{*} = 60 \  $ eV, and use the detection limit flux ($F_{X,\rm{lim}}$) of each survey (see Table \ref{tab:survey_LF}) to determine $z_{max*}$ 
for each survey. For the optical survey, we assume $F_{\rm X,lim}$ to be the flux limit of \swift/XRT in a stacked exposure time of 10ks. In eq. \ref{eq:N_TDE} $\dot{N}$ is the \textit{assumed} mean event rate which was chosen to be $\dot{N} = 2 
\times 10^{-7} \ \rm{Mpc}^{-3} \ \rm{yr}^{-1}$, following \citet{Sazonov2021}. The resulting $(A_{\rm survey} \times \tau_{\rm survey})^*$ for each survey is shown in Table \ref{tab:survey_LF}, while Table \ref{tab:append_X-TDE} show the detailed 
information from individual sources.

\begin{deluxetable}{cCCC}
\label{tab:survey_LF}
\tabletypesize{\scriptsize}
\tablecaption{Survey' Properties}
\tablewidth{0pt}
\tablehead{
\colhead{Survey} &  \colhead{$N_{TDE}$} & \colhead{$F_{X,lim}$} & \colhead{($A_{survey} \times \tau_{survey})^*$}$^a$ \\
\colhead{} &  \colhead{} & \colhead{(\ergcms)} & \colhead{(deg$^2 \ yr$)}
}
\startdata
ASASSN+\swift/XRT  &  6 (3)^b & 8\times10^{-14} & 138 \\
 ZTF+\swift/XRT     & 12 & 8\times10^{-14}   & 273 \\
 OGLE+\swift/XRT   &  1 &  8\times10^{-14}  & 22\\
 ROSAT   &  6 & 2\times10^{-13}  & 2009\\
 XMMLS   &  9 & 5\times10^{-13}  & 6484 \\
 eROSITA & 14 & 8\times10^{-14}  & 645\\
\enddata
\tablecomments{(a) Effective  $A_{survey} \times \tau_{survey}$ as measured from Eq. \ref{eq:N_TDE}. (b) Three were part of both in ZTF and ASASSN.}
\end{deluxetable}

\begin{figure}[!ht]
    \flushright
    \includegraphics[width=0.97\columnwidth]{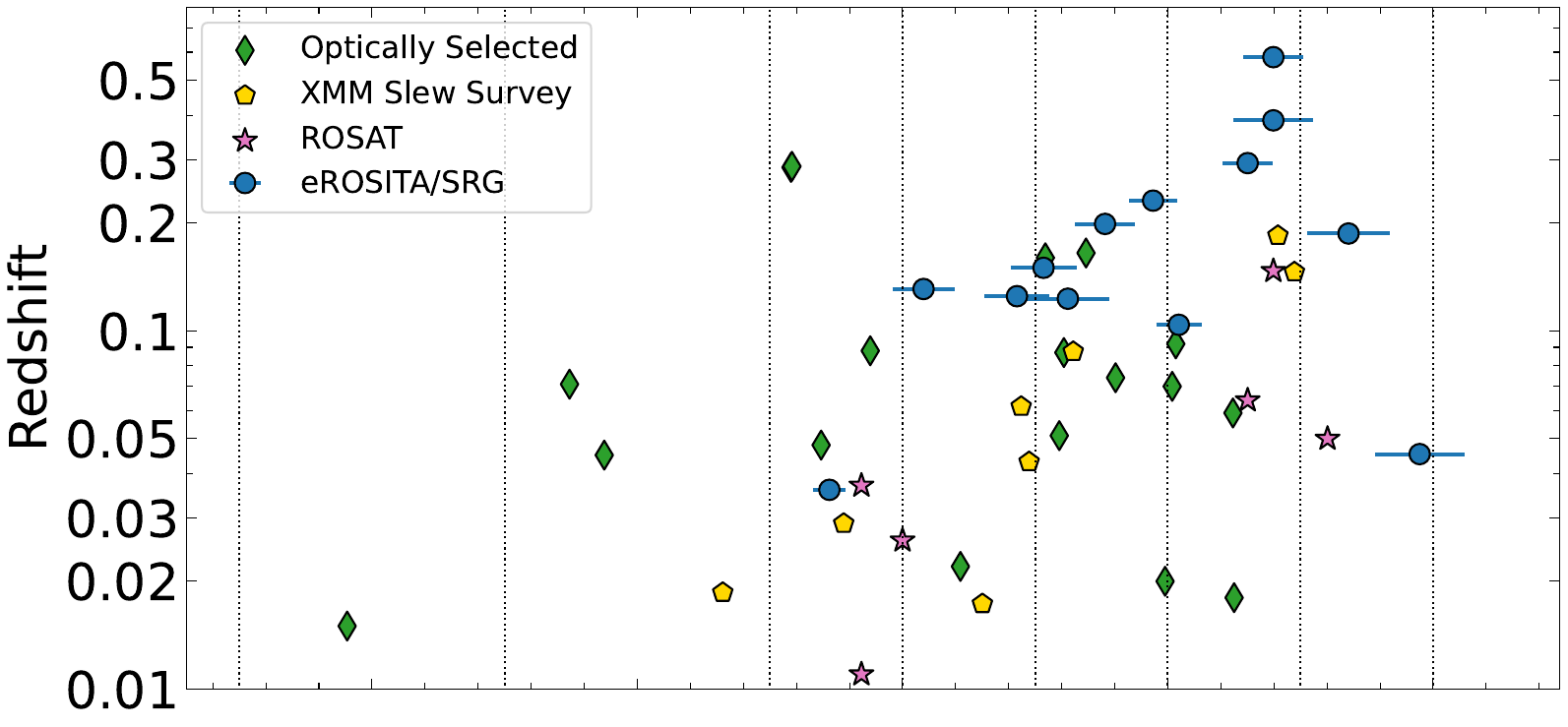}\\
    \includegraphics[width=0.99\columnwidth]{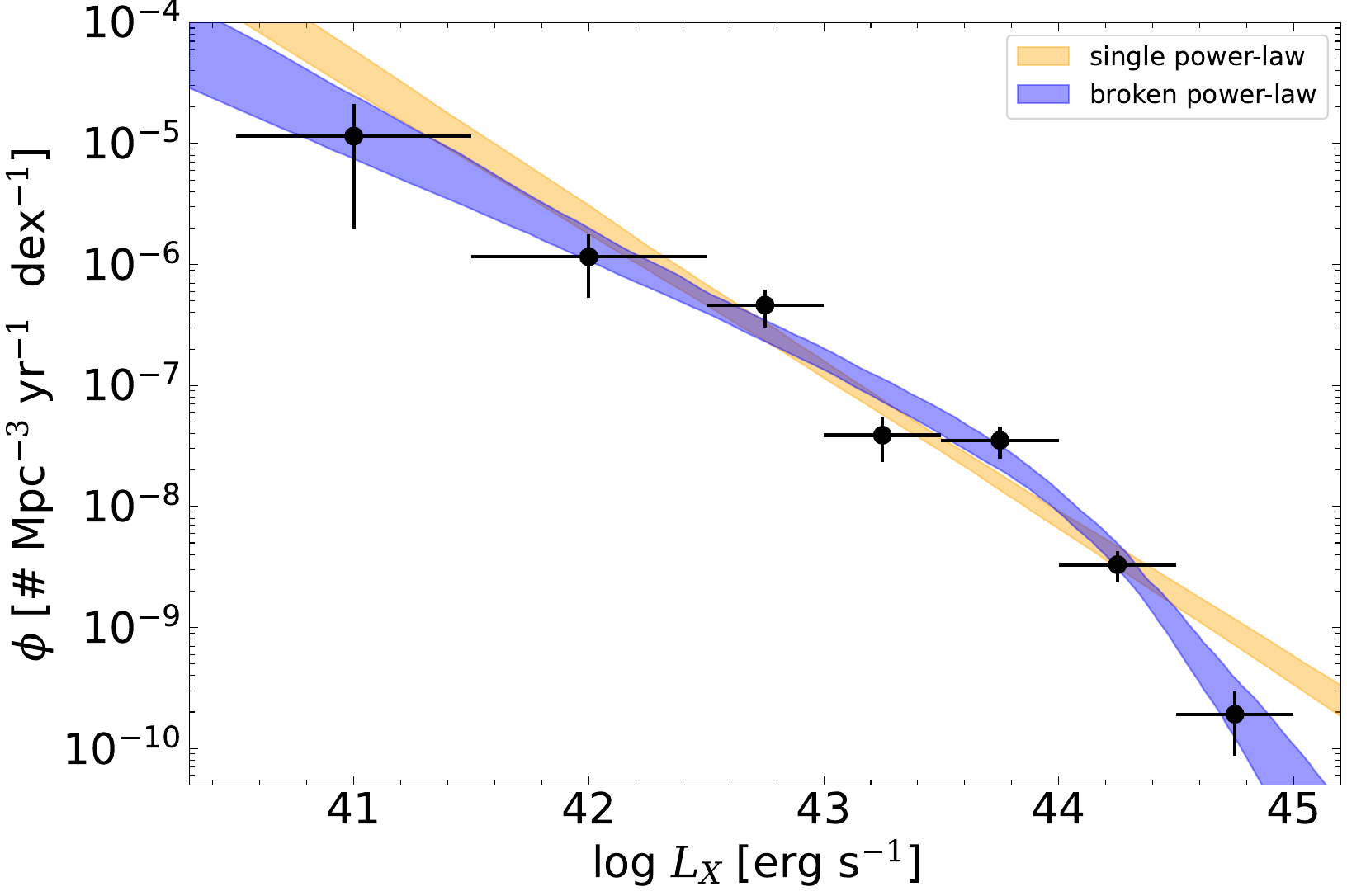}
    \caption{Top: Distribution of Redshift $\times$ peak X-ray luminosity (\lx) for all the sources included in the X-ray luminosity function analyses. Bottom: Derived luminosity function (black points), best-fitted single power-law (orange), best-fitted broken power-law (blue).} 
    \label{fig:xray_lum_func}
\end{figure}

In the upper panel of Fig. \ref{fig:xray_lum_func}, we show the distribution of the 44 TDEs in the redshift vs. peak \lx diagram, where boundaries of the nine $\Delta$log \lx bins are indicated with vertical lines. For a certain bin 
$j$ with $n_j$ TDEs and width $\Delta_j$log \lx, the rate luminosity function is  $  \phi_j = \left [ \sum_{i=1}^{n} \frac{1}{\mathcal{V}_{\rm max,i}} \right ] / \Delta_j {\rm log} L_X$, we compute the corresponding uncertainty of  
$\phi_j$ based in the Poisson error \citep{Gehrels1986}. For example when $n_j$ = 1, the upper and lower limits of $\phi_j$ are $\phi_j^u$ = $\phi_j \times$ 3.30/1 and $\phi_j^l$ = $\phi_j \times$  0.17/1, and when $n_j$ = 11, $\phi_j^u$ = $\phi_j \times$ 14.27/11 and $\phi_j^l$ = $\phi_j \times$  7.73/11. We show $\phi_j$ vs. log \lx in the bottom panel of Fig. \ref{fig:xray_lum_func}. 

First, we fit the seven LF data points with a single power-law of:

\begin{align}
    \phi(L_X) =  \dot{N}_0 \left ( \frac{L_X}{L_0} \right )^{-\gamma}
\end{align}

\noindent For $L_0 = 10^{43}$ \ergs, we obtained  $\gamma = 1.2 \pm 0.1$. The best-fit model, shown as an orange line in the bottom panel of Fig. \ref{fig:xray_lum_func}, is stepper than the power-law model with $\gamma = 0.6 \pm 0.2$ present by \citet{Sazonov2021}, however, the fit seems to slight over predict the number of low \lx sources and the number of sources with \lx $\geq 10^{44.5}$ \ergs. 

Next, we describe the luminosity function with a broken (or double) power-law in the form of:

\begin{align}
    \phi(L_{X}) = 
    \dot N_0 \left[ \left( \frac{L_X}{L_{\rm bk}}\right)^{\gamma_1 } + \left( \frac{L_X}{L_{\rm bk}}\right)^{\gamma_2 } \right]^{-1}
\end{align}

Performing a broken power-law fit to the luminosity function (LF) of tidal disruption events (TDEs), with $-\gamma_1$ representing the faint-end slope, $-\gamma_2$ the bright-end slope, and $L_{bk}$ the characteristic break luminosity, we utilize Markov Chain Monte Carlo (MCMC) to obtain $\gamma_1 = 0.96_{-0.24}^{+0.21}$, $\gamma_2 = 2.65_{-0.90}^{+1.1}$, and log $L_{\rm bk} = 44.1^{+0.4}_{-0.5}$~\ergs. The Bayesian information criterion (BIC) favors the broken power-law fit over the single power-law fit, with a smaller BIC value by 7.1. This suggests that the broken power-law LF provides a superior description of the data. Notably, our determined $\gamma_1$ below the break is steeper than the $\gamma=0.6\pm0.2$ reported by \citet{Sazonov2021}, indicating potential underestimation of the low \lx end or/and overestimation of the high end of the LF in their work. The selection bias in \citet{Sazonov2021}'s criteria, favoring sources with observed flux at least 10 times brighter than the previous upper limit, may contribute to underestimating the low luminosity end. Our findings reveal an extension of the LF to \lx values below $10^{42.5}$ \ergs (\citet{Sazonov2021}'s lowest \lx bin), reaching as low as \lx $\approx 10^{41}$ \ergs. This suggests a sizable population of X-ray-emitting TDEs that are too faint for current instruments unless occurring at very low redshifts, emphasizing that the absolute rate estimated by \citet{Sazonov2021} represents a lower limit on the rate of X-ray-emitting TDEs.

In the case that the X-ray luminosity of TDEs are Eddington
limited, and hence their fraction $l_x = L_X/L_{Edd}$ is $< 1$, the observed suppression of the TDE rate at \mbh $> 10^{8}$ 
\msun \citep{vanVelzen2018,Yao2023} can naturally explain the break in the X-ray LF at $\sim 10^{44}$ \ergs. Indeed, based on 
such arguments, \citet{Mummery2021} estimated a \textit{maximum} X-ray luminosity of $\sim 10^{44}$ \ergs for non-jetted TDEs \footnote{TDEs in which the jet is pointed towards us -- so-called jetted or relativistic TDEs -- have their luminosity beamed, hence those can reach \lx $\geq 10^{47}$ \ergs. This is a distinct physical scenario, than what is discussed in this section, therefore are not considered.}. Although a couple TDEs have shown a peak $L_X > 10^{44}$~\ergs (see top panel of Fig. \ref{fig:xray_lum_func}), the steep break, from $\gamma\approx1.0$ to $\gamma\approx2.7$ in the LF is still in qualitative agreement with the limiting luminosity expectation, given that a TDE with peak $L_X = 10^{45}$~\ergs should be $\sim$1000 times rarer than a TDE with peak $L_X = 10^{44}$~\ergs.

\section{Discussion}\label{sec:discussion}
\subsection{On the diversity of X-ray evolution: delayed accretion or variable optical depth? an orientation effect?}\label{sec:disc_lc}

As we have shown in \S\ref{sec:lc_div}, optically selected TDEs have a large diversity in X-ray evolution, with sources rarely showing prompt ($\Delta t \leq 100$ days) bright (\lx $\geq 10^{43}$ \ergs) X-ray emission that decays as a power-law with time as predicted from simple fallback accretion models. Instead, most sources show a faint X-ray emission at early times, with a subset showing a delayed increase in the observed X-ray luminosity, while others showing a faint and approximately constant X-ray luminosity during the UV/optical evolution.

From our X-ray spectral and SED analyses, a number of conclusions can be drawn:

\begin{itemize}
    \item The X-ray emission temperature ($T_p$) decreases with time (see Fig. \ref{fig:T_x_t});
    \item The decrease in $T_p$ is independent of the \lx, given that \textit{Late-time brightening} sources show highest $T_p$ at early times, when the \lx is minimum. 
    \item This independent $T_p$ and \lx evolution creates a decoupling between the two 
    parameters in these epochs/sources. Other epochs/sources show a \lx $\propto T^{\alpha}$ 
    relation (see Fig. \ref{fig:T_x_L}).
    \item The highest $T_p$ with lowest \lx results in an unphysical value for the X-ray emission radius ($R_p/R_g$) for these source/epochs. For other epochs/epoch $R_p$ is consistent with $R_{ISCO}$ (see Fig. \ref{fig:R_x_t}).
    \item The epochs with unphysical $R_p/R_g$ are also the epochs in which UV/optical to X-ray ratio (\lbblx) is too high to be produced by \textit{bare/unreprocessed} accretion disk (see Fig. \ref{fig:R_x_L}).
    \item There is a large range of observed \lbblx values at early times ($0.5 \leq $ \lbblx $\leq 3000$); at late times these values converge to ($0.5 \leq $ \lbblx $< 10$), see Fig. \ref{fig:SED_hist}.
    \item There is no \mbh dependence on the presence/absence of luminous X-ray emission (\lx $\geq 10^{42}$ \ergs).
\end{itemize}

\noindent Viable theoretical models for the TDE emission mechanism must be able to reproduce these observational findings.

A possible explanation for the late-time brightening of the X-ray emission, put forward by several authors \citep[e.g.,][]{Gezari2017,Liu2021}, is 
the delayed formation of the accretion disk in these sources. Some problems arise from this interpretation: 

\noindent {\it i}) soft X-ray emission (though faint, \lx $\leq10^{42}$
\ergs) is detected in the early times for all of the \textit{late-time brightening} sources in which an observation deep enough to detect such faint emission is available\footnote{The upper limits on AT2020ksf and OGLE16aa are higher than 10$^{42}$ \ergs, the two are also the highest redshift sources.}; which means that the structure responsible for X-ray emission in these sources is already present at very early times, i.e. is promptly formed;

\noindent {\it ii}) the temperature of 
these X-ray faint phases are as high or higher than the temperatures at the late-time; 

\noindent {\it iii}) the overall evolution of the temperature is consistent with a cooling accretion disk (see Fig. \ref{fig:T_x_t}), that would not necessarily be the case if the late time X-ray emission is tracing a different physical 
structure (disk) than the early time is tracing (stream-stream shocks), like proposed by \citet{Liu2021}; 

\noindent {\it iv}) if the presence of bright X-ray emission (\lx $> 10^{42}$ \ergs) is driven by the successful circularization of the debris streams, while in sources with \lx $\leq 10^{42}$ \ergs these do not 
circularize to form a disk; there should be a distinct \mbh distribution underlying the two populations -- given the strong \mbh dependence in the 
circularization time \citep{Bonnerot2016} --  as we have shown in Fig. \ref{fig:mbh_x}, there is not difference in their \mbh 
distribution.

\begin{figure*}
    
    \includegraphics[width=\textwidth]{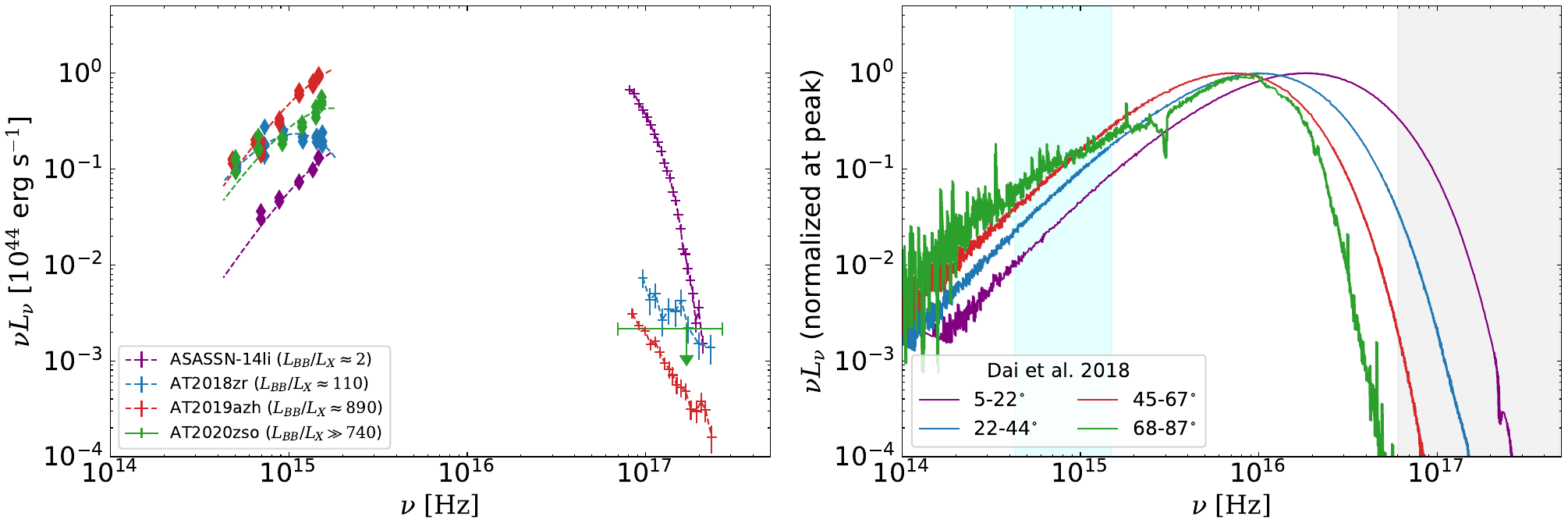}
    \caption{Comparison between observed SEDs and model SEDs by \citet{Dai2018}. In the right we show the early time (near UV/optical peak) observed SED for ASASSN14-li, AT2018zr and AT2019azh (this work) as well as for AT2020zso \citep{Wevers2022}, showing a large range of \lbblx values. The left panels show \citet{Dai2018}'s early time simulated SEDs as seen by distinct viewing angles from the disk pole: face-on $= 0^{o}$ and edge-on $= 90^{o}$.}
    \label{fig:SED_x_Dai}
\end{figure*}

As we argued in \S\ref{sec:T_x_L}, the observed properties seems to point towards the presence of prompt, but suppressed, X-ray emission in the early times of the \textit{Late-time brightening} and \textit{X-ray faint} sources. Suppression of the X-ray emission is an expected consequence of different TDE models that invoke the reprocessing of the high energy emission into the UV/optical wavelengths by an optionally thick material, these models \citep[e.g.,][]{Loeb1997,Ulmer1999,Coughlin2014,Metzger2016,Roth2016,Dai2018,Thomsen2022,parkinson22_disk_wind,Metzger2022}, assume different physical processes and geometries, but have the common property of the reprocessing of the X-ray emission at the highest accretion rates, i.e., early times, and its re-emission at lower energies. Although a large fraction of the optical TDEs in our sample shows suppression of the X-ray emission at early times, some do not, namely, ASASSN-14li, AT2019vcb, AT2019dsg, and AT2019ehz; this is also true for most X-rays discovered TDEs (comparison between these populations will be presented in \S\ref{sec:disc_ero_opt}). Instead, there is an at least three orders of magnitude range in observed \lbblx at early times (see Fig. \ref{fig:SED_hist}, and \ref{fig:SED_plot}). Some of this diversity was already previously known and had inspired a series of models where the presence/absence of strong reprocessing is orientation dependent \citep[e.g.,][]{Dai2018,Jonker2020,Thomsen2022}. 

In these models, the optical depth through the line-of-sight for the high energies photons has a strong dependence on the accretion rate and likely on the viewing angle, at the highest accretion rate, i.e., early times, if the source is seen at lower inclination angles with respect to the disk pole, the optical depth ($\tau $) and hence the reprocessing, are minimal, and the observed SED resembles the one of the underlying disk (i.e., \lbblx $\lesssim 10$, see Appendix \S\ref{app:disk_lbb_lx}), which would explain the \textit{power-law decaying} sources. At the largest inclination angles (towards edge-on), the system should be heavily optically thick ($\tau >> 10$) to the X-rays which should be reprocessed to the  lower energies, making the X-rays undetectable, and high lower limits on \lbblx to be measured. This should explain the TDE population with UV/optical emission only. When seen at intermediate angles, the system is not completely optionally thick, but the optical depth is still important ($\tau \sim \rm{few}$), and only a small fraction of the X-ray can escape unreprocessed; the SED is then UV/optical dominated (\lbblx between $\rm{few} \times 10^{1}$ to $\rm{few} \times 10^{3}$), but faint X-ray emission is still able to escape and be detected; as the accretion rate decreases, the optical depth of the system decreases, allowing for a larger fraction of the X-ray photons to escape, which would explain the \textit{Late-time brightening} sources \citep{Thomsen2022}, and perhaps the \textit{X-ray faint} sources given some fine tuning in the evolution of the parameters. 

In the left panel of Fig. \ref{fig:SED_x_Dai}, we show four early-time SEDs with a diverse range of \lbblx values, while in the right panel, we show the four simulated SEDs for distinct inclination angles as presented by \citet{Dai2018}. A direct comparison is, of course, not valid, given that in \citet{Dai2018} models, the only parameter changed between the SEDs is the viewing angle towards the system; every other parameter of the system is fixed, while in reality, our four example sources/SEDs 
may have distinct black hole masses, black hole spin, impact parameter ($\beta$), peak disk temperature, radial profile of the disk temperature and many other differences, that could also shape the SED, however, it is interesting to note that the large diversity of \lbblx observed can be, in principle, produce just by a change in viewing angle.

This scenario seems to be able to explain several of the observed properties:

\noindent {\it i}) The large range of \lbblx values at early times, and the convergence to disk-like values at late times;

\noindent {\it ii}) The diversity of X-ray light curves, including the suppressed (but still detected) prompt X-ray emission from the \textit{Late-time brightening} and \textit{X-ray faint} sources;

\noindent {\it iii}) The lack of \mbh dependence on the presence/absence of luminous X-ray emission.

As we pointed out in \S\ref{sec:T_x_L}, an important characteristic of the early suppression of the X-ray emission in the \textit{late-time brightening} and \textit{X-ray faint} sources is that such suppression seems to have minimal effect in the measured $T_p$, given that the expected decline in temperature is still observed, $T_p$ is the highest at early time and decay at late times. This would mean that although a large fraction of the X-rays are absorbed by this reprocessing-layer of ionized gas\footnote{The absorption by an ionized gas should not be confused with absorption by a neutral medium (modeled, e.g., by the \texttt{TBabs} in \texttt{xspec}) has a strong energy dependence in the soft X-rays, while (partially) ionized gas absorption, in contrast, has a higher optical depth in the hard X-rays \citep{Thomsen2022}.}, the output spectrum seems to have a similar temperature (shape) of the supposed underlying emitted spectrum.
This would require the absorption and re-emission process to have a \textit{`quasi-grey'} net effect in the X-ray 0.3-2.0 range. Such effect is, however, quite hard to be produced, for example, in \citet{Thomsen2022}'s simulations the early time output X-ray temperature can be up to $\sim 50\%$ colder than the injected spectrum, depending on the viewing angle and the ionization state of the gas. This effect does not seem to be present -- given the observed decline of $T_p$ with time for all classes of sources -- although the temperature of the underlying emission is, of course, not accessible for a direct comparison. Alternatively, if in these sources, the absorbing material is heavily optically thick but clumpy or has small holes so that a fraction of the source X-rays can get through unprocessed, then this could explain the apparent suppression of the observed X-ray emission -- given the inferred disk radius would be reduced by a factor of the square root of the transmitted over emitted fluxes -- while the emerging temperature would not be strongly modified \citep{Takeuchi2013,Kobayashi2018,Yao2022}. Independently of the driven mechanism, the X-ray emitting structure (consistent with an accretion disk) is promptly formed even in those sources showing a late-time brightening, and their early-time emission seems to point towards a partial absorption/reprocess scenario.

The \textit{flaring} source, AT2019ehz, has not been addressed yet; we first note that such short term flaring differs from the gradual late time X-ray increase of the \textit{late-time brightening} sources, the reprocessing scenario does not work in this case, given that this system is bright at early times, fade, and then re-brighten, which would disagree with the net brightening predicted under the reprocessing scenario. Furthermore, the flaring behavior is accompanied by an increase in $T_p$, a relation between $T_p$ and \lx is present (interestingly with the highest best-fitted $\alpha$ for a \lx $\propto T^{\alpha}$ relation), and no large variations in $R_p$ is observed, with $R_p/R_g$ always in the physically valid range. This behavior differs from the other sources with early X-ray faint emission -- as extensively discussed above -- and instead points towards a disk seen directly without much reprocessing, but with intrinsic variability. 

One possibility is that X-ray variability is produced by random short term fluctuations in the peak disk temperature \citep{Mummery2022}. Alternatively, the nascent accretion disk may be initially misaligned with respect to the MBH’s spin axis, which would induce relativistic torques on the disk and causes it to precess (Lense-Thirring precession), producing repeating flares, that should also modulated the observed $T_p$ \citep{Stone2012,Franchini2016,Pasham2024}. In this scenario relativistic torque effects aligns the disk and terminates precession and the flares; the flares are indeed not observed at late times in AT2019ehz, although the cadence of observation is not enough to confirm this. In both cases, a short term decoupling between the hotter (X-ray) and colder (UV/optical) emission regions, as observed, would be expected.

A definitive explanation for the flaring behavior of AT0219ehz is not within the scope of this 
work, it may not be possible at all, given the cadence/quality of the available data. A TDE discovered in 2022 \citep{Yao_22lri} has shown similar flaring behavior, for that source the cadence of the
observations is much higher (several per day for hundreds of days), and distinct models for 
the flaring behavior will be tested in a forthcoming study (Yao, Guolo, et al. in prep).

\subsection{On the large population of X-ray quiet TDEs}

As we have shown in \S\ref{sec:x_ratio}, most (up to $60\%$) discovered by optical surveys are X-ray quiet (\lx $\leq 10^{42}$ \ergs), the definitive picture for why that is the case is beyond the scope of this paper. However, a couple of insights can be made. 

First, TDEs with detectable X-ray emission, but having \lx $\leq 10^{42}$~\ergs do exist, as we have demonstrated, and should be common given: {\it i}) the derived LF 
(see \S\ref{sec:LF}), {\it ii}) the possible orientation effects (see \S\ref{sec:disc_lc}) that should make a large fraction of X-ray emission to be absorbed.

However, such X-ray luminosities can hardly be observed unless the TDE happens at  extremely low redshifts. Even with a modest $\sim$20~ks \xmm observation upper-limits of \lx $\approx 10^{41}$~\ergs can rarely be placed for the 
typical redshift range in which TDEs are observed (see the \lx upper limits for the non-detected TDE in Table \ref{tab:xmm_obs}). Sometimes these sources show a strong late-time X-ray rebrightening, but not always (e.g., AT2018zr, AT2018hyz, and 
AT2019qiz); besides, if a TDE is not detected at early times -- which for \lx $\leq 10^{42}$~\ergs only occurs for the 
nearest sources -- it is unlikely that this source will continue to be followed-up by an X-ray instrument, much less by \xmm.
 
Furthermore, to be able to produce observable radiation in the 0.3-2 keV range from an accretion disk, the inner temperature ($T_p$) must necessarily be $\geq 10^{5.1}-10^{5.2}$ K, since at lower peak temperatures the emission is shifted entirely to the extreme UV.

For a Galactic like gas-to-dust ratio ($N_H = 5.5\times 10^{21} \times E(B-V)$), soft X-rays are more absorbed than UV light. For instance, with $N_H = 5\times10^{20} \rm{cm}^{-2}$ ($E(B-V) = 0.09$), about 74\% of 0.3 keV X-rays get absorbed, while only 42\% of 2600 Å UV light is. With higher absorption of $N_H = 1\times10^{21} \rm{cm}^{-2}$ ($E(B-V) = 0.18$), 93\% of X-rays and 67\% of UV are absorbed \footnote{For this calculation a standard \citet{Wilms2000} abundance was assumed for the X-ray absorption, using \texttt{TBabs}, while the UV extinction was based on \citet{Calzetti2000} law.}. Therefore, soft X-rays photons are more affected by absorption along our line of sight than UV photons, making them harder to detect given same emitted flux. See for example the cases of AT2021mhg, AT2021sdu and AT2021yte in \S\ref{sec:x_ratio}.

Finally, it is still possible that the circularization of the stellar debris does not happen, and no accretion disk is formed in these sources. In contrast, another process, independent of accretion, such as shocks produced by the intersection between the streams, would be responsible for all the UV/optical emission. Although it is not clear why the proposed lack of circularization is not dependent on \mbh, see Fig. \ref{fig:mbh_x}.

The combination of the effects mentioned above is likely to be able to explain the large fraction of optically discovered TDE with no observable X-ray emission.

\subsection{On the bolometric luminosity of TDEs}

As pointed out by several authors \citep[e.g.,][]{Dai2018,Lu18,Thomsen2022,Mummery2023} a `missing energy' problem \citep{Piran2015} will only arise when the integrated UV/optical (fitted with a blackbody function) is incorrectly considered as the bolometric luminosity of the TDE.  This is an obvious statement for the sources with detected X-ray emission (see Fig. \ref{fig:Lbb_Lx_measurments}), but should also be true for those in which X-ray emission is not detected, since the bulk of the TDE emission should be emitted in the extreme UV (EUV) bands (see right panel of Fig. \ref{fig:SED_x_Dai}) which is not adequately modeled by the single temperature blackbody assumption for the UV/optical emission  \citep[see also right panel of Figure 5 in][]{Dai2018}.

When the full SED is consistent with a \textit{bare/unreprocessed} accretion disk, i.e., the extrapolated disk model fitted from the X-rays matches the observed UV/optical (usually when \lbblx $< 10$, see ASASSN-14li and late-time AT2019azh panel in Fig. \ref{fig:SED_plot}), the bolometric luminosity can be computed simply by integration over the disk SED.

We note, however, that if a \textit{bare/unreprocessed} disk SED is assumed in cases where there is X-ray suppression, e.g., early-times of AT2019azh or AT2018zr, where \lbblx $\gg 100$, 
the resulting SED will be based on an unphysical normalization ($R_p/R_g \leq$ 0.3) and hence will underestimate the true bolometric luminosity, even if considering the `disk bolometric luminosity'.

For the cases -- usually at early times -- in which the strong UV/optical excess (i.e., \lbblx $> 100$) is present (see again right panel of Fig. \ref{fig:SED_x_Dai}) the SED shape is strongly dependent on the radiative processes involved, and have large uncertainties because of the lack of constraints in the unobservable EUV waveband.

\subsection{On the unification of the TDE population: survey selection biases and the lack of an optical/X-ray dichotomy}\label{sec:disc_ero_opt}

\begin{figure}
    \centering
    \includegraphics[width=0.85\columnwidth]{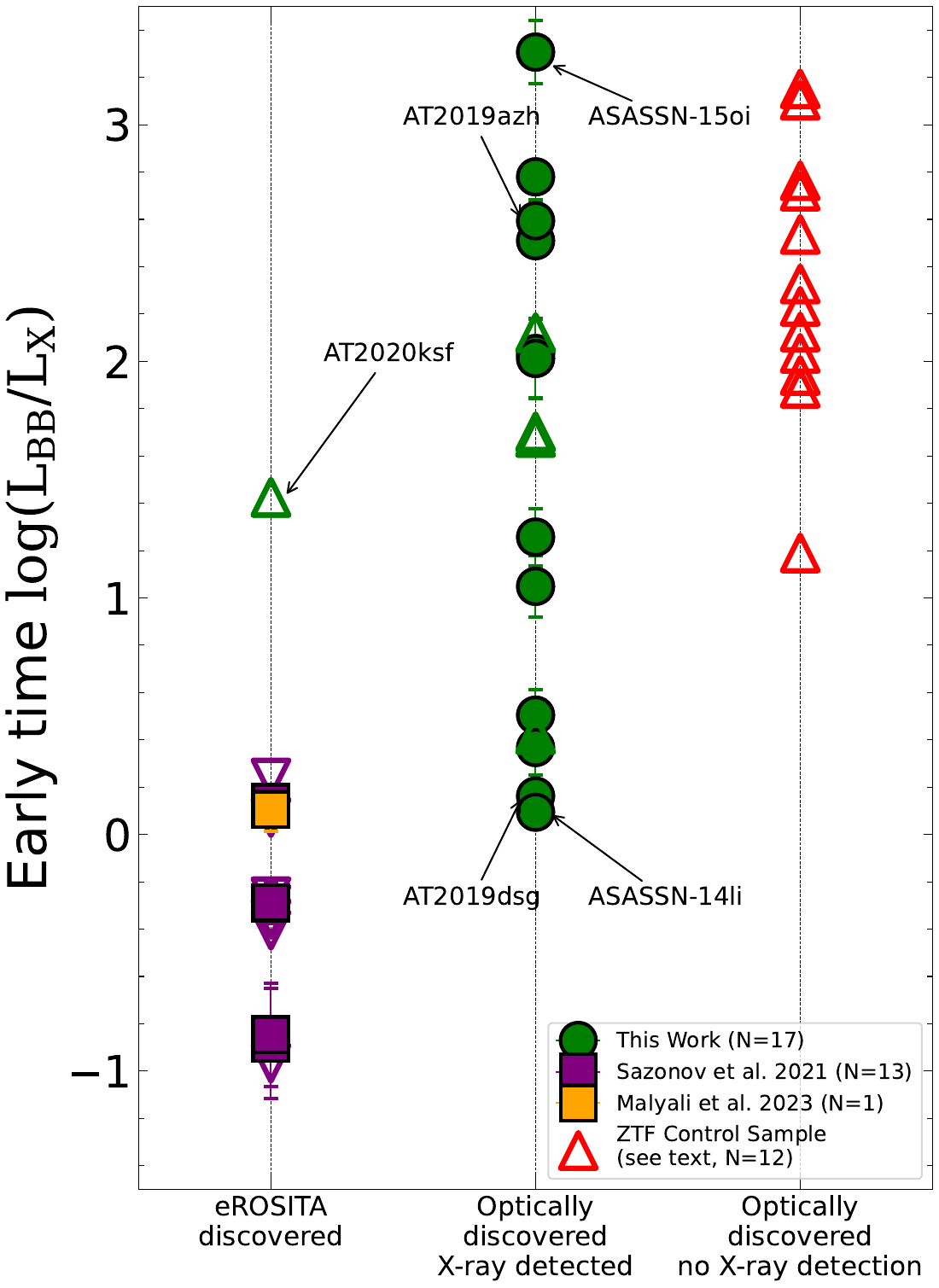}
    \caption{Distribution of early time \lbblx based on discovery wavelength/survey. Extreme sources are marked with arrow. Triangles are lower limits on \lbblx (i.e., no X-ray detection), inverse triangles are an upper limit on \lbblx (i.e., no UV/optical detection), filled markers represent detections in both UV/optical and X-rays. Color represent distinct samples or references. Sources with extreme values are marked with arrows. There is a continuous and wide distribution of \lbblx values, instead of a clear dichotomy between optically and X-ray discovered TDEs.}
    \label{fig:Lbb_Lx_survey}
\end{figure}

\begin{figure*}
    \centering
    \includegraphics[width=0.8\textwidth]{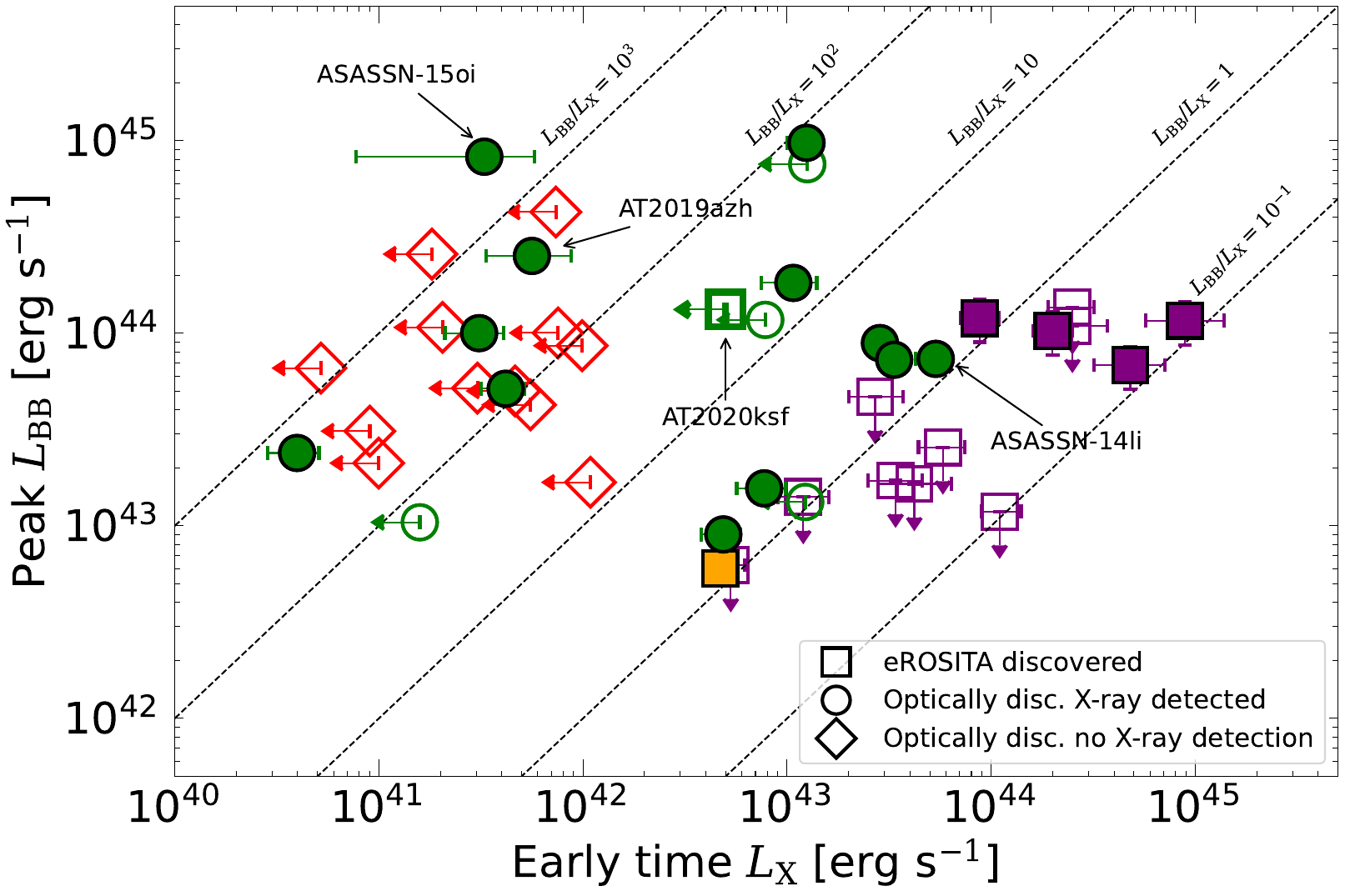}
    \caption{Distribution of peak \lbb $\times$ early time \lx for different TDE populations. Squares show \srge (X-ray) discovered sources, circles show optically discovered X-ray detected, while diamonds show optically discovered with no X-ray detection. Filled markers represent detections in both UV/optical and X-rays (early times), while hollow symbols represent upper limits in one of the two wavelength bands, where the arrows represent their 3$\sigma$ upper limit. The colors are the same as in Fig. \ref{fig:Lbb_Lx_survey}.}
    \label{fig:Lbb_x_Lx}
\end{figure*}

As we have shown, optical surveys can discover TDEs with a large range of \lbblx at early times, as we argue in \S\ref{sec:disc_lc} some (if not most) of this diversity can be explained by the viewing angle towards the system. We would like to compare the TDEs discovered by optical surveys and by X-ray surveys to look for differences (or lack thereof) in these populations and, if present, what could drive such differences.

TDEs discovered by most X-ray missions (e.g., \rosat and \xmm Slew Survey) had very little to no real-time UV/optical follow-up; hence accessing their \lbblx ratio is not possible. We thus focused our analyses on the more recent sources discovered by \srge, in particularly the 13 TDEs presented \citet{Sazonov2021} and eRASSt J074426.3 +291606 (hereafter J0744) presented by \citet{Malyali2023}. Unfortunately, the sources presented in \citet{Sazonov2021} had no UV follow-up, but some constraints on the optical emission from ground-base optical time-domain surveys were obtained. 

As reported by \citet{Sazonov2021}, all of their sources had $L_g/L_X$\footnote{Where $L_g$ is the optical \textit{g}-band luminosity.} ratios lower than 0.3. For a typical UV/optical blackbody temperature of $20,000$ K \citep{Hammerstein2022}, this translates into \lbblx $\leq$ 3. Such values, however, do not mean that these sources were \textit{`X-ray only'}.  For example, the early time \lbblx of ASASSN-14li and AT2019dsg were also lower than 3. This is also true for J0744, which had \lbblx $\approx$ 2 at early times. Most of the sources in \citet{Sazonov2021} had no optical counterpart detection, given the relatively high redshift range of their sample (all but one have $z > 0.1$, compare with our sample, in Table \ref{tab:sources}), ground-based optical surveys are not sensitive enough to detect the typical optical TDE luminosities at these redshifts. For the sources with detected optical counterparts, the range of \lbblx was $\sim 0.1 \leq L_{BB}/L_X \leq 2$. As discussed in \S\ref{sec:disc_lc}, for these values of \lbblx, no UV/optical excess is necessarily present; the shape of the SED shape is fully consistent with an \textit{bare/unreprocessed} accretion disk; hence these systems are likely to be seen at lower angles (i.e., towards face-on orientations), similar to ASASSN-14li and AT2019dsg. The small variations observed in the \lbblx can be fully explained by a range of values of $T_p$, that as shown in appendix \S\ref{app:disk_lbb_lx} can produce \lbblx as low as $\sim 0.05$ for the range of $T_p$ found in TDEs.

The distribution of \lbblx at early times for sources discovered by distinct 
methods/surveys is presented in Fig. \ref{fig:Lbb_Lx_survey}. Triangles are lower limits on \lbblx (i.e., no X-ray detection), inverse triangles are an upper limit on 
\lbblx (i.e., no UV/optical detection), filled markers represent detections in both
UV/optical and X-rays. In Fig. \ref{fig:Lbb_x_Lx} we show the peak \lbb vs. the early time \lx, where the distinct symbols shows: \srge (squares) TDEs, optically selected X-ray detected (circles) and optically selected with no X-ray detection (diamond); filled marker show detection in both optical/UV (y-axis) and X-ray (x-axis), while hollow markers either UV/optical or X-rays (at early times) were no detected, similarly to Fig.\ref{fig:Lbb_Lx_survey} distinct color represents the different references.

From Fig. \ref{fig:Lbb_Lx_survey}, it is clear that there is no obvious dichotomy between optical and X-ray discovered TDEs. Instead, there is a 
continuous distribution of \lbblx values at early times, that is at least as wide as \lbblx $\in 
(0.1, 32000)$. This can be explained by the fact that surveys at distinct wavelengths will be biased to discover sources  that are brightest in that wavelength range. Optical surveys will discover mostly sources with high \lbblx, up to $60\%$ of which with \lx $\leq 10^{42}$ \ergs (see \S\ref{sec:x_ratio}), but will still sometimes discover  sources with lower \lbblx, such as ASASSN-14li and 
AT2019dsg. X-ray surveys, on the other hand, will most likely discover sources with lower 
\lbblx, but not always; AT202ksf, for example, had no X-ray detection at early 
times (\lbblx $\geq 25$), was not identified in real-time as a TDE candidate by optical surveys, and 
later ($\Delta t \sim  200$ days) was discovered by \srge following its X-ray 
brightening. The same holds for their distribution in \lbb $\times$ \lx plane of Fig. \ref{fig:Lbb_x_Lx}, there is a continuous distribution of the X-ray and UV/optional selected populations, instead of clear dichotomy between them.

An \textit{unbiased}\footnote{This is still biased towards dust free host galaxies, for heavily dust obscured TDEs the UV/optical and soft X-ray are absorbed and re-emitted in the infrared, see for example \citet{Panagiotou2023} and \citet{Masterson2024}.} discovery of TDEs would be possible with either: {\it i}) the simultaneous operation of an wide field-of-view (FoV) X-ray survey telescope with UV follow-up capabilities \citep[e.g., STAR-X,][]{Hornschemeier2023} and a wide FoV optical survey \citep[e.g LSST,][]{Ivezic2019}, {\it ii}) or the simultaneous and coordinated operation of an wide FoV UV survey telescope  \citep[e.g, ULTRASAT and/or UVEX,][]{Sagiv2014,Kulkarni2021} and wide FoV X-ray survey telescope \citep[e.g., \srge and/or \textit{Einstein Probe},][]{Sunyaev2021,Yuan2015}.

\section{Conclusions}\label{sec:conclusions}
We have analyzed the \xmm and \swiftall X-ray and broad-band UV/optical emission of 17 optically selected X-ray detected TDEs discovered between 2014 and December 2021; we also compare our sample with the samples of optically discovered X-ray quiet TDEs and X-ray discovered TDEs, our main conclusions are:

\begin{itemize}
    \item The X-ray light curves show a large diversity, with sources rarely showing a power-law decay and a large fraction showing a late-time brightening. 
    \item The X-ray spectra are extremely soft in most sources and epochs, easily distinguishable from AGN X-ray spectra. 
    \item The overall behavior of the measured radius (normalization) and temperature (shape) resulting from the X-ray spectral fitting is in agreement with that expected for the innermost region of a newly formed and time-evolving accretion disk, including the cooling of the peak temperature and a radius (in most cases) consistent with innermost stable orbit.
    \item Sources with early-time faint X-ray emission show an unphysical radius for the X-ray emission at these epochs, while their temperature behaves as expected, indicating the apparent suppression/absorption of their intrinsic early X-ray emission.
    \item The spectral energy distribution (SED) shape, as probed by the ratio (\lbblx) between the UV/optical and X-ray luminosities, has a large range of values \lbblx $\in (0.5, 3000)$ at early times, at late times the range converges to disk-like values, \lbblx $\in (0.5, 10)$, for all sources.
    \item The combined X-ray spectral and SED properties and evolution favors a change in the optical depth (thick $\rightarrow$ thin)
    for the high energies photons through the line-of-sight, instead of the delayed formation of the accretion disk, in order to explain the late-time brightening observed in some sources. 
    \item Three sources show a soft $\rightarrow$ hard X-ray spectral transition, indicative of the formation of a hot corona akin to active galaxies, with the state transitions occurring at least 200 days after the UV/optical peak, but it is not sustained for more than a couple of months.
    \item We estimated that the fraction of optically discovered TDEs that are X-ray loud, with \lx $\geq 10^{42}$ \ergs, is at least $40\%$ and that X-ray loudness is not dependent on \mbh.
    \item We show that the TDE X-ray luminosity function from $10^{41}-10^{45}$ \ergs has a broken-power-law shape in the form of ${dN}/{{dL}}_X\propto {L}_X^{-1.0\pm0.2}$ at $L_X < L_{bk}$ and ${dN}/{{dL}}_X\propto {L}_X^{-2.7\pm1.0}$ at $L_X \geq L_{bk}$ with break luminosity of log$(L_{bk}) = 44.1^{+0.3}_{-0.5}$ \ergs. Reveling a large population of TDE with \lx $\leq 10^{42}$ \ergs (and high \lbblx), for which the X-ray emission can not be detected with current instruments unless occurs at very low $z$.
    \item We show that there is no dichotomy between optical and X-ray discovered TDEs, and instead there is a continuous range of early time \lbblx, at least as wide as \lbblx $\in (0.1, 3000)$, with X-ray/optical surveys discovering preferentially, but not exclusively, from the lower/higher portion of the distribution, in agreement with unification models for the overall TDE population.
   
\end{itemize}

\clearpage

\begin{acknowledgments}
\textit{Acknowledgments} -- M.G. thanks T. Wevers, J. Krolik, P. Jonker and E. Coughlin for comments, suggestions and discussions. M.G. and S.G. are supported in part by NASA XMM-Newton grants 80NSS23K0621 and 80NSSC22K0571. Y. Y. acknowledges support by NASA XMM-Newton grant 80NSSC23K0482. E.H. acknowledges support by NASA under award number 80GSFC21M0002. This work is based on observations obtained with the Samuel Oschin Telescope 48-inch and the 60-inch Telescope at the Palomar Observatory as part of the Zwicky Transient Facility project. 
ZTF is supported by the National Science Foundation under grant No. AST-2034437 and a collaboration including Caltech, IPAC, the Weizmann Institute of Science, the Oskar Klein Center at Stockholm University, the University of Maryland, Deutsches Elektronen-Synchrotron and Humboldt University, the TANGO Consortium of Taiwan, the University of Wisconsin at Milwaukee, Trinity College Dublin, Lawrence Livermore National Laboratories, IN2P3, University of Warwick, Ruhr University Bochum, and Northwestern University. Operations are conducted by COO, IPAC, and UW. The ZTF forced-photometry service was funded under the Heising-Simons Foundation grant \#12540303 (PI: Graham). This work has made use of data from the Asteroid Terrestrial-impact Last Alert System (ATLAS) project. The Asteroid Terrestrial-impact Last
Alert System (ATLAS) project is primarily funded to
search for near earth asteroids through NASA grants
NN12AR55G, 80NSSC18K0284, and 80NSSC18K1575. This work made use of data supplied by the UK Swift
Science Data Centre at the University of Leicester.

\end{acknowledgments}

\software{
\texttt{astropy} \citep{Astropy2013},
\texttt{emcee} \citep{Foreman-Mackey2013},
\texttt{heasoft} \citep{HEASARC2014},
\texttt{matplotlib} \citep{Hunter2007},
\texttt{Prospector} \citep{Johnson2021},
\texttt{scipy} \citep{Virtanen2020},
\texttt{xspec} \citep{Arnaud1996}.
}
\facilities{
XMM,  Swift, eROSITA, PO:1.2m., OGLE.} 
\bibliographystyle{aasjournal}
\bibliography{tde}

\appendix
\newcommand*\SimPL{{\texttt{\textsc{simPL}}}}
\section{Modeling of corona emission in TDEs with \SimPL}\label{app:simpl}

In the study of active galactic nuclei (AGN), the corona is the primary X-ray emitter, while the cold accretion disks mainly emit in the UV bands. AGN X-ray spectra are often described using the \texttt{powerlaw} model, despite some attempts to develop more comprehensive models like \texttt{Optxagnf} \citep{Done2012}, these face challenges due to numerous parameters and degeneracies. Therefore, the \texttt{powerlaw} model remains prevalent in AGN X-ray literature.

In X-ray binaries (XRBs), the accretion disks are hotter and emit mainly in the X-ray bands. In the soft state of XRBs, the accretion disk dominates the X-ray emission, but as they transition to the hard state, an emergent corona with a power-law spectrum becomes dominant. The \texttt{powerlaw} model is widely used in XRB modeling but suffers from the drawback of rising infinitely at low energies, which is inconsistent with Comptonization.

\begin{figure}[!ht]
    \centering
    \includegraphics[width=0.9\columnwidth]{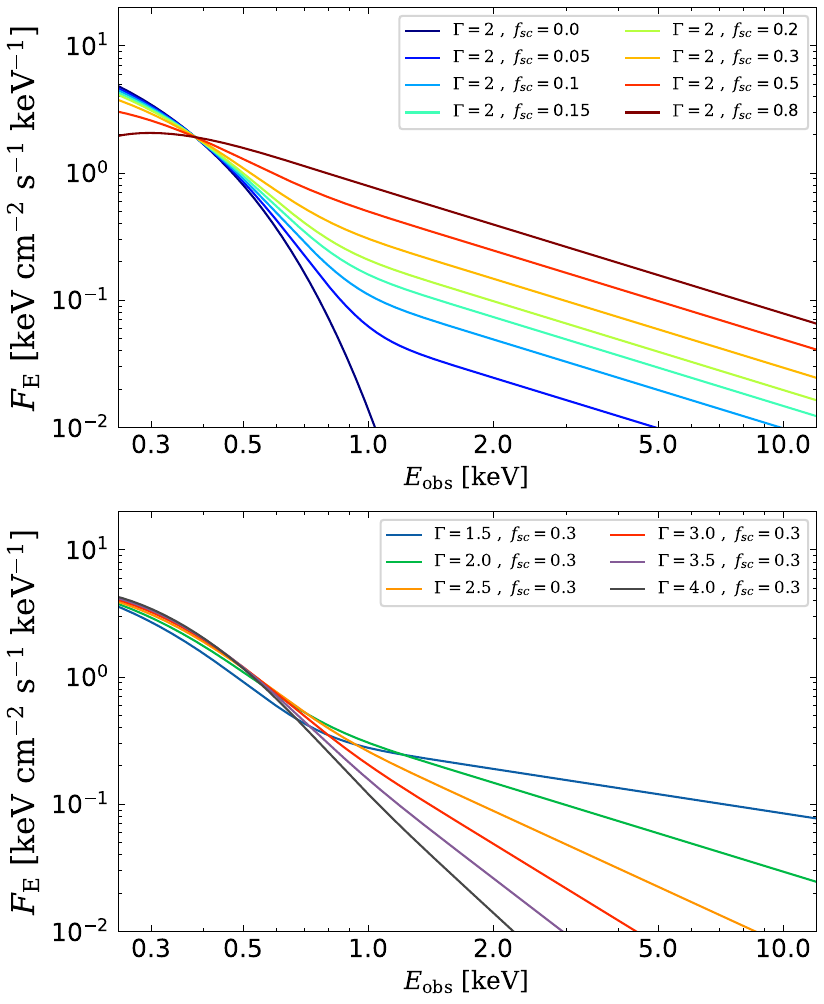}
    \caption{Simulation of \texttt{SimPL} model spectra. Upper panel: Fixed power law index ($\Gamma_{sc}$) varying fraction of up-scattered photons ($f_{sc}$). Bottom panel: Fixed $f_{sc}$ varying $\Gamma_{sc}$.} 
    \label{fig:simpl}
\end{figure}

\begin{figure}[!ht]
    \centering
    \includegraphics[width=0.9\columnwidth]{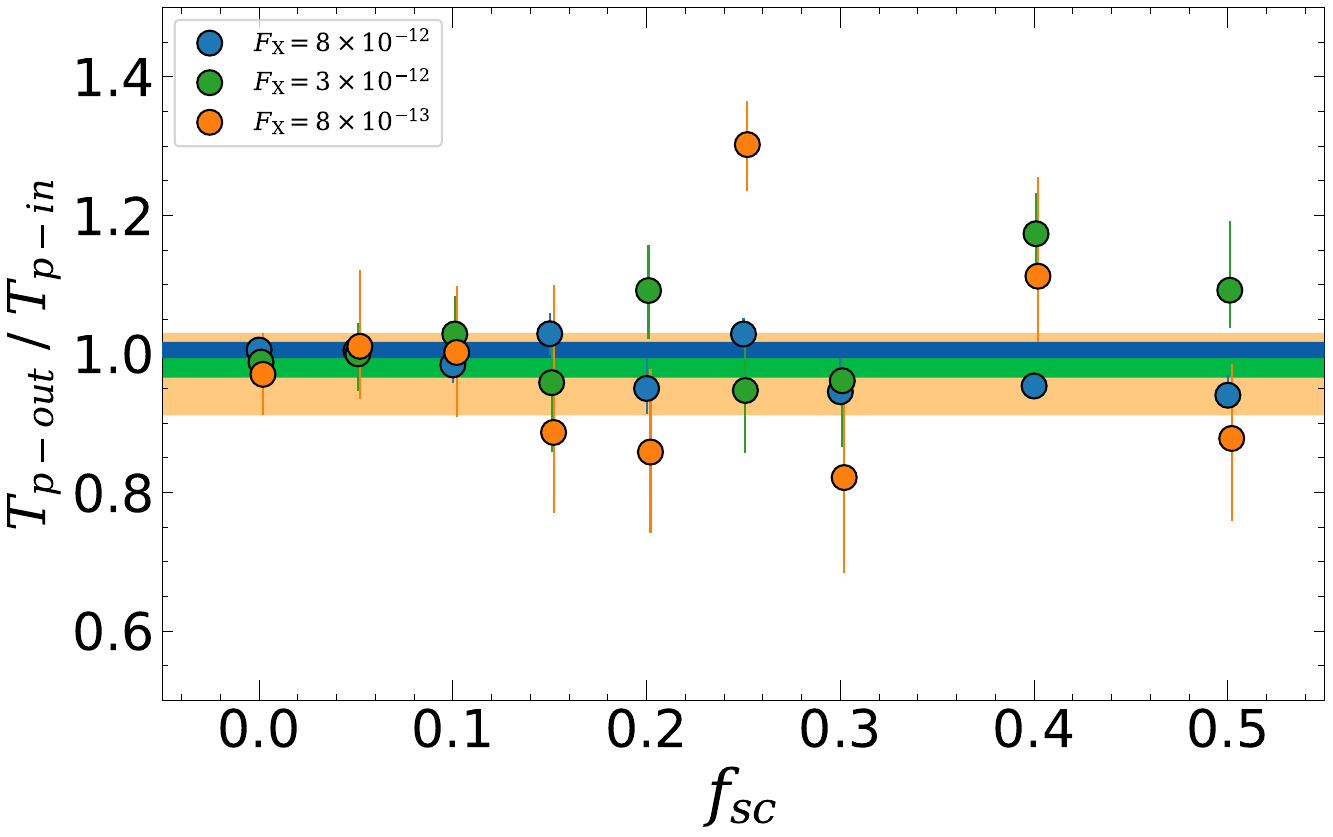}
    \caption{Simulation on the measurement of the underlying disk properties in the presence of a corona comptonization. The y-axis show the ratio between the input and output $T_p$ as a function of the corona strength ($f_{sc}$). The color show distinct fluxes of the mock spectra, and the shaded regions the uncertainty in the $f_{sc}=0$ spectra, i.e. the uncertainty from the instrumental S/N.}
    \label{fig:simpl2}
\end{figure}

To overcome the limitation of the \texttt{powerlaw} model in accurately describing Comptonization, \citet{Steiner2009} introduced \texttt{simPL}, a flexible convolution model for fitting X-ray spectra of XRBs. \texttt{simPL} captures Comptonization effects using any seed photon spectrum and shares parameters such as the photon index ($\Gamma$) with the \texttt{powerlaw} model. However, it employs the scattered fraction ($f_{sc}$) as the normalization factor instead of photon flux, simplifying the model by omitting specific details of the Comptonizing medium while maintaining a physically consistent approach. Unlike the \texttt{powerlaw} model, \texttt{simPL} directly links the power-law component to the input photon energy distribution, resulting in a power-law tail at higher energies without extending indefinitely to lower energies. This behavior aligns with Compton-scattering expectations and is commonly observed in physical Comptonization models. Notably, in the soft X-ray bands, \texttt{simPL} exhibits a natural cutoff consistent with Comptonization, whereas the \texttt{powerlaw} model continues to rise without limit \citep{Yao2005}.

Similarly to XRBs, TDEs also have a portion of their continuum disk emission in the X-rays. Therefore, a similar approach can be applied to model the X-ray spectra of TDEs. Figure \ref{fig:simpl} demonstrates the effects of using \texttt{simPL} through a series of simulations where a thermal model with $T_{p} \approx 70$ eV and $R_p \approx 10^{12}$ cm is convoluted with \texttt{simPL} using different values of $f_{sc}$ (top panel) and $\Gamma$ (bottom panel). With $f_{sc} = 0$, there is no corona emission, the resulting model corresponds to the input disk spectrum. As $f_{sc}$ increases from 0.05, the spectrum remains dominated by the disk, but with a faint hard excess. Between $0.5 \leq f_{sc} \leq 0.20$, the source enters an intermediate state, where the thermal and non-thermal spectra have similar fluxes. When $f_{sc} \approx 0.3$, the total spectrum becomes almost indistinguishable from a pure power-law, and the spectrum remains consistent with a power law for higher values.

An important question to our X-ray spectral analyses (\ref{sec:X-ray_fit} and \ref{sec:T_x_L}) is up to what corona strength (as probed by $f_{sc}$) the underlying thermal continuum can be recovered from the fitting. To answer this question we simulate mock X-ray spectra using the \texttt{fakeit} command in \texttt{xspec} for a stacked 30 ks observation with \swift/XRT, assuming \texttt{\texttt{simPL$\otimes$tdediscspec}}, a random $T_p$ in the range of observed in our sample, for three fluxes ($F_{\rm X} = 8\times10^{-12}, 3\times10^{-12}, 8\times10^{-13}$ \ergcms) levels, and varying $f_{sc}$ between 0.0 (no corona) and 0.5. We then fit the mock spectra with the same model in order to measure the best-fit $T_p$, and compare with the input  
value. Fig. \ref{fig:simpl2} summarized our findings. The underlying (input) temperature can be recovered (within the error-bar) considering the uncertainties related with the S/N of the instrument (shaded regions) up to $f_{sc} \approx 0.2$, at higher values the underlying information on the temperature of the disk is lost by the emergency of the corona in the higher energies of the disk spectrum, and can not be uniquely recovered.

\section{Simulation of expected \lbblx for standard disk}\label{app:disk_lbb_lx}

\begin{figure}
    \centering
    \includegraphics[width=0.9\columnwidth]{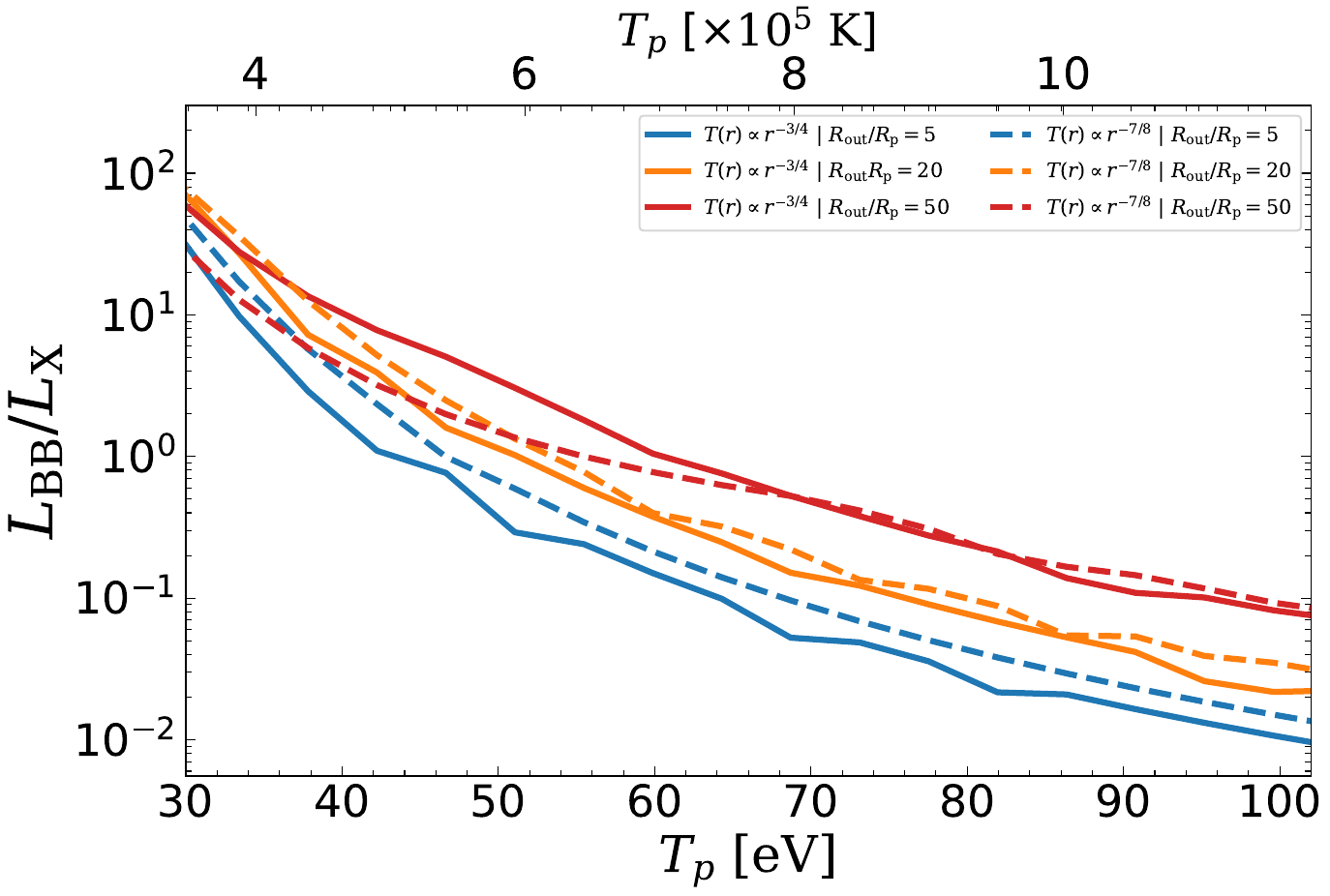}
    \caption{Simulation of expected UV/optical to X-ray luminosity ratio (\lbblx) for standard disk. The y-axis  show the expected \lbblx from a \textit{bare/unreprocessed} accretion disk with two distinct temperature profiles (solid and dashed lines), as a function of the peak temperature of the disk ($T_p$).}
    \label{fig:disk_sim2}
\end{figure}

An important observational probe of TDE emission is the ratio \lbblx between the UV/optical luminosity (as fitted 
by a blackbody) and the 0.3-10 keV X-ray luminosity. Although additional emission processes should be involved, 
the formation of accretion disk is a natural prediction of a tidal disruption event. In this section, we aim to 
probe the range of \lbblx that can be produce by an \textit{bare/unreprocessed} accretion disk, given the range of inner peak
temperatures ($T_p$) we observe from the X-ray fitting. To obtain that we simulate accretion disk with varying $T_p$ between $5.5 \leq \rm{log} \ T_p \leq 6.1$ Kelvin, we test two distinct disk solutions for the the temperature 
profile of the disk: the standard  vanishing ISCO stress solution \citep{Shakura1973,Makishima_86}, $T(R) \propto R^{-3/4}$, and the finite ISCO stress \citep{Agol2000}, $T(R) \propto R^{-7/8}$. We also explore distinct disk outer radius, from $R_{out}/R_p \in (5,50)$.  We pass the synthetic SED into the 
sensibility curve of the six UVOT and two ZTF filters, and fitted the resulting broad-band UV/optical with a 
blackbody to obtain \lbb, the same way it was done in the observed data (see Fig. \ref{fig:Lbb_Lx_measurments} 
for illustration). The ratio \lbblx as a function of $T_p$ for the two distinct disk solution is shown in 
Fig.~\ref{fig:disk_sim2}. For the entire range \lbblx is between $5 \times 10^{-2}$ and a maximum of $\sim$70. At late
times, most our sample shows disk cooling, and all TDE have $\rm{log } T_p \leq 5.8$, this limits \lbblx to values $> 0.3$, but rarely higher 
than 10, which is in agreement with our findings in Fig. \ref{fig:SED_hist}.

\section{The BASS AGN sample}\label{app:bass}

The Swift/BAT 70-month AGN catalog consists of 858 nearby ($z\lesssim0.3$ for unbeamed) AGN \citep{kossBASSXXIIBASS2022}, and using soft X-ray observations by XMM-Newton, Swift, ASCA,
Chandra, and Suzaku, their broadband X-ray spectra were characterized and presented by \cite{Ricci2017}.
Some of the properties they constrain are the intrinsic X-ray luminosity (in the 2--10, 20--50, and 14-150 keV bands), the line-of-sight column density of obscuring material ($N_\mathrm{H}$), the slope of the X-ray power-law continuum, and the temperature of the thermal plasma for obscured sources.
The many phenomenological models used are broadly classified into four groups: unobscured (352), obscured (386), blazars (97), and other non-AGN models (2). The remaining details of the X-ray modeling of the BASS sample can be found in \cite{Ricci2017}.

From among these sources, we selected those that were either obscured or unobscured (which excludes beamed and non-AGN sources), and that had spectroscopic redshift measurements from optical counterparts.
This resulted in 617 sources.
To calculate the HR, the X-ray spectra were simulated from the models using \texttt{XSPEC}, and the count rate was recorded for the soft (0.3--2.0 keV) and hard (2.0--10.0 keV) bands.
Spectra were measured with a long response time (1 Ms) to minimize the effects of statistical noise on the HR.
This was done twice: with the response files for Swift/XRT photon counting grade 0-12 and XMM-Newton Epic PN. 
We also measured the 0.3-10.0 keV intrinsic luminosities by setting all $N_\mathrm{H}$ parameters of all model components to zero (or, in certain cases, the minimum nonzero value allowed by the model) and using the \texttt{calcLumin} command.

\section{Supplementary Data}

\begin{deluxetable}{cCcC}[!h]
\label{tab:appendix_ratio}
\tabletypesize{\scriptsize}
\tablecaption{ZTF control sample, described in \S\ref{sec:x_ratio}.}
\tablewidth{0pt}
\tablehead{
\colhead{Source} &  \colhead{z} & \colhead{$L_X \geq 10^{42}$ \ergs}  & \colhead{log $(M_{\rm{BH}}\rm{/}M_{\odot})^{a}$}
}
\startdata
\hline
AT2018zr  & 0.075 & False & 5.83 \pm 0.51 \\
 AT2018bsi & 0.051 & False & 7.46 \pm 0.47 \\
 AT2018hco & 0.088 & False & 6.44 \pm 0.48 \\
 AT2018iih & 0.212 & False & 7.93 \pm 0.48 \\
 AT2018hyz & 0.046 & False & 6.12 \pm 0.46 \\
 AT2018lna & 0.091 & False & 5.21 \pm 0.54 \\
 AT2019azh & 0.022 & True  & 6.68 \pm 0.46 \\
 AT2019dsg & 0.051 & True  & 7.04 \pm 0.45 \\
 AT2019ehz & 0.074 & True  & 5.75 \pm 0.59 \\
 AT2019mha & 0.148 & False & 6.41 \pm 0.49 \\
 AT2019meg & 0.152 & False & 5.81 \pm 0.52 \\
 AT2019qiz & 0.015 & False & 6.49 \pm 0.49 \\
 AT2019teq & 0.087 & True  & 6.32 \pm 0.49 \\
 AT2020pj  & 0.068 & False & 6.43 \pm 0.49 \\
 AT2019vcb & 0.088 & True  & 5.59 \pm 0.52 \\
 AT2020ddv & 0.160 & True  & 6.09 \pm 0.55 \\
 AT2020ksf & 0.092 & True  & 5.92 \pm 0.48 \\
 AT2020ocn & 0.070 & True  & 6.77 \pm 0.49 \\
 AT2020mot & 0.070 & False & 7.04 \pm 0.47 \\
 AT2020wey & 0.027 & False & 5.38 \pm 0.51 \\
 AT2020zso & 0.057 & False & 6.12 \pm 0.48 \\
 AT2020vwl & 0.033 & False & 5.80 \pm 0.48 \\
 AT2021ehb & 0.018 & True  & 7.04 \pm 0.46 \\
 AT2021nwa & 0.047 & False & 7.21 \pm 0.46 \\
 AT2020ksf & 0.092 & True  & 5.92 \pm 0.48 \\
 AT2021yzv & 0.286 & True  & 7.45 \pm 0.47 \\
\enddata
\tablecomments{a) Black hole masses. When a $\sigma_*$ measurement is available it is estimated using the \citet{Gultekin2019} $\sigma_*-$\mbh relation, when $\sigma_*$ is not available, this is estimated from the $M_*$-\mbh relation presented in \citealt{Yao2023}.}
\end{deluxetable}
\vspace{-1cm}
\begin{figure}[!b]
    \centering
    \includegraphics[width=0.99\columnwidth]{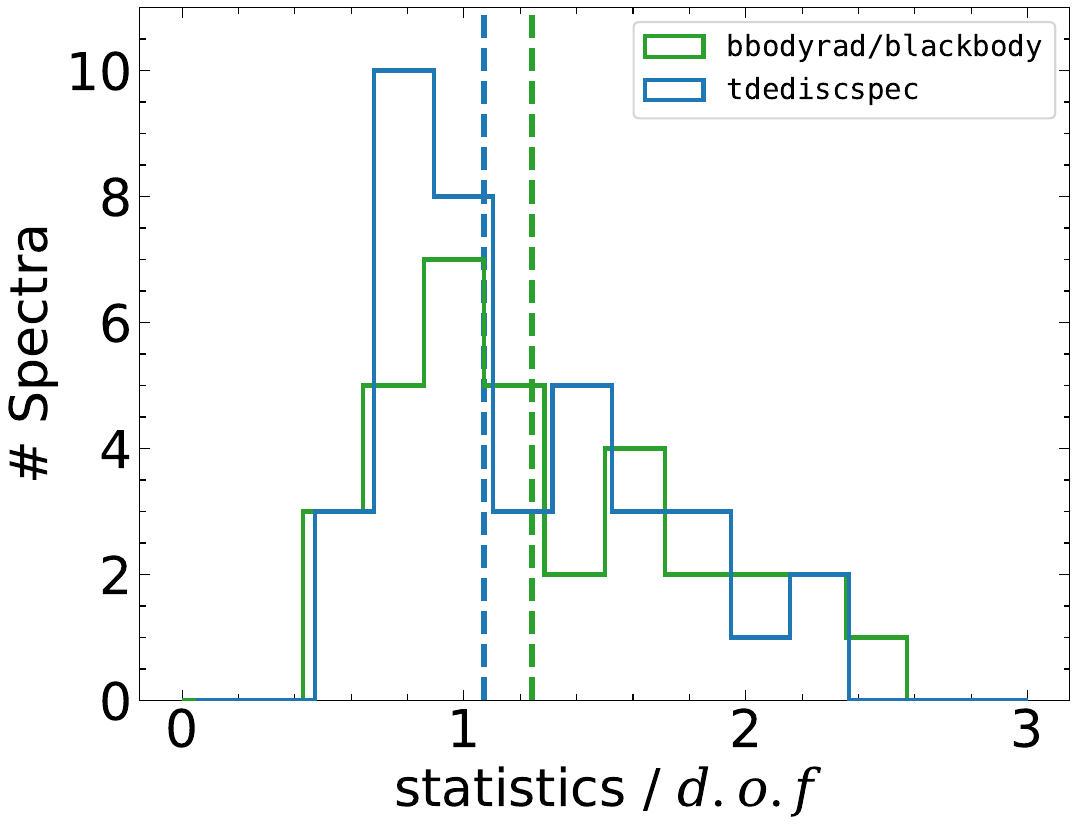}
    \caption{Distribution of ratio of the statistics ($\chi^2$ for \xmm and `c-statistics' \citep{Cash1979} for \swift/XRT) over degrees of freedom ($d.o.f$) for spectra with negligible `hard excess' (i.e. $f_{sc} \sim$ 0.0) for \texttt{tdediscspec} and \texttt{blackbody}, \texttt{xspec} models. Dashed line show the median values for the models.}
    \label{fig:model_stat}
\end{figure}

\begin{deluxetable}{cCCc}[h]
\label{tab:append_X-TDE}
\tabletypesize{\scriptsize}
\tablecaption{Luminosity function data}
\tablewidth{0pt}
\tablehead{
\colhead{Source} &  \colhead{z} & \colhead{`peak' \ \lx } & \colhead{Survey}\\
\colhead{} &  \colhead{} & \colhead{(\ergs)} & \colhead{}
}
\startdata
\hline
 AT2018zr         & 0.071 & 5.55\times10^{41} & ZTF        \\
 AT2018hyz        & 0.045 & 7.51\times10^{41} & ZTF/ASASSN \\
 AT2019azh        & 0.022 & 1.65\times10^{43} & ZTF/ASASSN \\
 AT2019dsg        & 0.051 & 3.90\times10^{43} & ZTF        \\
 AT2019ehz        & 0.074 & 6.35\times10^{43} & ZTF        \\
 AT2019qiz        & 0.015 & 8.06\times10^{40} & ZTF/ASASSN \\
 AT2019teq        & 0.087 & 4.06\times10^{43} & ZTF        \\
 AT2019vcb        & 0.088 & 1.66\times10^{43} & ZTF        \\
 AT2020ddv        & 0.160 & 3.45\times10^{43} & ZTF        \\
 AT2020ksf        & 0.092 & 1.07\times10^{44} & ZTF/eROSITA       \\
 AT2020ocn        & 0.070 & 1.04\times10^{44} & ZTF        \\
 AT2021ehb        & 0.018 & 1.78\times10^{44} & ZTF        \\
 ASASSN-14li      & 0.020 & 9.77\times10^{43} & ASASSN     \\
 ASASSN-15oi      & 0.048 & 4.93\times10^{42} & ASASSN     \\
 AT2018fyk        & 0.059 & 1.76\times10^{44} & ASASSN     \\
 OGLE16aaa        & 0.165 & 4.92\times10^{43} & OGLE       \\
 AT2021yzv        & 0.288 & 3.83\times10^{42} & ZTF        \\
 J135514.8+311605 & 0.199 & 5.80\times10^{43} & eROSITA    \\
 J013204.6+122236 & 0.123 & 4.20\times10^{43} & eROSITA    \\
 J153503.4+455056 & 0.231 & 8.80\times10^{43} & eROSITA    \\
 J163831.7+534020 & 0.581 & 2.50\times10^{44} & eROSITA    \\
 J163030.2+470125 & 0.294 & 2.00\times10^{44} & eROSITA    \\
 J021939.9+361819 & 0.387 & 2.50\times10^{44} & eROSITA    \\
 J161001.2+330121 & 0.131 & 1.20\times10^{43} & eROSITA    \\
 J171423.6+085236 & 0.036 & 5.30\times10^{42} & eROSITA    \\
 J071310.6+725627 & 0.104 & 1.10\times10^{44} & eROSITA    \\
 J095928.6+643023 & 0.045 & 8.90\times10^{44} & eROSITA    \\
 J091747.6+524821 & 0.187 & 4.80\times10^{44} & eROSITA    \\
 J133053.3+734824 & 0.150 & 3.40\times10^{43} & eROSITA    \\
 J144738.4+671821 & 0.125 & 2.70\times10^{43} & eROSITA    \\
 NGC5905          & 0.011 & 7.00\times10^{42} & ROSAT      \\
 RXJ1624+7554     & 0.064 & 2.00\times10^{44} & ROSAT      \\
 RBS 1032         & 0.026 & 1.00\times10^{43} & ROSAT      \\
 RXJ1420+5334     & 0.147 & 2.50\times10^{44} & ROSAT      \\
 RXJ 1242-1119    & 0.050 & 4.00\times10^{44} & ROSAT      \\
 TDXF1347-3254    & 0.037 & 7.00\times10^{42} & ROSAT      \\
 NGC 3599         & 0.003 & 1.20\times10^{41} & XMMLS      \\
 SDSS J1323+48    & 0.087 & 4.40\times10^{43} & XMMLS      \\
 SDSS J0939+37    & 0.184 & 2.60\times10^{44} & XMMLS      \\
 2MASX J0203-07   & 0.062 & 2.80\times10^{43} & XMMLS      \\
 2MASX J02491-04  & 0.019 & 2.10\times10^{42} & XMMLS      \\
 SDSS J1201+30    & 0.146 & 3.00\times10^{44} & XMMLS      \\
 2MASX 0740-85    & 0.017 & 2.00\times10^{43} & XMMLS      \\
 XMMSL2 J1446+68  & 0.029 & 6.00\times10^{42} & XMMLS      \\
 XMMSL1 J1404     & 0.043 & 3.00\times10^{43} & XMMLS      \\
\enddata
\end{deluxetable}

\begin{deluxetable*}{cCcCCCCC}
\tabletypesize{\tiny}
\label{tab:best_fit_ta}
\tablecaption{Best-fit parameters for final X-ray models: \texttt{tdediscspec} or \texttt{simPL$\otimes$tdediscspec}.}
\tablewidth{5pt}
\tablehead{
\colhead{Source} &  \colhead{MJD} & \colhead{Instrument}  & \colhead{$T{_p} \ [{\rm K}]$}  &\colhead{$R{_p} \ [{\rm cm}]$} & \colhead{$f_{sc}$} & \colhead{$\Gamma_{sc}$} & \colhead{statistics$^{a}$/d.o.f}
}
\startdata
\hline
 \multirow{3}{*}{AT2018zr}  & 58220 & \xmm & 1.1^{+0.1}_{-0.1}\times10^{6} & 2.5^{+1.1}_{-0.7}\times10^{10} & \nodata & \nodata & 6.3/8  \\
  & 58241 & \xmm & 1.0^{+0.2}_{-0.1}\times10^{6} & 2.0^{+1.0}_{-0.6}\times10^{10} & \nodata & \nodata & 6.0/6  \\
  & 58569 & \xmm & 5.2^{+0.6}_{-0.4}\times10^{5} & 2.0^{+1.0}_{-0.6}\times10^{11} & \nodata & \nodata & 11.3/7 \\
\hline
 AT2018hyz & 58463^{+30}_{-28} & \swift/XRT & 1.2^{+0.3}_{-0.2}\times10^{6} & 1.1^{+0.9}_{-0.5}\times10^{10} & \nodata & \nodata & 13.0/22 \\
\hline
 \multirow{3}{*}{AT2019azh} & 58579 & \xmm & 5.7^{+1.4}_{-0.7}\times10^{5} & 8.9^{+4.2}_{-1.9}\times10^{10} & 0.03^{+0.06}_{-0.02} & 2.2^{+1.2}_{-0.7} & 10.3/16 \\
  & 58760 & \xmm & 4.1^{+0.1}_{-0.3}\times10^{5} & 2.4^{+0.3}_{-0.3}\times10^{12} & 0.01^{+0.02}_{-0.01} & 1.3^{+2.3}_{-1.3} & 30.2/18 \\
  & 58788 & \xmm & 3.5^{+0.1}_{-0.1}\times10^{5} & 4.0^{+0.2}_{-0.1}\times10^{12} & 0.01^{+0.01}_{-0.01} & 4.3^{+0.5}_{-0.2} & 34.4/21 \\
  & 58971 & \xmm & 2.8^{+0.1}_{-0.1}\times10^{5} & 6.2^{+6.0}_{-1.6}\times10^{12} & \nodata              & \nodata           & 16.1/7  \\
\hline
 \multirow{3}{*}{AT2019dsg} & 58625^{+4}_{-5} & \swift/XRT & 6.7^{+1.4}_{-1.0}\times10^{5} & 2.3^{+3.7}_{-1.3}\times10^{12} & \nodata & \nodata & 26.0/41         \\
  & 58633           & \xmm       & 5.7^{+0.1}_{-0.6}\times10^{5} & 2.6^{+0.2}_{-0.9}\times10^{12} & \nodata & \nodata & 33.3/22         \\
  & 58641^{+7}_{-7} & \swift/XRT & 5.1^{+0.9}_{-0.7}\times10^{5} & 2.8^{+5.1}_{-1.5}\times10^{12} & \nodata & \nodata & 36.4/34         \\
\hline
  \multirow{3}{*}{AT2019ehz} & 58628^{+1}_{-1}       & \swift/XRT & 1.1^{+0.1}_{-0.1}\times10^{6} & 2.5^{+0.4}_{-0.3}\times10^{11} & \nodata & \nodata & 84.8/78 \\
  & 58633           & \xmm       & 8.2^{+0.6}_{-0.5}\times10^{5} & 8.3^{+1.9}_{-1.4}\times10^{10} & \nodata & \nodata & 6.5/9   \\
  & 58682^{+2}_{-2} & \swift/XRT & 1.1^{+0.1}_{-0.1}\times10^{6} & 1.9^{+0.4}_{-0.3}\times10^{11} & \nodata & \nodata & 52.4/58 \\
\hline
 \multirow{2}{*}{AT2019teq} & 58841 & \xmm & 1.9^{+0.1}_{-0.1}\times10^{6} & 2.5^{+0.6}_{-0.1}\times10^{10} & \nodata & \nodata & 34.2/41 \\
  & 58915 & \xmm & \nodata                       & \nodata                        & \nodata & \nodata & 82.5/48 \\
\hline
 \multirow{2}{*}{AT2019vcb} & 58991 & \xmm & 5.4^{+0.4}_{-0.3}\times10^{5} & 3.9^{+2.4}_{-1.2}\times10^{11} & \nodata & \nodata & 25.1/18 \\
  & 59764 & \xmm & 4.0^{+1.0}_{-0.5}\times10^{5} & 2.9^{+9.5}_{-0.5}\times10^{11} & \nodata & \nodata & 9.1/8   \\
\hline
\multirow{1}{*}{AT2020ddv}  & 58967 & \xmm & 7.0^{+0.2}_{-0.1}\times10^{5} & 8.4^{+0.7}_{-0.5}\times10^{11} & \nodata & \nodata & 29.8/16 \\
\hline
 \multirow{4}{*}{AT2020ksf}   & 59185^{+5}_{-5} & \swift/XRT & 8.0^{+0.4}_{-0.4}\times10^{5} & 7.0^{+1.2}_{-0.9}\times10^{11}  & \nodata & \nodata & 48.9/57 \\
  & 59202^{+3}_{-3} & \swift/XRT & 6.5^{+0.5}_{-0.4}\times10^{5} & 1.1^{+0.3}_{-0.2}\times10^{12}  & \nodata & \nodata & 38.8/35 \\
  & 59311^{+3}_{-3} & \swift/XRT & 6.8^{+0.5}_{-0.4}\times10^{5} & 1.0^{+0.3}_{-0.2}\times10^{12} & \nodata & \nodata & 39.6/42 \\
  & 59725           & \xmm       & 4.3^{+0.2}_{-0.1}\times10^{5} & 2.1^{+0.3}_{-0.1}\times10^{12}  & \nodata & \nodata & 21.4/13 \\
\hline
 \multirow{11}{*}{AT2020ocn}   & 59041^{+1}_{-1}   & \swift/XRT & 8.0^{+0.7}_{-0.6}\times10^{5} & 4.5^{+1.4}_{-1.0}\times10^{11} & \nodata              & \nodata           & 29.7/42   \\
  & 59048             & \xmm       & 4.9^{+0.1}_{-0.4}\times10^{5} & 2.3^{+0.1}_{-0.4}\times10^{12} & \nodata              & \nodata           & 35.2/15   \\
  & 59051^{+2}_{-3}   & \swift/XRT & 5.4^{+0.4}_{-0.4}\times10^{5} & 8.2^{+2.8}_{-1.9}\times10^{11} & \nodata              & \nodata           & 27.8/34   \\
  & 59067^{+4}_{-6}   & \swift/XRT & 9.1^{+0.5}_{-0.4}\times10^{5} & 3.2^{+0.5}_{-0.4}\times10^{11} & \nodata              & \nodata           & 62.0/59   \\
  & 59094^{+12}_{-9}  & \swift/XRT & 8.9^{+0.2}_{-0.2}\times10^{5} & 3.8^{+0.3}_{-0.3}\times10^{11} & \nodata              & \nodata           & 89.4/84   \\
  & 59131^{+18}_{-18} & \swift/XRT & 5.1^{+0.3}_{-0.3}\times10^{5} & 1.1^{+0.2}_{-0.2}\times10^{12} & \nodata              & \nodata           & 30.7/36   \\
  & 59189^{+18}_{-18} & \swift/XRT & 7.9^{+0.5}_{-0.4}\times10^{5} & 3.1^{+0.6}_{-0.5}\times10^{11} & \nodata              & \nodata           & 39.4/51   \\
  & 59278^{+9}_{-8}   & \swift/XRT & 7.9 \times10^{5}({\rm fixed})  & 2.7^{+0.2}_{-0.2}\times10^{11} & 0.31^{+0.69}_{-0.12} & 3.0^{+0.4}_{-0.4} & 70.4/90   \\
  & 59349             & \xmm       & 7.9 \times10^{5} ({\rm fixed}) & 1.5^{+0.8}_{-0.2}\times10^{11} & 0.44^{+0.18}_{-0.24} & 2.9^{+0.1}_{-0.2} & 56.4/56   \\
  & 59373^{+12}_{-11} & \swift/XRT & 7.9 \times10^{5} ({\rm fixed}) & 2.0^{+0.2}_{-0.2}\times10^{11} & 0.48^{+0.25}_{-0.15} & 2.6^{+0.2}_{-0.2} & 117.4/126 \\
  & 59712             & \xmm       & 7.9 \times10^{5} ({\rm fixed}) & 2.9^{+3.3}_{-0.2}\times10^{11} & 0.12^{+0.07}_{-0.06} & 3.1^{+0.8}_{-0.5} & 12.2/16   \\
\hline
 \multirow{12}{*}{AT2021ehb} & 59407^{+10}_{-10} & \swift/XRT & 1.3^{+0.2}_{-0.1}\times10^{6} & 2.5^{+0.6}_{-0.3}\times10^{10}  & 0.04^{+0.04}_{-0.01} & 1.4^{+1.3}_{-0.4} & 78.8/98   \\
  & 59430             & \xmm       & 6.0^{+0.2}_{-0.2}\times10^{5} & 1.4^{+0.2}_{-0.1}\times10^{12}  & 0.06^{+0.02}_{-0.01}  & 4.4^{+0.2}_{-0.2} & 48.7/45   \\
  & 59441^{+15}_{-16} & \swift/XRT & 6.4^{+1.1}_{-0.6}\times10^{5} & 3.8^{+1.8}_{-1.3}\times10^{11}  & 0.21^{+0.10}_{-0.12}  & 4.1^{+0.3}_{-0.5} & 153.2/156 \\
  & 59475^{+12}_{-12} & \swift/XRT & 9.3^{+0.5}_{-0.5}\times10^{5} & 1.5^{+0.2}_{-0.2}\times10^{11}  & 0.13^{+0.03}_{-0.02}  & 2.7^{+0.2}_{-0.2} & 262.8/250 \\
  & 59509^{+12}_{-16} & \swift/XRT & 1.2^{+0.1}_{-0.1}\times10^{6} & 7.6^{+1.4}_{-1.0}\times10^{10}  & 0.20^{+0.04}_{-0.03}  & 2.2^{+0.1}_{-0.1} & 285.4/306 \\
  & 59539^{+10}_{-10} & \swift/XRT & 1.5^{+0.1}_{-0.1}\times10^{6} & 4.0^{+0.8}_{-0.6}\times10^{10}  & 0.27^{+0.04}_{-0.04}  & 1.9^{+0.1}_{-0.2} & 285.4/359 \\
  & 59565^{+8}_{-9}   & \swift/XRT & 1.0^{+0.1}_{-0.1}\times10^{6} & 1.3^{+0.7}_{-0.3}\times10^{11}  & 0.59^{+0.04}_{-0.05}  & 2.0^{+0.1}_{-0.1} & 401.2/467 \\
  & 59585^{+7}_{-7}   & \swift/XRT & 1.3^{+0.2}_{-0.2}\times10^{6} & 9.7^{+4.1}_{-2.5}\times10^{10}  & 0.59^{+0.07}_{-0.06}  & 1.9^{+0.1}_{-0.1} & 455.9/476 \\
  & 59604             & \xmm       & 1.2^{+0.1}_{-0.1}\times10^{6} & 8.6^{+0.7}_{-0.6}\times10^{10}  & 0.13^{+0.02}_{-0.02}  & 2.8^{+0.1}_{-0.1} & 77.6/79   \\
  & 59640^{+7}_{-7}   & \swift/XRT & 1.1^{+0.1}_{-0.1}\times10^{6} & 5.6^{+1.5}_{-0.9}\times10^{10}  & 0.20^{+0.05}_{-0.04}  & 2.3^{+0.2}_{-0.2} & 183.5/214 \\
  & 59661^{+7}_{-7}   & \swift/XRT & 3.9^{+1.3}_{-1.2}\times10^{5} & 8.3^{+22.9}_{-5.1}\times10^{11} & 0.31^{+0.15}_{-0.09}  & 3.5^{+0.3}_{-0.3} & 66.5/85   \\
  & 59825             & \xmm       & 1.6^{+0.2}_{-0.1}\times10^{6} & 2.8^{+0.6}_{-0.5}\times10^{10}  & 0.22^{+0.04}_{-0.04}  & 2.2^{+0.1}_{-0.1} & 77.2/71   \\
\hline
 AT2021yzv & 59654 & \xmm & 1.0^{+0.6}_{-0.3}\times10^{6} & 1.5^{+13.2}_{-0.9}\times10^{11} & \nodata & \nodata & 2.9/3 \\
\hline
\multirow{6}{*}{ASASSN-14li}& 56997 & \xmm & 4.1^{+0.1}_{-0.1}\times10^{5} & 3.7^{+0.1}_{-0.1}\times10^{12} & \nodata & \nodata & 66.3/31 \\
 & 57023 & \xmm & 4.0^{+0.1}_{-0.1}\times10^{5} & 3.8^{+0.1}_{-0.1}\times10^{12} & \nodata & \nodata & 98.9/25 \\
 & 57213 & \xmm & 3.3^{+0.1}_{-0.1}\times10^{5} & 4.2^{+0.1}_{-0.1}\times10^{12} & \nodata & \nodata & 21.6/19 \\
 & 57399 & \xmm & 3.0^{+0.1}_{-0.1}\times10^{5} & 4.9^{+0.2}_{-0.2}\times10^{12} & \nodata & \nodata & 26.5/18 \\
 & 57726 & \xmm & 2.7^{+0.1}_{-0.1}\times10^{5} & 4.8^{+0.3}_{-0.3}\times10^{12} & \nodata & \nodata & 22.7/12 \\
 & 58092 & \xmm & 2.3^{+0.1}_{-0.1}\times10^{5} & 6.0^{+0.6}_{-0.5}\times10^{12} & \nodata & \nodata & 18.3/13 \\
\hline
 \multirow{2}{*}{ASASSN-15oi} & 57324 & \xmm & 5.2^{+0.1}_{-0.5}\times10^{5} & 5.4^{+0.4}_{-1.6}\times10^{11} & \nodata & \nodata & 12.7/7  \\
 & 57482 & \xmm & 3.3^{+0.3}_{-0.1}\times10^{5} & 3.3^{+1.8}_{-0.1}\times10^{12} & \nodata & \nodata & 15.8/10 \\
\hline
  \multirow{9}{*}{AT2018fyk}  & 58404^{+13}_{-21} & \swift/XRT & 1.3^{+0.1}_{-0.1}\times10^{6} & 1.1^{+0.2}_{-0.1}\times10^{11} & 0.01^{+0.04}_{-0.01} & 1.9^{+1.5}_{-0.9} & 101.5/108 \\
 & 58436^{+23}_{-16} & \swift/XRT & 1.2^{+0.1}_{-0.1}\times10^{6} & 1.4^{+0.4}_{-0.2}\times10^{11} & 0.05^{+0.06}_{-0.03} & 2.4^{+0.6}_{-0.6} & 128.1/117 \\
& 58461             & \xmm       & 1.2^{+0.1}_{-0.1}\times10^{6} & 1.5^{+0.1}_{-0.1}\times10^{11} & \nodata              & \nodata           & 32.5/31   \\
 & 58476^{+15}_{-12} & \swift/XRT & 1.3^{+0.1}_{-0.1}\times10^{6} & 1.1^{+0.3}_{-0.2}\times10^{11} & 0.01^{+0.02}_{-0.01} & 1.4^{+0.7}_{-0.5} & 112.9/103 \\
 & 58598^{+34}_{-35} & \swift/XRT & 1.7^{+0.1}_{-0.1}\times10^{6} & 9.1^{+0.9}_{-0.7}\times10^{10} & 0.19^{+0.03}_{-0.02} & 2.1^{+0.1}_{-0.1} & 428.0/470 \\
 & 58681^{+30}_{-45} & \swift/XRT & 1.4^{+0.1}_{-0.1}\times10^{6} & 1.2^{+0.3}_{-0.2}\times10^{11} & 0.27^{+0.05}_{-0.04} & 2.3^{+0.1}_{-0.1} & 372.8/354 \\
 & 58745^{+34}_{-28} & \swift/XRT & 1.8^{+0.3}_{-0.3}\times10^{6} & 4.7^{+2.2}_{-1.1}\times10^{10} & 0.13^{+0.09}_{-0.05} & 1.8^{+0.3}_{-0.3} & 165.6/194 \\
 & 58783             & \xmm       & 1.3^{+0.1}_{-0.1}\times10^{6} & 1.1^{+0.1}_{-0.1}\times10^{11} & 0.16^{+0.01}_{-0.01} & 2.2^{+0.1}_{-0.1} & 160.4/136 \\
 & 58822^{+35}_{-38} & \swift/XRT & 2.1^{+0.2}_{-0.2}\times10^{6} & 4.0^{+1.0}_{-0.6}\times10^{10} & 0.14^{+0.06}_{-0.04} & 1.7^{+0.2}_{-0.2} & 221.4/282 \\
\hline
  \multirow{3}{*}{OGLE16aaa} & 57548           & \xmm       & 5.6^{+0.9}_{-0.8}\times10^{5} & 4.1^{+4.6}_{-1.6}\times10^{11} & \nodata & \nodata & 11.1/16 \\
  & 57558^{+1}_{-1} & \swift/XRT & 5.6^{+0.9}_{-0.7}\times10^{5} & 2.5^{+2.1}_{-1.0}\times10^{12} & \nodata & \nodata & 9.2/15 \\
  & 57722           & \xmm       & 5.5^{+0.2}_{-0.2}\times10^{5} & 1.4^{+0.2}_{-0.1}\times10^{12} & \nodata & \nodata & 12.0/15 \\
\enddata
\tablecomments{a) $\chi^2$-statistics for \xmm spectra, and `c-statistics' \citep{Cash1979} for \swift/XRT spectra.}
\end{deluxetable*}
\end{document}